\documentclass[preprints,review,accept,oneauthor,pdftex,10pt,a4paper]{mdpi} 

\firstpage{1} 
\pubvolume{xx}
\issuenum{1}
\articlenumber{1}
\pubyear{2018}
\copyrightyear{2018}
\externaleditor{Academic Editor: name}
\history{Received: date; Accepted: date; Published: date}

\usepackage[T1]{fontenc}
\usepackage{ae,aecompl}

\usepackage{graphicx}	
\usepackage{amsmath}	
\usepackage{amssymb}	
\usepackage{bm}
\usepackage{subfig}
\usepackage{times}

\usepackage{adjustbox}
\usepackage{array}

\newcolumntype{R}[2]{%
    >{\adjustbox{angle=#1,lap=\width-(#2)}\bgroup}%
    l%
    <{\egroup}%
}

\Title{Binary neutron star and short gamma-ray burst simulations in light of GW170817}

\Author{Antonios Nathanail $^{1,\dagger,\ddagger}$\orcidA{}}

\AuthorNames{Antonios Nathanail}

\address{%
$^{1}$ \quad Institut f\"ur Theoretische Physik, Goethe Universit\"at Frankfurt,
Max-von-Laue-Str.1, 60438 Frankfurt am Main, Germany; nathanail@th.physik.uni-frankfurt.de\\
}

\abstract{
In the dawn of the multi-messenger era including gravitational waves, 
which was marked by the first ever
coincident detection of gravitational waves and electromagnetic radiation
it is important to lay back and think about established knowledge. 
Numerical simulations 
of binary neutron star mergers and simulations of short GRB jets have to combine
efforts in order to understand such complicated and phenomenologicaly rich
explosions. We review the status of numerical relativity simulations with 
respect to any jet  or  magnetized outflow produced after merger. 
We compare what is known from such simulations, with what is used and obtained 
from short GRB jet simulations propagating through the BNS ejecta. 
We point out facts that are established and can be considered known, 
and things that need to be further revised and/or clarified.
}


\begin{document}

\setcounter{section}{-1} 

\section{Introduction}

The detection of GW170817 marked the dawn of the multi-messenger 
gravitational-wave era \citep{Abbott2017, Abbott2017b}. 
The subsequent observation of a  short 
gamma-ray burst (GRB) almost $\sim 1.7$ seconds after merger 
\citep{Goldstein2017,Savchenko2017}, showed that a least a subset 
of short GRBs is produced by binary neutron star (BNS) mergers. Hours 
after merger a precise localization could be established through 
optical observations of GW170817 \citep{Coulter2017,Soares-Santos2017}, 
identifying the host as galaxy NGC 4993, which is 
at a distance of 40 megaparsecs. Further detection in 
UV/optical/Infrared established perennially the connection of 
BNS mergers with a kilonova (macronova) 
\citep{Soares-Santos2017,Arcavi2017,Nicholl2017,Pian2017,
Smartt2017,Tanvir2017,Ustumi2017,
Kilpatrick2017,Kasliwal2017,Covino2017,Cowperthwaite2017,
Buckley2018,Drout2017,Evans2017,Arcavi2018}. 

A coincident detection of GW and a short GRB from a BNS merger
 was  long ago conjectured that short-duration GRBs 
come from BNS mergers \citep{Eichler89,Narayan92,Mochkovitch93}.
These unprecedented observations open new windows and insights for 
the in depth study of such objects and such events. 
The observation opened the  possibility of 
constraining the maximum mass of neutron stars  
and the equation of state (EOS)
\citep{Annala2017,Bauswein2017b,Margalit2017,Radice2017b,
Rezzolla2017,Ruiz2017,Shibata2017c,De2018,Tews2018a,Tews2018,Alsing2017,
Burgio2018,Raithel2018,Paschalidis2017}.

It was long proposed that a BNS merger would give rise to 
emission powered by the radioactive decay of 
r-process nuclei \citep{Lattimer74,Li:1998}.
Several groups concluded that this was the case 
for the optical/NIR emission that followed GW170817 
\citep{Kasen2017,Drout2017,Tanaka2017,Kasliwal2017,Murguia-Berthier2017, 
Waxman2017,Villar2017, Tanvir2017, Utsumi2017,Perego2017,Metzger2018}.
This observation triggered  further modeling for the actual components that give rise to
this emission and how these components were produced. 

The prompt gamma-ray emission was reported in \citep{Goldstein2017,Savchenko2017}.
It was the most faint (short or long) GRB ever detected \citep{Goldstein2017}.
The first X-ray afterglow observations came nine days after merger 
\citep{Evans2017,Troja2017,Margutti2017}. 
The first radio counterparts came later, sixteen 
days after merger \citep{Hallinan2017,Alexander2017}.  
 Every information that would come from 
the afterglow observations would be invaluable to reveal the 
nature of the outflow and its structure. 
A relativistic outflow from a BNS merger was indeed observed
 \citep{Alexander2017,Haggard2017}.
Was that the most peculiar short GRB ever detected \citep{Kasliwal2017,Granot2017}?
The continuous rising of the 
afterglow the first 100 days suggested that a simple top-hat\footnote{A top-hat jet is one 
with constant Lorentz factor and emissivity  within the jet that goes sharply to zero 
outside of jet opening angle. It is the simplest model to explain GRBs
have been widely used to explain GRB properties.} jet model 
seen off-axis is not adequate for explanation\cite{Alexander2017,Mooley2018,Ruan2018,Pooley2018,
Margutti2018,Lyman2018}.
However, at a 100 days after merger the data could not exclude other jet 
structure or cocoon models. Energy injection was evident at that time 
\citep{Li2018}.
Then, a turnover in the light curve appeared close to 200 days 
\citep{Dobie2018,Alexander2018,Nynka2018}. 
This emission 
is well understood and comes  from the interaction of the outflow as it smashes 
into the inter-stellar medium producing a shock which accelerates 
electrons that radiate synchrotron radiation and can give a 
great insight in the whole structure of the initial outflow.

In order to digest all these new insightful observations, 
and the yet to come in the next years we have  to combine 
all pieces available. What has been achieved  from BNS numerical 
relativity simulations has to be part of any adequate modeling 
of short GRB outflows. These are:
the ejected matter and the production of neutrino driven winds,
the enormous magnetic field evolved in the merger process, 
and its amplification during merger, 
the actual possibility of launching a 
relativistic outflow after merger are the starting points 
given by numerical relativity.
Is it a stable  magnetar or the collapse to a black hole (BH)
torus system that powers an outflow?  In what follows we try 
to present results from numerical relativity BNS simulations 
relevant for short GRBs. Afterwards, we turn our attention to 
efforts in short GRB jet simulations propagating through 
the BNS ejecta.

This is a rather focused review on what we know from numerical 
relativity concerning short GRBs and and how this knowledge is applied 
to short GRB simulations. It will not at all 
follow the path of excellent reviews that exist in the subject of BNS 
mergers. For the interested reader we cite several detailed 
reviews of subjects relevant to the detection of a BNS merger, 
a short GRB and a kilonova.
Detailed reviews of all the aspects of numerical relativity and 
its applications to BNS mergers \citep{Faber2012:lrr,Baiotti2016},
a focused review on BH - neutron star binaries 
\citep{Shibata06c}, a review discussing the connection between BNS mergers and short GRBs 
in numerical relativity results \citep{Paschalidis2016}, 
observational aspects of short GRBs and connection 
to BNS mergers \citep{Berger2013b,Fong2015}, regarding BNS merger 
and electromagnetic counterparts from kilonova 
\citep{Rosswog2015,Fernandez2015b,Thielemann2017b,Metzger2017}, 
a review on rotating stars in relativity with applications on the 
post merger phase \citep{Paschalidis2017b},
reviews for short GRBs \citep{Nakar:2007yr,Lee:2007js} and detailed 
reviews on all aspects of GRBs \citep{Piran:2004ba,Meszaros:2006rc,Kumar2014}.
In section \ref{sec:BNS} we review the relevant knowledge 
from BNS simulations. We mainly follow results from 
magnetohydrodynamic (MHD) simulations in BNS studies. 
At the end of section \ref{sec:BNS} we show the different 
paths that a BNS may follow after merger with respect 
to the achieved magnetic energy growth during merger. This translates 
to the total  mass of the binary. In section \ref{sec:shortGRB} 
we follow the studies that focus on  the interaction of a BNS 
relativistic outflow passing through the matter that has been 
ejected during merger. 
In section \ref{sec:conc} we conclude.

\section{BNS numerical simulations}
\label{sec:BNS}

Sixteen systems of double neutron stars have been observed in our galaxy. 
The observational data for the total mass of double neutron stars from our galaxy 
show a narrow distribution in the range 
$2.58-2.88 \, M_{\odot}$ \citep{Tauris2017}. 
A double neutron star system will inspiral and emit 
gravitational waves that result in the orbital decay, shrinking 
their separation. When they come close enough, tidal forces 
result in deformation of the shape of the two neutron stars. 
Only numerical relativity can adequately describe the 
inspiral process from then on.

When the two neutron stars  contact each other a merger 
product is formed. If this is massive enough that  
cannot support itself from gravitational collapse, 
a BH is formed in the first millisecond after merger 
surrounded by a negligible disk. If the configuration 
is less massive, it does not collapse straight away.
The merger product is differentialy 
rotating and thus it can support  more mass than 
the limit for a uniformly rotating star. At this stage the 
merger product is called a hyppermassive neutron star
(HMNS) \citep{Baumgarte00bb}. Gravitational-wave emission and magnetic field 
instabilities can remove angular momentum and 
make the HMNS unstable triggering its collapse 
and leaving a BH with a surrounding disk. The loss of 
thermal pressure due to neutrino cooling could also trigger 
the collapse of the HMNS \citep{Sekiguchi2011,Paschalidis2012},
however see \citep{Kaplan2014}.
However, if the total mass of the object is smaller than 
the mass that can be supported when allowing for maximal
uniform rotation -- the supramassive limit -- it can also 
lose differential rotation and not collapse. 
This would result to a uniformly rotating super massive neutron star
(SMNS) surrounded by a disk. The SMNS will continue to loose angular momentum 
through magnetic spin down and also accrete mass from the surrounding 
disk. Its lifetime varies from a second to millions of seconds, in the latter 
case it can be considered as a stable configuration. 

A robust picture regarding the ejected matter during and 
after merger has been drawn from numerical simulations. 
These include dynamical ejecta during merger and secular 
ejecta that follow, like neutrino driven winds and magnetic winds
\cite{Rosswog1999,Aloy:2005,Dessart2009,Rezzolla:2010,Roberts2011,
Kyutoku2012,Rosswog2013a,Bauswein2013b,
Hotokezaka2013,Foucart2014,Siegel2014,
Wanajo2014,Sekiguchi2015,Radice2016,Sekiguchi2016,Lehner2016,
Siegel2017,Dietrich2017,Bovard2017,Fujibayashi2017,Fujibayashi2017b}.
Other important properties of the merger product 
such as the spin and the rotation profile have been studied
\citep{Kastaun2014,Kastaun2017,Hanauske2016}.  
We continue focusing on the properties of the magnetic field, 
its amplification during merger and all the variety of observational outcomes 
that depend on the collapse time of the merger product and are dictated 
by the magnetic field.

\textbf{Magnetic Field Amplification}
\begin{figure*}
  \begin{center}
    \includegraphics[width=0.6\columnwidth]{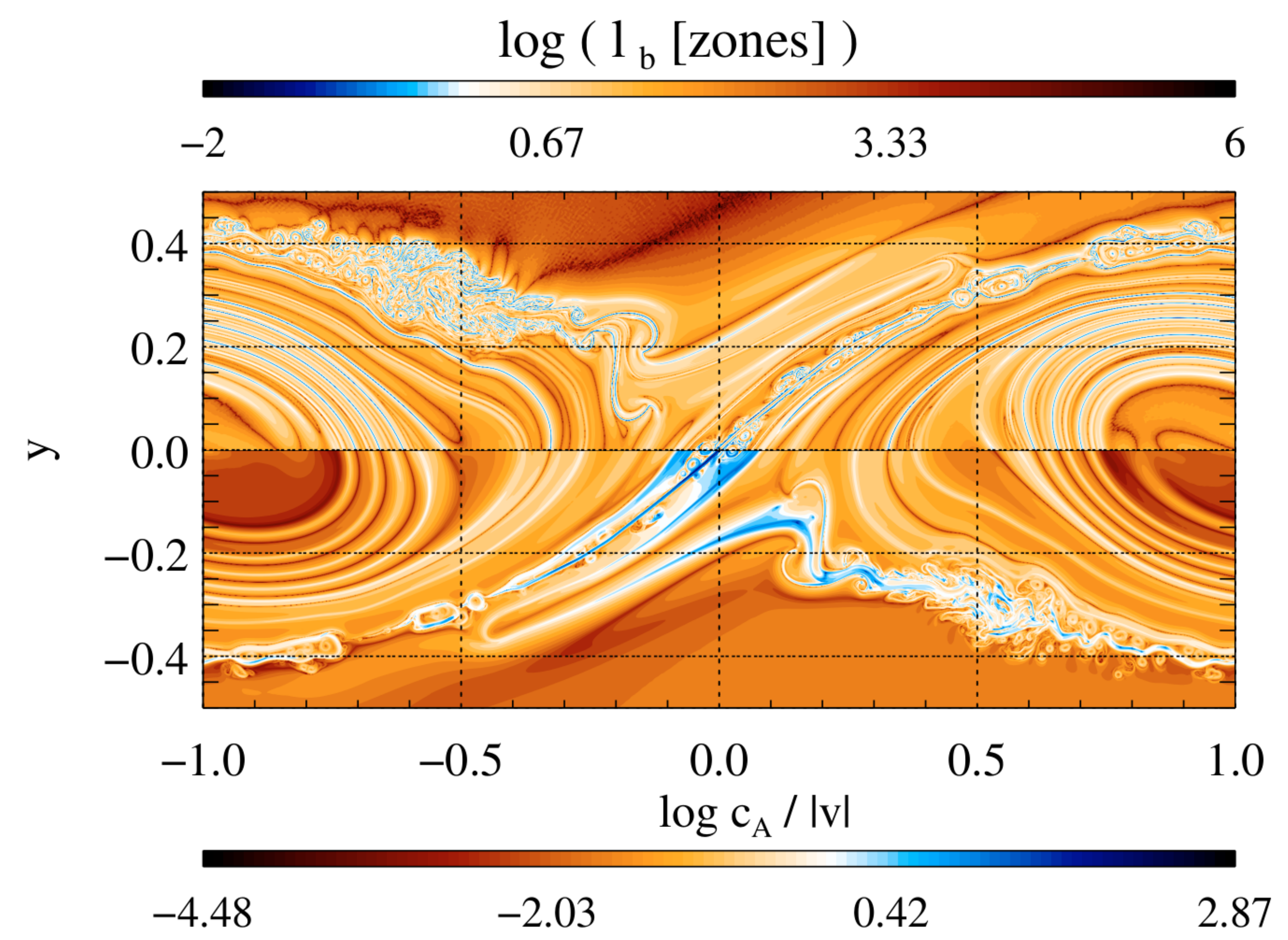}
  \end{center}
 \caption{Snapshots of a certain model  from \citep{Obergaulinger10}. It is taken 
 shortly after  the termination of the kinetic amplification phase
The top panel shows the characteristic length scale of the magnetic field, 
$|B|/|\nabla \times B|$ in units of the zone size. Regions where 
magnetic structures are larger than one computational zone are depicted 
in orange-red colors and blue colors where they are smaller. 
The bottom panel shows the  ratio between the Alfvén
velocity and the modulus of the fluid velocity. Strongly magnetized regions are 
depicted in blue, whereas weakly magnetized in orange-red.
(Reproduced with permission from \citep{Obergaulinger10}, 
$\copyright$ ESO, 2010)}
  \label{fig:ober}
\end{figure*}
The importance of the  Kelvin-Helmholtz instability in BNS mergers was pointed out 
by Rasio \& Shapiro \citep{Rasio99}. As the stellar surfaces come into contact, 
a vortex sheet (shear layer) is developed which 
is Kelvin-Helmholtz (KH) unstable. The first simulation reporting on the KH 
instability for BNS was \citep{Rosswog02}. 
It was reported in \citep{Price06} that the KH 
instability could amplify the magnetic field beyond the magnetar level. 
They reported a lower value of $2\times 10^{15} \, {\rm G}$. However, 
they mentioned that numerical difficulties do not allow to reach the realistic values 
of amplification, which could be far above this limit.
To address the full problem in numerical relativity is not so easy, 
the reason is that high-resolution simulations are necessary, since 
the KH instability growth rate is proportional to the wave number of 
the mode, the shortest wavelengths grow the most rapidly. 
Studies of BNS mergers tried to clarify the picture 
and indeed showed some amplification, yet the saturation level 
was not pinpointed \citep{Liu:2008xy,Anderson2008,Giacomazzo2011b,
Giacomazzo:2014b,Dionysopoulou2015,Palenzuela2015,Kiuchi2014}. 
Another approach  is local simulations that can imitate 
the conditions of shear layers and study in detail the different phases of 
such procedure. The growth phase, where the KH vortex is formed, 
the amplification phase where the magnetic field is wound up by the 
evolving KH vortex and the last phase where the magnetic field has 
locally reached equipartition that results in the KH vortex to lose 
its energy. In figure \ref{fig:ober} such configuration is depicted 
after the end of the amplification phase. The blueish regions in the 
lower panel of figure \ref{fig:ober} indicate strongly magnetized 
regions that occur after amplification.
 Local simulations 
 do not have such stringent resolution 
limitations as the global ones \citep{Obergaulinger10,Zrake2013b}.

%
\begin{figure*}
  \begin{center}
    \includegraphics[width=0.6\columnwidth]{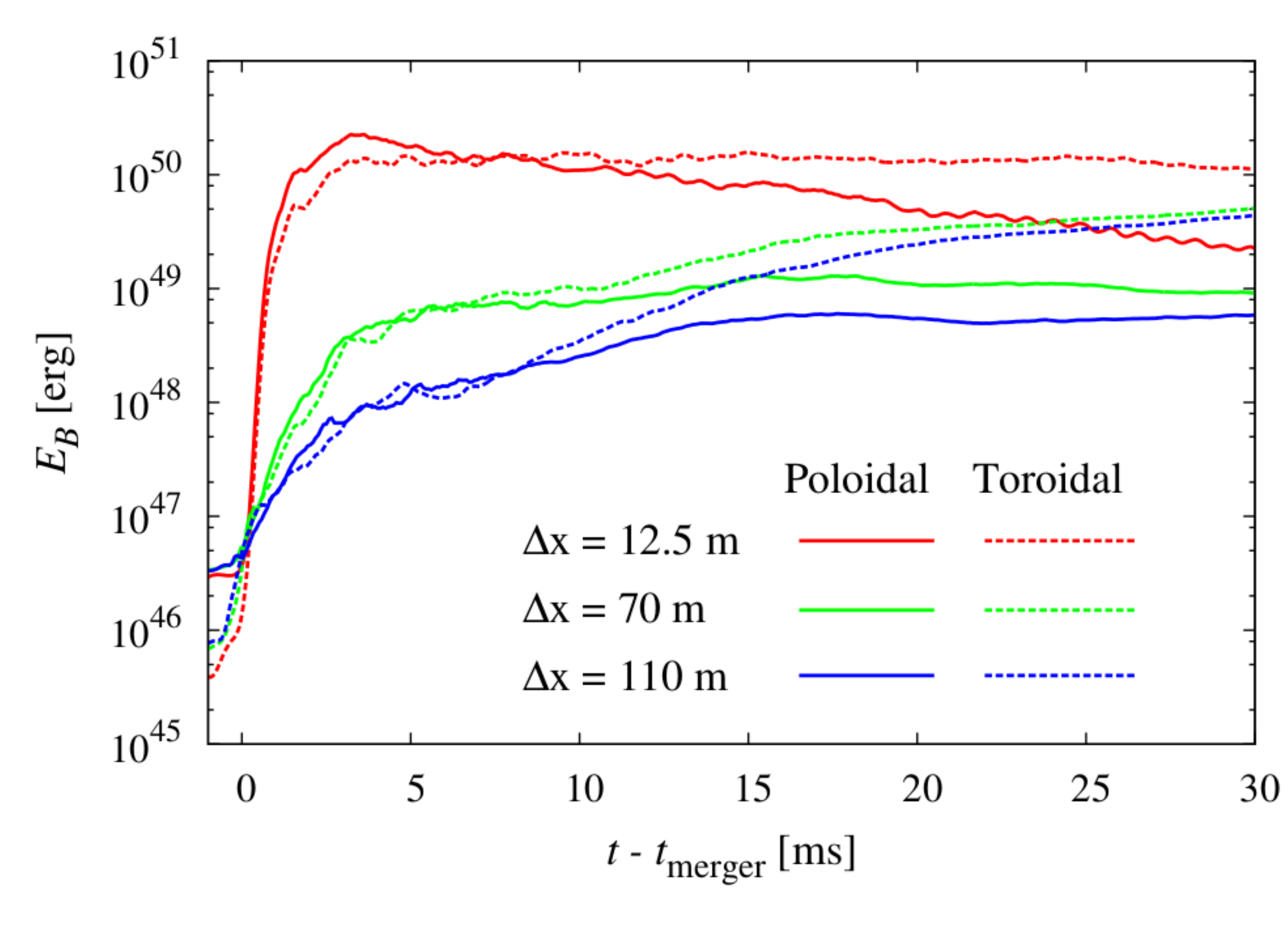}
  \end{center}
 \caption{The evolution of the magnetic-field energy  as a function of time 
 from \citep{Kiuchi2017}. The growth of the magnetic field is evident in the first 
 five milliseconds. However, the strong dependence to resolution indicates 
 that the upper limit of amplification is not yet known. 
Solid and dashed curves indicate the poloidal and toroidal magnetic field 
components, respectively.
(Reprinted with permission from \citep{Kiuchi2017}. 
$\copyright$ (2018) by the American Physical Society.)}
  \label{fig:kiu}
\end{figure*}
A high resolution study by Kiuchi \citep{Kiuchi2015a} showed that for an initial 
maximum magnetic field of $ 10^{13} \, {\rm G}$, the maximum magnetic field 
during merger and in the first $4-5 \, {\rm ms}$ can reach $ 10^{17} \, {\rm G}$.
They showed that the saturation magnetic energy is above $\gtrsim 4\times 
10^{50}\, {\rm erg}$, which is $\gtrsim 0.1\%$ of the bulk kinetic energy. 
Going to even higher resolution and running for  longer time,  
the upper bound for the amplified magnetic energy has not been reached yet.
Higher values of the amplified magnetic energy live in denser regions 
\citep{Kiuchi2017}. This may indicate that the higher values  
of the magnetic field are either trapped in the dense core, 
or that they need a diffusion timescale to diffuse out from the 
core and reorder \citep{Harutyunyan2018}.
These results have built stable foundations that magnetic field
amplification is an integral part of the BNS merger 
and happen in the first millisecond after merger as seen in 
figure \ref{fig:kiu}. Another 
point to make here is that this is true only if the binary does not 
experience a prompt collapse, in which case there is no time to 
amplify the magnetic field and the EM output of the remnant follows a
different path, we focus on this later in more detail. 

Another cause for magnetic field amplification is magnetic 
winding due to differential rotation which continues 
to function even after the KH may saturate. Furthermore,
there are indications from studies of core-collapse supernovae 
that also the magneto-rotational instability (MRI) 
can be important \citep{Moesta2013}.
From such simulations it 
has been shown that the MRI can amplify the magnetic field 
by a factor of 100. The importance of parasitic instabilities 
that may quench such mechanisms have also to be taken 
into account \citep{Rembiasz2016}.

\textbf{Observational signatures during magnetic field amplification}
Are there direct observational signatures of the field amplification?
The magnetic energy increases in extreme values. It has been proposed that if 
only a fraction of this energy  dissipates by reconnection 
it yields a EM counterpart at the time of merger. This 
could be observable to a distance of 200 Mpc \citep{Zrake2013}.
This radiation can only escape if produced in an 
optically thin surface layer. However, the higher values for 
amplification were reported in the dense core of the 
merger remnant \citep{Kiuchi2017}. The evolution of this 
turbulent magnetic field is not yet fully understood,  
it may take  a much larger amount of time 
than the merger timescale to diffuse from the dense 
core \citep{Harutyunyan2018}. If the merger 
remnant lives for at least a second, then the Hall effect becomes 
important, and  would depict the structure of the 
magnetic field at late times \citep{Harutyunyan2018}.  
\begin{figure*}
  \begin{center}
    \includegraphics[width=0.4\columnwidth]{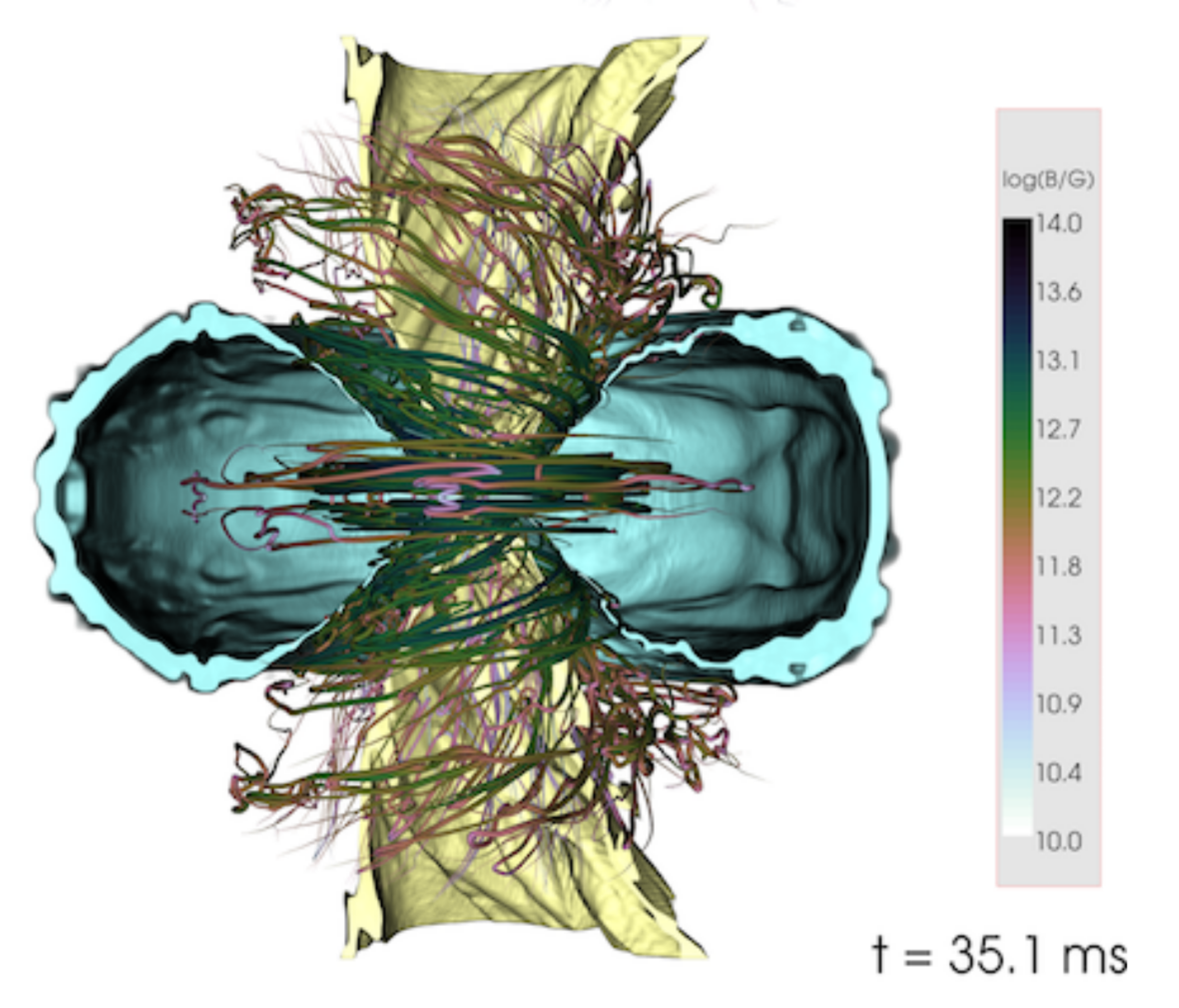}
  \end{center}
 \caption{The magnetic field structure for a model from \citep{Kawamura2016}.
 It is depicted at $35.1 {\rm ms}$ after merger. Two isosurfaces of density 
 are shown in yellow ($10^{8} {\rm g/cm^3}$) and cyan $10^{10} {\rm g/cm 3})$. 
 The field lines are colored indicating magnetic field strength. The toroidal field 
 inside the torus is easily seen, together with a poloidal funnel 
 above the BH. This model collapsed to a BH at $t_{BH}\sim 8.7 {\rm ms}$ after 
 merger. Due to resolution the KH instability is not entirely 
 accounted for in these simulations. (Reprinted with permission from
 \citep{Kawamura2016}. 
$\copyright$ (2016) by the American Physical Society.)}
  \label{fig:kawa}
\end{figure*}

\textbf{BH torus from BNS in MHD}
Strong magnetic fields are present during and after the 
merger of a BNS. The next meaningful ingredient is 
the outcome and lifetime of the merger remnant. Due to numerical 
limitations, existing studies cover the collapse 
of the merger remnant to a BH only if it happens 
before $\sim 100 {\rm ms}$ after merger. 
It was long ago proposed that BNS mergers could launch 
a short GRB. This  connection was made clear  by the  recent observations 
\citep{Alexander2017,Haggard2017}.
However, it is still something that should be 
achieved by global simulations. 
The first attempts  in a magnetohydrodynamic (MHD) concept in 
full GR, did not show any signs of a jet production
following merger and the collapse of the merger remnant 
\citep{Anderson2008,Liu:2008xy}.
In subsequent studies a magnetic jet structure, as called by the authors,
was reported, this is a low density funnel with ordered poloidal magnetic 
field above the BH \citep{Rezzolla:2011}. This is 
indeed the first step in order to 
imagine the production of a relativistic magnetized jet. Another 
important aspect, is that an ordered poloidal magnetic field 
is needed in order to account for energy extraction from 
the BH in a Blandford-Znajek way \citep{Blandford1977}.
However, even if the magnetic field is not poloidal there could be other 
outcomes for an outflow. Another simulation by a different group 
did not find such a structure, instead the picture drawn from their 
simulations indicated in an expanding toroidal field \cite{Kiuchi2014}, 
which is also capable of producing a kind of a jet with a different 
mechanism \citep{Contopoulos1995}.
%
%

It was further showed and confirmed in a resistive MHD framework that, 
at least for merger remnants that 
collapse in the first $\sim 10 {\rm ms}$, the BH-torus system 
produces a low density funnel above the BH \citep{Dionysopoulou2015}. 
This is usually reported as a low density region above the 
newly formed BH. But, how low is low?  In order to 
launch an outflow, it is at least needed that the magnetic pressure 
from the jet can push and accelerate this material in the polar 
region. The studies in \citep{Rezzolla:2011, Dionysopoulou2015} 
used an ideal fluid equation of state (EOS), whereas in 
\citep{Kiuchi2014} a piece-wise polytrope, as was pointed by 
\citep{Ruiz2016,Paschalidis2016}  the jet structure depends on 
the EOS.
Also other studies have reported the production of a 
magnetic structure even using a different 
EOS \citep{Kawamura2016,Endrizzi2016}, 
including also a neutrino treatment \citep{Palenzuela2015}. 
In figure \ref{fig:kawa} such a configuration 
with a BH-torus system is depicted. In this specific model the merger 
product collapsed to a BH at $t_{BH}\sim 8.7 {\rm ms}$. The 
snapshot is taken at $t\sim 35.1 {\rm ms}$ after merger. In the low density 
funnel above the BH the magnetic structure is clearly seen.

\begin{figure*}
  \begin{center}
    \includegraphics[width=0.4\columnwidth]{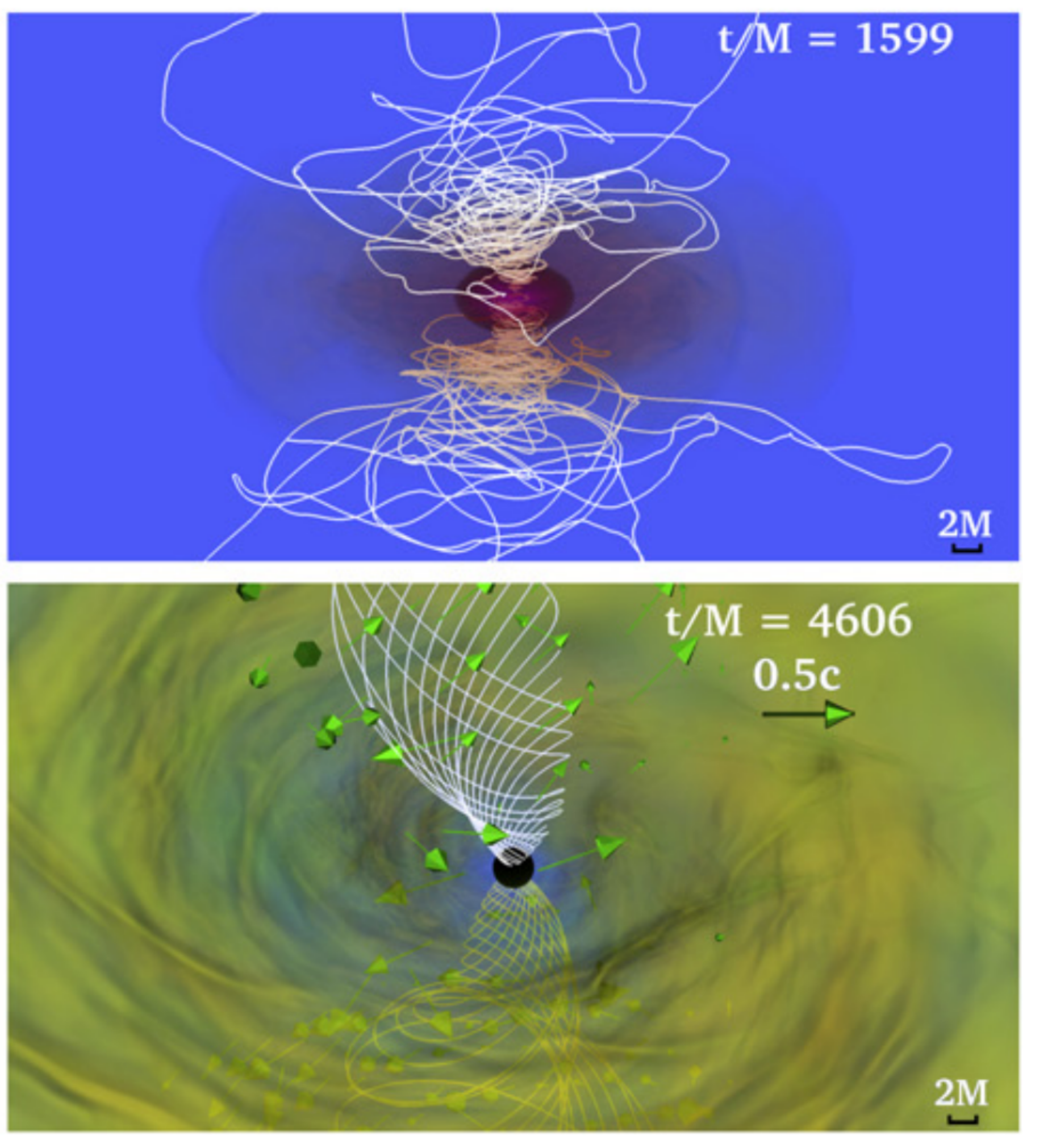}
  \end{center}
 \caption{Snapshots of the rest-mass density of a model from 
 \citep{Ruiz2016}. Magnetic field lines are depicted as white 
 lines and arrows indicate plasma velocities. In this model the 
 merger remnant collapses to a BH at $t_{BH}\sim 1215 M = 18 {\rm ms}$ 
 after merger. The upper panel is at a slightly later time after collapse, 
 whereas the lower panel is at $t\sim 67.7 {\rm ms}$. We have to point out that while the density contours are selected far from the 
 magnetic jet structure, the funnel is filled with low density matter which 
 supports the collimation of the magnetic structure.
 The length scale of the plots is $M=4.43 {\rm km}$.
 (Reprinted  from \citep{Ruiz2016}. 
 $\copyright$ AAS. Reproduced with permission.)}
  \label{fig:ruiz}
\end{figure*}

Lately,  a production of an incipient jet (as called by the authors)
was reported, which attained a Poynting luminosity of 
$\sim 10^{51} {\rm erg/s}$ and a maximum Lorentz factor 
of $\Gamma = 1.25$ \citep{Ruiz2016}. Towards the end of the
simulation they reported a magnetically dominated funnel above 
the BH, which can be seen in the lower panel 
of figure \ref{fig:ruiz}. The snapshot is taken 
at $t\sim 67.7 {\rm ms}$, whereas the merger product 
collapsed to a BH at $t_{BH}\sim 18 {\rm ms}$ after merger. 
It is clear that at late times the low density funnel above 
the BH is decreasing even more in density. That  allows
a magnetically dominated region to evolve. Using the magnetization of 
the outflow, they  estimated the half opening angle of 
the jet funnel to be $\sim 20^{\circ}-30^{\circ}$ \citep{Ruiz2016}.
%


\textbf{EM luminosity}
But why so much discussion about magnetic field and its activity 
on the production of jets. Other mechanisms have been proposed, 
such as neutrino annihilation \citep{Eichler89,Ruffert99b}. However 
recent studies in which neutrinos are also treated 
to study a BNS merger and the evolution
of accretion to a BH, it was found that  due to a very baryon-loaded 
environment such a mechanism alone does not suffice 
\citep{Just2016,Perego2017}.
In the other hand the electromagentic energy extraction from a BH 
(known as the Blandorf-Znajek mechanism) has been widely studied 
(numerically \citep{Komissarov2001,Komissarov2007c}, semi-analytically
\citep{Nathanail2014} and analytically \citep{Gralla2016}) and mostly 
understood and accepted. It needs only two ingredients, a rotating 
BH and an ordered poloidal magnetic field to extract
this rotational energy. 
\begin{equation}
\begin{aligned}
L_{BZ}\sim \frac{1}{6 \pi^2 c} \Psi_m^2 \Omega_{BH}^2
\sim B_p^2 R_{BH}^2 \left(\frac{\alpha}{M_{BH}}\right)^2 \\
\sim 10^{51} \left(\frac{B_p}{2\times 10^{15} {\rm G}}\right)^2 
\left(\frac{M_{BH}}{2.8 M_{\odot}}\right)^2 
\left(\frac{\alpha}{0.8M_{BH}}\right)^2
\end{aligned}
\label{eq:BZ}
\end{equation}
where $\Psi_m$ is the magnetic flux accumulated on the BH horizon, 
$\Omega_{BH}$ the angular velocity of the BH, $B_p$ the poloidal magnetic 
field on the BH horizon, $\alpha$ the spin parameter of the BH and 
$M_{BH}$ the mass of the BH \citep{Shapiro2017}.

In a BNS merger you have both, when
the merger remnant collapses, a  BH is formed and in all cases 
reported it attains a spin 
of $\sim 0.8$, and magnetic field is a known ingredient 
of a neutron star and as we have already discussed it is further amplified 
during merger. In a baryon polluted environment 
like the one that  exists around the remnant after merger, there are also 
other things to worry about. The ram pressure from the material 
of the polar regions, or even fall-back material in this region
may not allow this outflow to form and evolve. Maybe this is 
the reason, together with a low magnetic field, that in some studies with limited amount of 
evolution time no outflow was formed \citep{Kawamura2016}.
If this is the case, then it is expected that some hundreds 
of milliseconds later, the overall pressure of the funnel could 
decrease significantly to allow for a magnetically dominated 
outflow to emerge. 
\begin{figure*}
  \begin{center}
    \includegraphics[width=0.52\columnwidth]{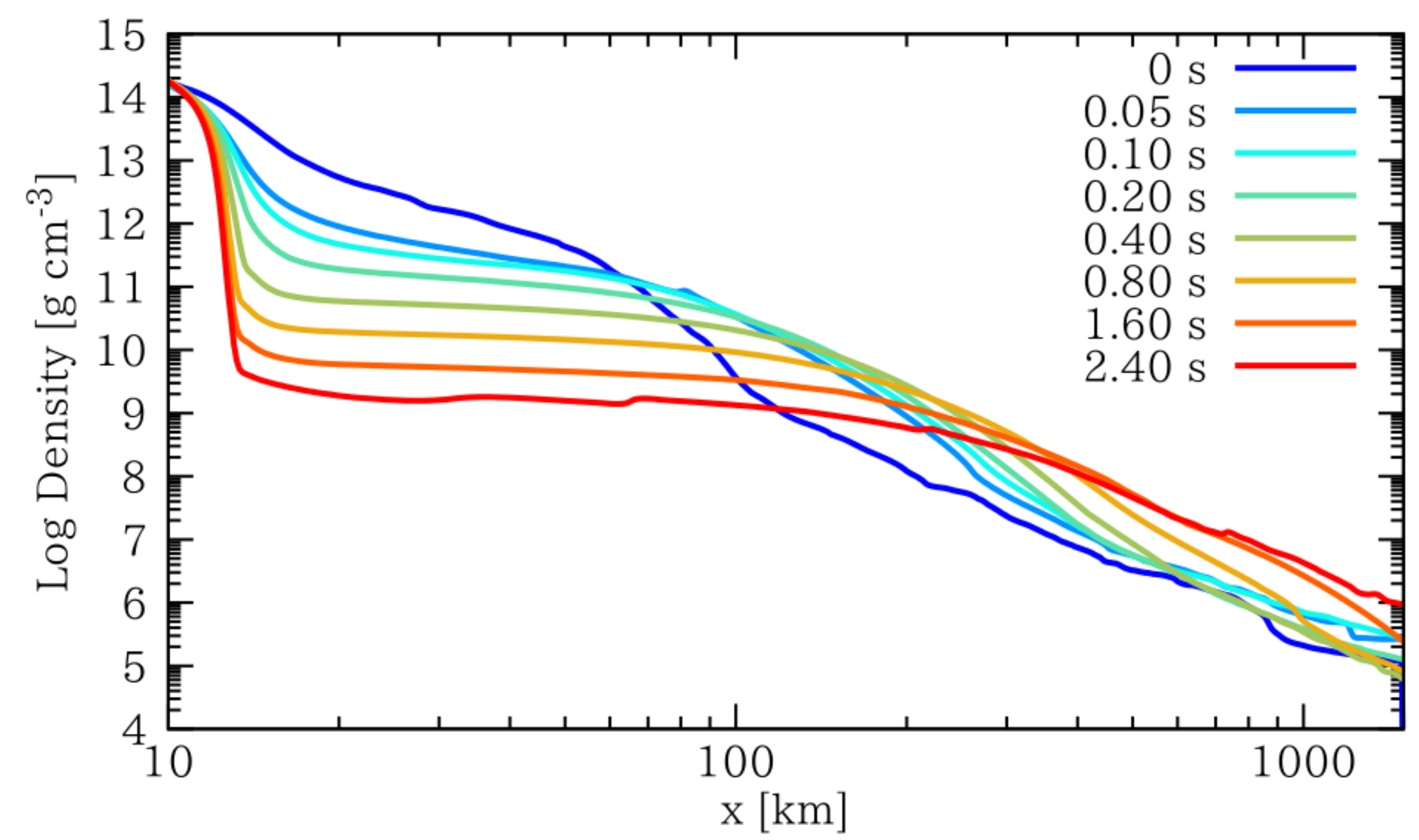}
  \end{center}
 \caption{Density profiles on the equatorial plane at
different time slices $t\sim 0,
0.05, 0.1, 0.2, 0.4, 0.8, 1.6,$ and $2.4 {\rm s }$. The torus gradually
expands with time and  the torus density decreases. This
is due to the  the viscous angular momentum transport. Even at one second after
merger the SMNS lives in a low density torus, compared to its nuclear densities. 
(Reprinted from \citep{Fujibayashi2017b}. 
 $\copyright$ AAS. Reproduced with permission.)}
  \label{fig:dens1}
\end{figure*}

\textbf{Duration of a BH torus}
Following the above discussion it is evident to ask how long 
this configuration will last. This is indicated by the mass of 
the surrounding disk plus the mass accretion rate.
We briefly discuss the duration  connected with mass of the torus. 
 It is usually assumed that the duration 
of the short GRB ($< 2 {\rm s}$) is due to accretion timescale 
of the surrounding torus. Studies have shown that the mass of the torus 
can be as large as $M_{T} \sim 0.001-0.2M_{\odot}$ 
\citep{Shibata:2003ga, Shibata06a, Baiotti08,Liu:2008xy, 
Rezzolla:2010df,	Kawamura2016, Fujibayashi2017b}.   Through 
numerical simulations, a simple phenomenological expression 
can be derived that reproduces the mass from the surrounding torus
\citep{Shibata06a,Rezzolla:2010}. A general result is that unequal mass 
binaries have a more massive torus around the BH formed. 
Whereas, equal mass binaries acquire less massive torus.
Of course, in the case of a prompt collapse the surrounding 
torus is negligible, but this is something we discuss after 
the accretion timescale comment. Furthermore, in the case of a late 
collapse the surrounding disk is expected to be negligible 
\citep{Margalit2015,Camelio2018}.

The duration of any event coming from the BH torus depends 
on the lifetime of the torus, and the torus will live on an accretion 
timescale. A rough estimate for the viscous 
accretion timescale can be given as:
\begin{equation}
\begin{aligned}
t_{accr} \simeq \, 1 \, \left( \frac{R_T}{50\,{\rm  km }}
\right)^2 
\left(\frac{H_T}{25 \,{\rm  km }}\right)^{-1}  
\left( \frac{\alpha}{0.01}\right)^{-1}\left(\frac{c_s}{0.1c}\right)^{-1}
{\rm  s } \,
\end{aligned}
\label{tac}
\end{equation}
where $R_T$ and  $H_T$ are the radius and the 
typical vertical scale height of the torus, $c_s$ the speed of sound and the 
$\alpha$ parameter \citep{Shakura1973}.
As such, if the BNS merger produces a BH torus system the accretion 
timescale sets the duration of the outflow, if any is produced. 
However, we should point here that it is also 
relevant to discuss the duration  of a gamma-ray pulse produced 
by a relativistic outflow in a different way. 
The photosphere  is defined as the radius that the outflow first 
becomes transparent and the first photons are emitted. 
If  an outflow has attained   a Lorentz factor $\Gamma$, 
then  photons emitted at any point on the jet are beamed 
within a $1/\Gamma$ cone, as seen in the lab frame. Thus, 
assuming that the outflow has a conical shape  with opening
angle $\theta_j$, initially when $\Gamma > 1/\theta_j$, 
an observer can see only  radiation from a small fraction of the jet.
The duration of the pulse can be interpreted as photons coming 
from this cone that the observer is able to see, the $1/\Gamma$ cone. 
For a mildly relativistic outflow with $\Gamma > 1/\theta_f$, 
the relevant timescale of the pulse is
\begin{equation}
\begin{aligned}
dt\sim \, 1-2 \, \left(\frac{r_{em}}{10^{12} {\rm cm}}\right)   
\left(  \frac{\Gamma}{6-10}\right)^{-2}  \, s \,
\end{aligned}
\end{equation}
where $r_{em}$ is the emission radius \citep{Piran:2004ba}.
The point here is that even if the accretion timescale is smaller
and a relativistic outflow is produced, then the duration can be 
provided also by other robust physical arguments. For an ultra relativistic 
outflow the duration of the pulse is very small and then 
the accretion timescale can enter as a justification of the 
duration of the event.

The discussion so far was mainly for a merger remnant that collapses 
to a BH after $10 ms$ or more. The effect of the collapse 
of the merger remnant when it occurs in the first milliseconds is different.
The general thinking in the community leads to no expectations 
for an EM counterpart, if the BNS merger undertakes 
a prompt collapse to a BH. This is based on results of 
simulations that showed some robust features of this evolution 
track, for the case of an equal mass binary. 
These features are, limited amount of mass is dynamically ejected and
no expectation for a kilonova what soever. Another feature is the limited amount 
of time between merger and collapse which does not allow for 
a significant  magnetic field amplification, as a result 
the magnetic energy will not 
reach such high values. However, a detailed high resolution study 
of a prompt collapse does not exist. 

Lastly, the limited amount of 
mass left around the BH can not sustain any magnetic structure 
for an amount of time more than few milliseconds. This means that 
whatever is formed after merger will be lost on this timescale.
However, the magnetic field that stays outside the BH will dissipate 
away on this timescale. Most of the matter will be lost behind the BH horizon, 
but the magnetic field lines will snap violently. This will produce 
a magnetic shock that will dissipate a significant fraction of the 
magnetic energy by accelerating electrons and produce a massive burst, 
similar to a blitzar \citep{Falcke2013}. 
This can produce an EM counterpart on such a timescale. 
Prompt collapse events produce less massive 
accretion disks than those arising from delayed collapse.
Studies have shown that the result of a prompt collapse is 
a spinning BH and a accretion disk with 
a negligible mass of $M_T \sim 0.0001-0.001 M_{\odot}$ 
\citep{Shibata02a, Shibata:2003ga,  Shibata06a,Liu:2008xy,
Hotokezaka2011,Bauswein2013}. 
Of course negligible mass for the surrounding torus 
in the delayed collapse scenario can also be 
due to the EOS \citep{Shibata06a,Rezzolla:2010}.

\textbf{Prompt Collapse}
The prompt collapse has also an impact on the magnetic field
evolution, since the HMNS lifetime is limited also the 
magnetic field amplification is limited \citep{Ruiz2017}. 
However, the exact level of such limitation is not known.
The mass threshold at which the HMNS prompt collapses to a BH
strongly depends on the EOS \citep{Shibata05c, Shibata06a, Oechslin07a,
Hotokezaka2011,Studzinska2016}. It is clear that a soft EOS, meaning that 
matter can be compressed in a more effective way, is more compact and 
the threshold mass to collapse to a BH is smaller. The other way around, 
a stiff EOS does not allow for such compression and a star is less compact, 
thus allowed to have a larger threshold mass \citep{Shibata06a}.
A BNS with a total mass of $M_{tot}\sim 2.8 M_{\odot}$ can in principle 
promptly collapse to a BH, whereas for a slightly less mass it can lead 
to a delayed collapse some milliseconds after merger  \citep{Hotokezaka2011}.
Reducing even further the total mass to be less than $\lesssim 2.7 M_{\odot}$, 
a stable configuration can be achieved.
Interestingly, from the known double neutron star systems observed in our galaxy 
the total mass is around $\sim 2.7 M_{\odot}$ \citep{Zhang2011b}. 
This means that we could expect all outcomes, meaning prompt, delayed collapse 
or a stable configuration.
\begin{figure*}
  \begin{center}
    \includegraphics[width=0.6\columnwidth]{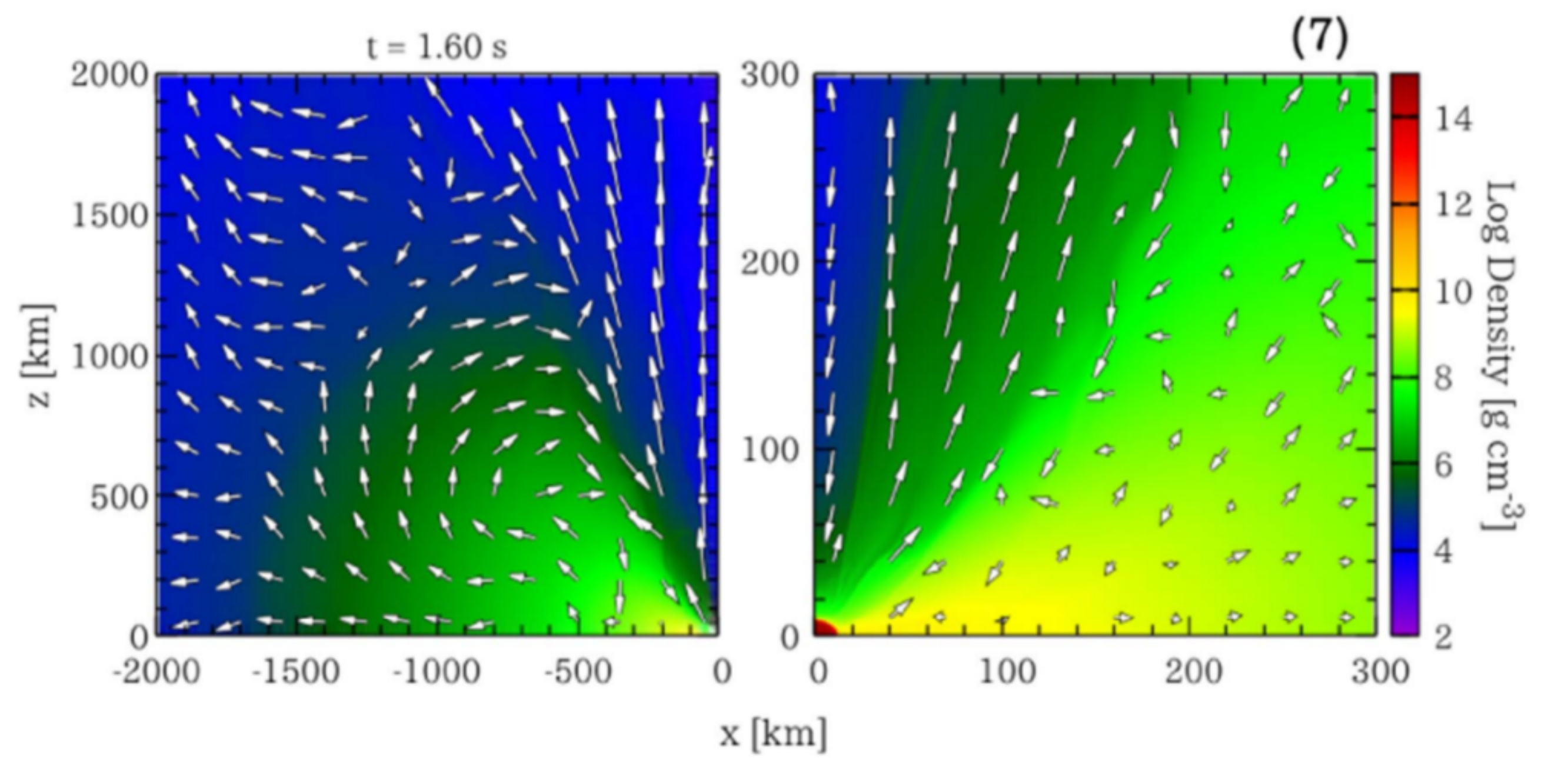}
  \end{center}
 \caption{Snapshots of the density and poloidal velocity field for a model 
 from \citep{Fujibayashi2017b}. The velocity vectors and their 
 length, correspond to the logarithm of the velocity in the poloidal plane. 
 The left panel shows a region of $2000 {\rm km}$, whereas the right 
 panel a narrow region of $300 {\rm km}$. This profile corresponds 
 to $t\sim 1.6 {\rm s}$ after merger. 
 (Reprinted from \citep{Fujibayashi2017b}. 
 $\copyright$ AAS. Reproduced with permission.)}
  \label{fig:dens2}
\end{figure*}

It was reported that following a prompt collapse to a BH no kind of jet 
can be formed \citep{Ruiz2017a}. The system does not have the time to develop 
a jet structure. However it posses a magnetic field, for which we do not 
know exactly the level of amplification. So, when the negligible torus is 
eventually accreted, all this magnetic energy will dissipate away. 
As we discussed before, prompt collapse leads also to a very small torus.
The torus lifetime can be as small as  
$t_{T} \sim 5 \left(\frac{M_T}{0.001M_{\odot}}\right) 
\left(\frac{\dot{M}}{0.2M_{\odot}s^{-1}}\right)^{-1} \, {\rm ms}$
 \citep{Ruiz2017a}. 
We may estimate the energy stored in the near by magnetosphere to be 
\begin{equation}
\begin{aligned}
E_{_{\rm EM}} \simeq  10^{40} \; b_{12}^2 \, r_{10}^{3}
\ \ {\rm erg}\,.
\end{aligned}
\label{eq:frb}
\end{equation}
assuming no amplification took place. It has a millisecond 
duration and an energy close to the requirement for a 
fast radio burst (FRB, \cite{Lorimer2007,RaneLorimer2017}).
Overall, this could be similar 
to the model proposed for FRBs where a supramassive neutron star 
collapses to a BH \citep{Falcke2013,Most2017}.
Thus, a prompt collapse is lacking of   many interesting features
coming from the delayed collapse, but could give 
answers to other mysterious EM signals.

\textbf{SMNS spin down}
A stable neutron star configuration can also be the end of a BNS merger. 
If the total mass of the binary is below a certain limit, 
then even significant accretion of the surrounding matter 
cannot trigger its collapse. This may have distinct observational 
features and could explain X-ray plateaus in the afterglow 
of short GRBs \citep{Rowlinson2013}.  It has been suggested 
that a long-lived magnetar as a BNS merger product 
can power such emission by its spin down dipolar radiation
\citep{Zhang2001,Gao2006,Fan2006,Metzger2008a,Metzger:2010}.
Such simulations showed that a stable neutron star with a surrounding 
disk can be a BNS merger product and 
the luminosity from such configuration is significant 
\citep{Giacomazzo2013,DallOsso2014}. However, the  first 
gamma-rays from the  short GRB  could not be explained.
To overcome this drawback, different scenarios have been proposed. 
The production of the gamma-rays is attributed to the 
collapse of this long-lived object to a BH which happens after the 
production of of the X-ray radiation, the observational features of such 
model together with the prompt gamma-rays of a short GRB comes from 
diffusion arguments \citep{Rezzolla2014b,Ciolfi2014}.
\begin{figure*}
  \begin{center}
    \includegraphics[width=0.67\columnwidth]{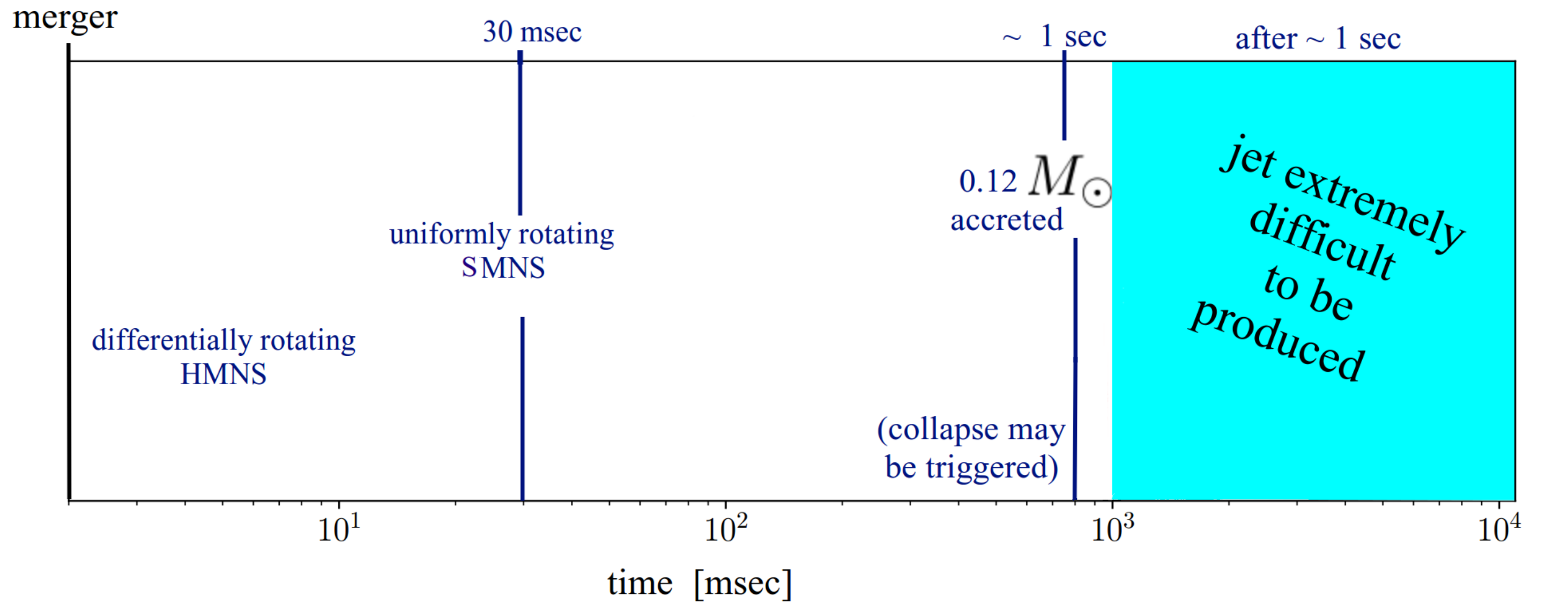}
  \end{center}
 \caption{The lifetime of the merger remnant of mass close to  the
 maximum for uniform rotation. The remnant does not collapse when 
 differential rotation is lost, but collapse may be triggered when 
 almost $0.1 M_{\odot}$ has been accreted.  If collapse is triggered 
 after one second, then the production of a jet may not be favored. In 
 such case an explosion is triggered releasing the enormous 
 amounts of magnetic energy stored in the close magnetosphere of the 
 SMNS as discussed in \citep{Nathanail2018}.
 (Reprinted from \citep{Nathanail2018}. 
 $\copyright$ AAS. Reproduced with permission.)}
  \label{fig:life}
\end{figure*}

In most studies mentioned the long-lived merger product is 
presumably loosing angular momentum due to 
magnetic spin down and the production of dipolar 
radiation where energy is lost in a rate 
\begin{equation}
\begin{aligned}
\dot{E}_{mag} = \frac{\mu^2 \Omega^4}{c^3}(1+\sin^2\chi),
\end{aligned}
\label{eq:sd}
\end{equation}
where $\mu = B r_{NS}^3$ is the magnetic dipole moment, $B$ the 
dipole magnetic field, $r_{NS}$ the neutron star radius, $\Omega$ the 
angular velocity and $\chi$ the inclination angle between the 
magnetic and the rotation axis \citep{Spitkovsky2006, Contopoulos2006a}. 
However, if a stable object is produced, it 
entirely lives in the environment of a surrounding torus 
starting exactly at the surface of the star \cite{Giacomazzo2013}.
This means that this neutron star is  impossible to acquire a dipolar 
magnetic field, since the magnetic loops cannot close through 
the torus but they have rather opened up during merger or 
they will open up due to differential 
rotation \citep{Siegel2014}. Also if any closed field lines 
are left, they are influenced by neutrino heating  
\citep{Thompson03,Komissarov2007,Thompson2018}, however this effect
will be lost in $1-2$ seconds. The last, but  most significant argument 
is that field lines which thread the disk will open up, due to the differential 
rotation of the two footpoints of the magnetic field line, one anchored on the SMNS
and one footpoint threading the disk, similar to the BH case
\citep{Uzdensky2005,Parfrey2015,Contopoulos2015}. 
Even if most of the mass of the disk is accreted or expelled, the remaining 
negligible mass will not allow the field lines to close.
Thus, the structure of the magnetosphere of the merger remnant 
can be approximately a split-monopole configuration
\citep{Margalit2017}. A neutron star with a split-monopole configuration 
spins down with a different dependence on rotation, similar 
to a BH spin down where also all field lines are open 
\citep{Nathanail2015,Nathanail2016}. The spin down follows an 
exponential decrease 
\begin{equation}
\begin{aligned}
\dot{E}_{mag} = - \frac{2}{3\pi c} B^2r_{NS}^4 \Omega^2 \exp^{t/\tau_B}
\end{aligned}
\end{equation}
where $\tau_B = 67 \left( B/10^{15} G\right)^{-2}
\left(r_{NS}/12 km \right)^{-2} s$. Essentially, the spin down of such configuration 
goes faster than the dipolar one, since all the field lines are open 
and contribute to the spin down process.

\textbf{SMNS and the surrounding disk}
Next we want to discuss about the evolution of the surrounding
disk of the remnant and the outcome of the collapse 
of the remnant after one second from merger. Due to 
transfer of angular momentum the disk expands over time 
and due to accretion onto the compact remnant, its mass decreases 
over time \citep{Fujibayashi2017b}.  As in \eqref{tac} 
the viscous accretion timescale estimated
for the torus: 
\begin{equation}
\begin{aligned}
t_{accr} \simeq \, 1 \,{\rm  s } \left( \frac{R_T}{50\,{\rm  km }}
\right)^2 
\left(\frac{H_T}{25 \,{\rm  km }}\right)^{-1} \\ \times
\left( \frac{\alpha}{0.01}\right)^{-1}\left(\frac{c_s}{0.1c}\right)^{-1} \, ,
\end{aligned}
\end{equation}
where $H_T$ is the typical vertical scale height of the torus and $R_T$ is 
its radius. Then, the mass accretion rate 
onto the SMNS yields 
\begin{equation}
\begin{aligned}
\dot{M}_{SMNS} \simeq \frac{M_T}{t_{accr}} 
\sim 0.2 \,  M_{\odot} s^{-1} \left( \frac{\alpha}{0.01}\right)
\left( \frac{M_T}{0.2 M_{\odot}}\right) \\
\times
\left( \frac{R_T}{50\,{\rm  km }}
\right)^{-2} \left(\frac{H_T}{25 \,{\rm  km }}\right)\,,
\end{aligned}
\end{equation}
where $M_T$ is the mass of the torus. The mass of the torus 
decreases in time and the torus expands, thus,  this accretion 
rate is not stationary.  This effect is seen in figure 
\ref{fig:dens1}, where the density profile is shown in the 
equatorial plane at different time slices. As time goes by, 
the torus expands and the torus density decreases 
significantly. The radius of the 
torus may reach $140\,{\rm  km }$ in $1{\rm  s }$. 
The sum of the mass accreted can be estimated to be
$\sim 0.12 M_{\odot}$ in  $1{\rm  s }$ \citep{Fujibayashi2017b}.

If the SMNS is close to its maximum mass limit, this significant 
mass accretion in one second may trigger its collapse.
Furthermore, the expansion of the torus  is also significant 
during this time. The density of the torus in the vicinity of 
the SMNS could designate the outcome of 
the collapse to an induced magnetic explosion. The  
estimation for the density of the torus at $1{\rm  s }$ 
yields:
\begin{equation}
\begin{aligned}
\rho_T \simeq \frac{M_T}{2 H_T \pi R_T^2} 
\sim 9.2\times 10^{9} g/cm^3 \left( \frac{M_T}{0.08 M_{\odot}}\right)
\\ \times 
\left( \frac{R_T}{140\,{\rm  km }}\right)^{-2}  
\left(\frac{H_T}{70\,{\rm  km }}\right)^{-1}\,,
\end{aligned}
\end{equation}
quantities are for the expanded torus at $1{\rm  s }$ 
after merger. The density in the poloidal plane 
is shown in figure \ref{fig:dens2} at time $t\sim 1.6 {\rm s}$ 
after merger. The density drops around $5-6$ orders 
of magnitude in the first $1300-1500 {\rm km}$. 
The possibility that no debris disk is formed
at all, has also been discussed  \citep{Margalit2015,Camelio2018}.

\begin{table}
\caption{\textbf{Outcome of the collapse of the merger remnant}, the 
different columns indicate the different possible outcomes for 
the merger remnant. The different outcomes depend on the collapse time 
to a BH. Different rows are: the collapse time to a BH $t_{BH}$, 
if magnetic field amplification occurs or not, the amount of 
the magnetic energy $E_B$, if there is ejected matter, the amount 
of mass surrounding the BH when it is formed, the lifetime of this 
disk around the BH, the EM outcome will be produced either by the 
collapse or by the absence of the collapse and the estimated energy 
that is released during the collapse or the absence of collapse.}
\centering
\begin{tabular}{|r|c|c|c|c|}
\hline 
\textbf{Possibilities for the}&
\textbf{Prompt collapse } &
\textbf{Delayed collapse} &
\textbf{"further" }           & 
\textbf{no collapse}  \\
\textbf{merger remnant} &&&\textbf{Delayed collapse}&
    \\ \hline \hline
collapse to BH, $t_{BH}$ &$1-2 \,{\rm ms}$& $7-500\,{\rm ms}$& $1-3\,{\rm s}$& $\infty$\\    
B-amplification   &not significant& yes&yes&yes \\
Magnetic energy, $E_{B}$ & $10^{40-44}\,{\rm erg}$&$10^{51} \,{\rm erg}$&
$10^{51} \,{\rm erg}$&$10^{51} \,{\rm erg}$  \\  
ejecta     &not significant& yes&yes&yes       \\
BH surrounding disk   &  negligible &  $0.05-0.2 \, M_{\odot}$ & $0.01-0.05 \, M_{\odot}$ 
& no BH disk  \\ 
disk lifetime &$2-8 \,{\rm ms}$& $0.2-1 \,{\rm s}$ &$0.1-0.2 \,{\rm s}$ & $0$  \\
EM outcome &magnetic energy  & magnetic jet & magnetic explosion & magnetic wind  \\
&dissipation&&&(spin down) \\
Estimated energy & $10^{40-44}\,{\rm erg}$ &$10^{51}\,{\rm erg}$& $10^{51} 
\,{\rm erg}$ & $10^{50}\,{\rm erg}$ \\
\hline
\end{tabular}
 \label{table:outc}
\end{table}

\textbf{Jet or magnetic explosion}
Observational signatures of magnetic fields
Previously, we discussed about the production of a low density
funnel that appears after the collapse of the merger remnant to a BH.
All results from simulations so far, describe such evolution 
in the case that the collapse occurred  
in the first milliseconds after merger. Here, we describe 
the conditions and the outcome of the collapse to a BH,
if this happens after $1{\rm  s }$ from merger. The foremost point, 
is the condition for the establishment of 
a magnetic jet. A stable magnetic jet configuration needs 
the torus pressure to balance the magnetic pressure from 
the jet itself. Due to magnetic field amplification discussed 
earlier, we assume that the mean magnetic field  
of the SMNS is $B \simeq  10^{16} \, G$. This yields:

\begin{equation}
\begin{aligned}
\frac{B^2_{SMNS}}{8 \pi}  \simeq 4 \times 10^{30}\, dyn/cm^2
\left( \frac{B_{SMNS}}{10^{16} \, G }\right)^2 \\
\gg 9.2 \times 10^{29} \, dyn/cm^2  \left( \frac{\rho_T}{9.2\times
  10^{9} \, g/cm^3 }\right)
   \simeq \rho_T \, c^2\,.
\end{aligned}
\end{equation}
At later times that the torus has expanded even more, 
the establishment of a magnetic jet becomes more problematic, due to the 
imbalance between the magnetic pressure and the disk ram pressure. 
We may use also  the accretion rate at  $1{\rm  s }$ as reported in 
\cite{Fujibayashi2017b}, which is  $\sim 0.02 \, M_{\odot} s^{-1}$.  
This yields:
$$B^2_{SMNS}/8 \pi \gg 2.6 \times 10^{28} \, dyn/cm^2  
 \sim 
\dot{M}c/4\pi r_{BH}^2.$$

Figure \ref{fig:life} summarizes the above discussion.
The main point is that, if the collapse is triggered around or 
after  $\sim 1{\rm  s }$ after merger, the magnetic energy 
of the SMNS is released and induce a powerful explosion of 
$E_{exp} \sim 10^{51} {\rm erg}$, contrary to a magnetic jet expected
\citep{Nathanail2018}.

We may summarize our own understanding of the outcome of the 
collapse of the merger remnant, which strongly depends on the time 
that the collapse is triggered. Of course, the triggering of the collapse
depends on the EOS and the total mass of the binary, 
however here we will not go to that depth 
and only characterize the outcome with respect to the collapse time. 
The possible outcomes are summarized in Table \ref{table:outc}.
The four columns represented four different types for the outcome of 
a BNS merger. The different rows show characteristics that are 
essential to the observable outcome of a BNS merger. 

The prompt collapse that is characterized by the collapse of the merger 
product in the first $1-2 \, {\rm ms}$ (first column of Table \ref{table:outc}) 
does not have an effective magnetic field amplification phase and also 
no significant ejecta, but due 
to the negligible disk that surrounds the newly formed BH, the lifetime 
of this disk is on the order of few milliseconds. As a result, 
all its magnetic energy will dissipate in that timescale. 
The energetics of such an explosion (depicted in Eq. \eqref{eq:frb}) 
and its timescale points to an event similar to FRBs.
In all cases that the remnant lives longer than the first milliseconds, 
it is certain that magnetic field is amplified to high values.
The case where the merger product (a HMNS at this stage) collapses in 
a few milliseconds to tens of milliseconds, is the most discussed case. 
This  is expected to produce a canonical magnetic jet that interacts 
with the merger ejecta. If the collapse is delayed for a second (or more) 
then the low density of the torus may be insufficient to act as 
a boundary for a magnetic jet and a magnetic explosion is triggered.

At the end of this section we list some interesting and critical points 
known from numerical simulations of BNS mergers and provide 
some comparison with points known from short GRBs.

\bigskip

\textbf{Critical points}
\begin{itemize}
\item If the merger product does not collapse in the first millisecond, 
then the magnetic energy is amplified to values higher than $10^{50} {\rm erg}$
\cite{Kiuchi2017}.
\item the saturation level of magnetic field amplification is not yet known
\cite{Kiuchi2017}.
\item amplified magnetic field is turbulent, it needs time (more than a second)
 to rearrange in a coherent large scale structure \citep{Harutyunyan2018}.
\item after the  collapse to a BH in $10{\rm ms}$, 
a magnetic jet structure is produced \citep{Rezzolla:2011,Kawamura2016,Ruiz2016}.
\item ordered poloidal magnetic field above $10^{15} {\rm G}$ is needed 
for a BZ luminosity of $\sim 10^{51} {\rm erg/s}$ \citep{Blandford1977}.
\item the production of an ultra relativistic outflow has never been reported in BNS 
simulations \citep{Rezzolla:2011,Kiuchi2014,Dionysopoulou2015,Palenzuela2015,
Kawamura2016,Ruiz2016}.
\item the magnetic jet funnel reported in BNS simulations has an opening angle 
of $\gtrsim 20^{\circ}-30^{\circ}$, and a maximum Lorentz factor 
reported $\Gamma = 1.25$ \citep{Ruiz2016}.
\item if the collapse of the SMNS to a BH occurs late enough, the mass of the 
surrounding disk is negligible \citep{Margalit2015,Camelio2018}.
\end{itemize}

All these critical points should be taken into account for 
the understanding for any magnetized outflow (relativistic or 
non-relativistic) that emerges from the merger remnant or 
the collapse of the merger remnant to a BH. In order to help 
comparisons with observations, we should also mention here that 
there are short GRBs observed with a lower limit on the opening angle 
$\gtrsim 15^{\circ}$
and some observed short GRBs that have jets with opening angles of $7^{\circ}-8^{\circ}$, 
\citep{Fong2015}. However, the opening angle given from 
numerical relativity simulations at the base of the jet may (most probably) 
change through the interaction with the BSN ejecta, this is discussed in the 
next section.

\section{Short GRB Jet simulations}
\label{sec:shortGRB}

It is understood that if the merger does not follow a quick prompt collapse 
then significant mass is  ejected following the BNS merger. 
Mass can be ejected dynamically,  
by winds driven from the newly
formed hypermassive neutron star (HMNS) and from the debris
disk that forms around it  
\cite{Rosswog1999,Aloy:2005,Dessart2009,Rezzolla:2010,Roberts2011,
Kyutoku2012,Rosswog2013a,Bauswein2013b,
Hotokezaka2013,Foucart2014,Siegel2014,
Wanajo2014,Sekiguchi2015,Radice2016,Sekiguchi2016,Lehner2016,
Siegel2017,Dietrich2017,Bovard2017,Fujibayashi2017,Fujibayashi2017b}. 
As a result, any outflow that emerges from the merger remnant or 
the collapse of the merger remnant has to pass through this dynamical ejecta.

To continue further in the discussion of the interaction 
between the BNS ejecta and a (maybe mildly) relativistic 
outflow that emerges after merger,  we need to 
define  characteristic names widely used  in the literature. 
We follow the terminology, as nicely 
given by Nakar \& Piran  \citep{Nakar2018}:

It is important to define the angle that the observer is looking 
to the emission produced from the outflow, with respect to the motion 
of the outflow itself. Assume that an emitting region 
moves relativisticly with a Lorentz factor $\Gamma$, then the emission 
is called:  \\
\textbf{\emph{On-axis emission:}} if the angle $\theta$ between the 
line-of-sight and the velocity of the emitting material 
satisfy $\theta \lesssim 1/\Gamma$. This emission 
is Lorentz boosted for relativisticly moving  material. \\
\textbf{\emph{Off-axis emission:}} if the angle $\theta$ between the 
line-of-sight and the velocity of the emitting material 
satisfy $\theta \gtrsim 1/\Gamma$. In the case for a 
relativisticly moving emitting material, then this appears 
fainter than being on-axis. 
It is clear that emission, which originally is observed off-axis, 
will become on-axis when the emitting material decelerates 
significantly and expands sideways. Originally, on-axis emission, 
stays always on-axis. We should also point that the observer angle 
is usually defined as the angle between the jet axis (the symmetry axis)
and the line-of-sight. For BNS mergers it is generally supposed 
that the jet axis coincides with angular momentum axis of the BNS system. 
Next, we define characteristic names  concerning the intrinsic properties and 
structure of the emitting material. 
\begin{figure*}
  \begin{center}
    \includegraphics[width=0.4\columnwidth]{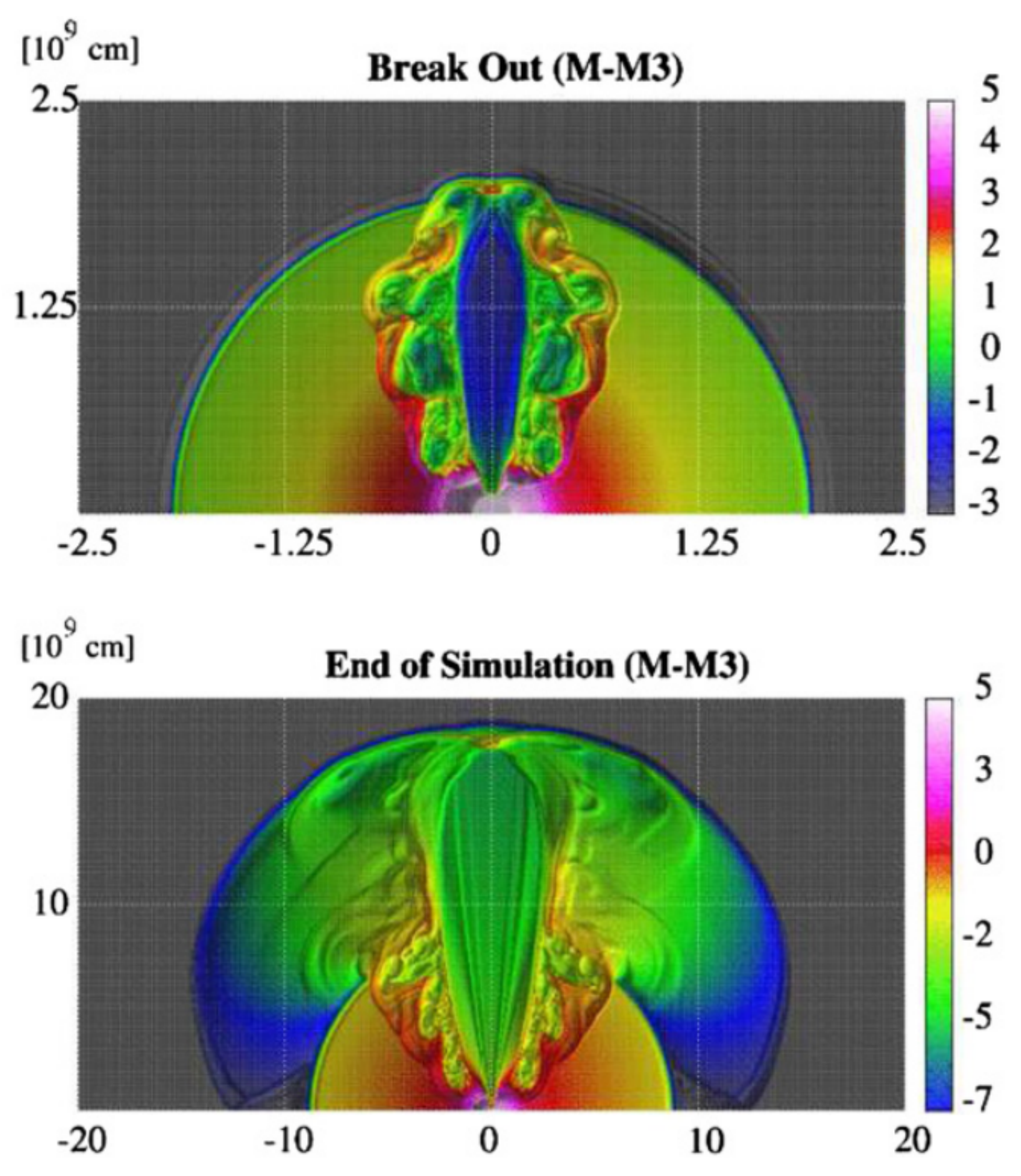}
  \end{center}
 \caption{Two snapshots from a model of \citep{Nagakura2014}. 
 The top panel is at the time where the jet breaks out and the lower 
 panel at the end of the simulation. The jet was injected at $50 {\rm ms}$ after 
 merger with an opening angle of $15^{\circ}$. The average opening angle 
 of the jet after break out is $12.6^{\circ}$.  
 (Reprinted from \citep{Nagakura2014}. 
 $\copyright$ AAS. Reproduced with permission.) }
  \label{fig:nag}
\end{figure*}

\textbf{\emph{Structured relativistic jet:}} 
as the name indicates, this is a relativistic jet 
along the symmetry axis that  acquires a certain structure. 
This structure can be angular and/or radial. 
A simple example can be a "top-hat" jet, a blast wave where 
the energy and radial velocity are uniform inside a cone 
(Blandford-McKee \cite{Blandford-McKee1976}). Another example, 
usually inferred  for short GRBs, is a successful jet with a
cocoon, where the cocoon term is defined below. 
In general, a jet can be composed by a fast core at small polar angles  
surrounded by a slower, underluminous sheath.
The presence of a spine-sheath structure can be independent from that of a cocoon.

\textbf{\emph{Cocoon:}} If a jet propagates within a dense 
medium, then the jet transfers energy and shocks this material.
There is also a reverse shock that goes down to the jet itself. 
The resulting configuration is called a cocoon. 
In the case of BNS mergers the dense medium is the ejected 
material (dynamical and secular ejecta). So, if a jet is
produced after merger, then also a cocoon is. 
There is a differentiating factor of whether the jet was 
successful or not.

\textbf{\emph{Choked jet with cocoon:}} The jet that produced 
a cocoon from the interaction with a dense medium did not 
have enough energy to break out of the medium and it 
is choked. The jet transfers all its energy to the medium 
and the shocked material may acquire a certain angular structure. 
The reverse shock may produce also a radial structure inside the 
cocoon in the region of the choked jet. 
In the case of a BNS mergers, a choked jet would mean that 
no usual short GRB was produced. However, a mildly relativistic 
outflow may be produced. 

\textbf{\emph{Successful jet with cocoon:}}The jet that produced 
a cocoon from the interaction with a dense medium had 
enough energy to break out of the medium. An ultra relativistic 
outflow passed through the medium and eventually decelerates 
through the interaction with the inter-stellar medium (ISM). 
The jet transferred some of its energy to the medium 
and a cocoon was produced.
In the case of a BNS merger, a successful jet would mean that 
a usual short GRB was produced, pointing along the 
jet (BNS) axis. However, a mildly relativistic 
outflow may also be produced. In this case two components can be 
identified, an ultra relativistic core which is surrounded 
by a mildly relativistic cocoon.

Assuming the possibility discussed in the previous section, 
that a jet is never formed, we could rephrase the last
case to a successful explosion with a cocoon. Meaning, that no 
jet was formed but rather an instantaneous 
explosion occurred which followed the delayed (over a second) 
collapse of the remnant \citep{Nathanail2018}. 
In such case the core is not ultra relativistic, 
but just a bit faster than the surrounding cocoon itself.

It is important to note that there exist previous studies that have 
discussed the formation of cocoons in a slightly different context. 
Namely, for 
long GRBs where the jet has to propagate through the stellar 
envelopes and not the BNS ejecta. The main differences should be 
in the density profile and how it falls off. 
Cocoons in long GRBs have been discussed by 
\citep{Ramirez-Ruiz2002}.
The mixing of the cocoon components has been discussed 
in \citep{Morsony2007, Lazzati2010,Mizuta:2008a,
LopezCamara2013,Mizuta2013}

In what follows we review studies that have developed a robust 
picture regarding the outcome of a BNS merger with respect 
to the prompt emission which is a short GRB and the afterglow 
emission which can give a physical insight on the 
outflow that produced it. The common understanding for the prompt emission 
is that it is powered by some internal dissipation mechanism within the jet. 
 The common interpretation 
of the late afterglow is the interaction of the produced outflow 
with the ISM, during which the outflow sweeps up matter 
from the ISM and that results on the eventual deceleration of the outflow.
\begin{figure*}
  \begin{center}
    \includegraphics[width=0.4\columnwidth]{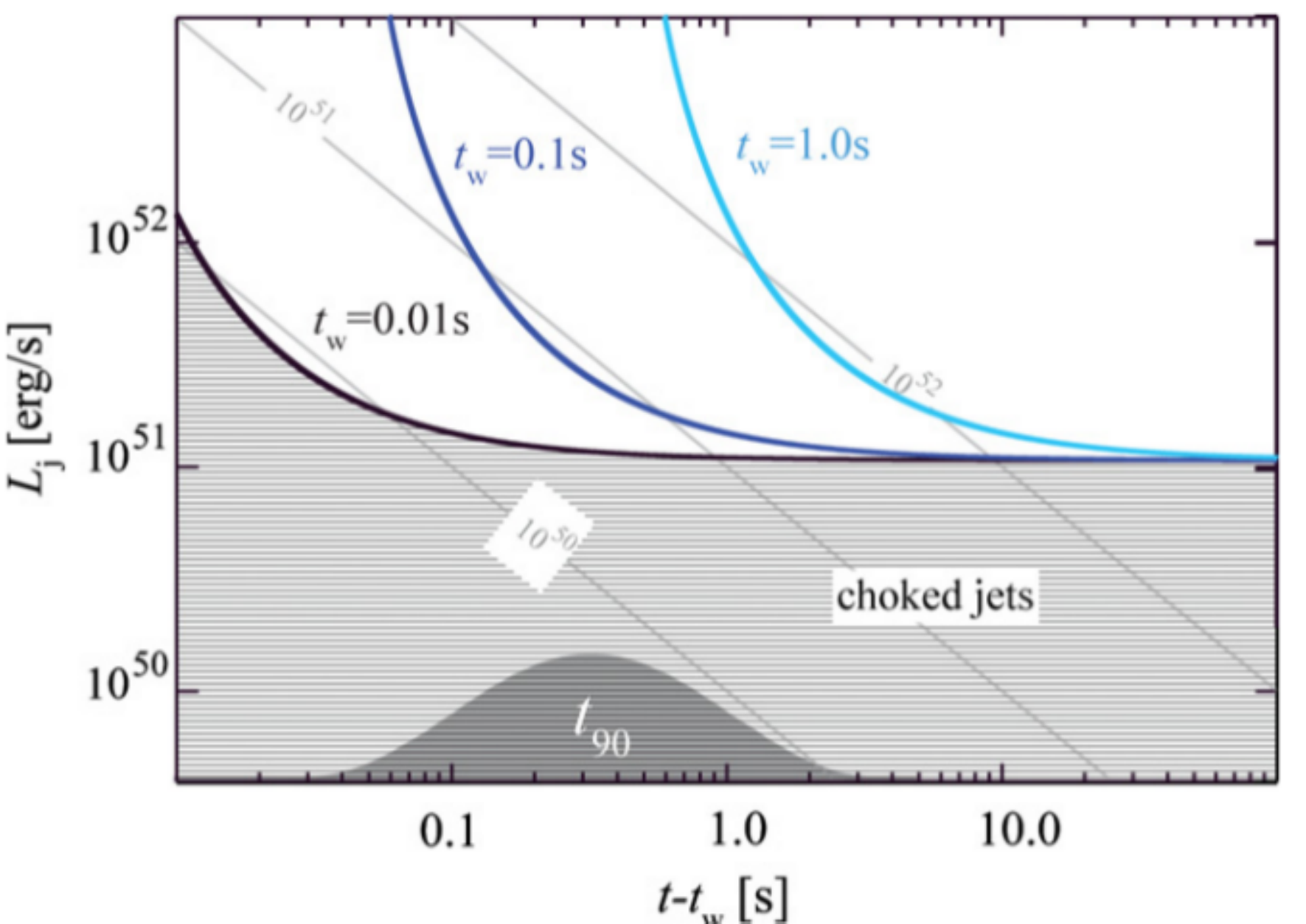}
    \includegraphics[width=0.4\columnwidth]{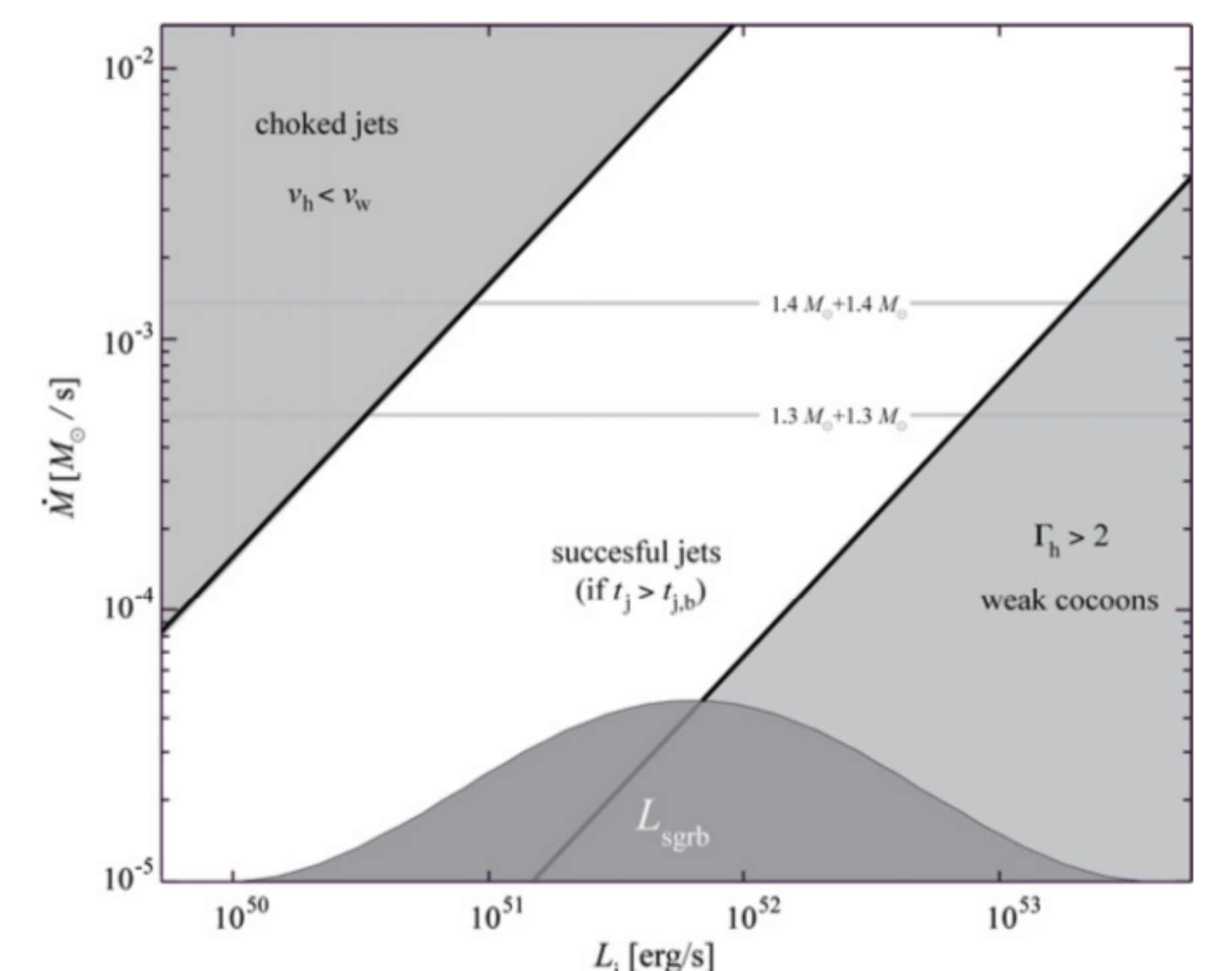}
  \end{center}
 \caption{The left panel shows luminosity versus time, where $t_w$ 
 is the lifetime of the neutrino wind and the time that the jet begins 
to expand. The colored lines indicate a model from \citep{Murguia-Berthier2014},
that has a wind injection rate of $\dot{M}_w \sim 10^{-3} M_{\odot}s^{-1}$ 
with a velocity of $0.3c$, each line indicate a different termination time 
$t_w \sim 0.01, \, 0.1,\, 1\, {\rm s}$. For such a heavy wind, the luminosity 
of the jet has to be above $10^{51}{\rm ergs^{-1}}$ and operate for at least the same 
time as the wind. 
(Reprinted from \citep{Murguia-Berthier2014}. 
 $\copyright$ AAS. Reproduced with permission.)}
  \label{fig:Murg_1}
\end{figure*}

\textbf{Jet through the BNS ejecta}
In that respect, \citep{Nagakura2014} took into account 
the density profile of a BNS numerical relativity simulation 
\citep{Hotokezaka2013}, in order to study the propagation of a 
hydrodynamical jet through such ejecta and develop a picture of whether 
the jet could break out from them or not. 
Such studies built a consensus that even if the outflow 
emerging from the BNS has a wide opening angle, it will be subsequently 
collimated as it tries to pass through the ejecta\citep{Nagakura2014,
Murguia-Berthier2014,Murguia-Berthier2016}.
These works described a density distribution that the jet should pass 
through, this density distribution of matter has been ejected 
primarily during merger. Any outflow produced in the base 
of the merger configuration has to pass through these ejecta 
and may change its shape through collimation or loose some energy 
by the interaction with the ejecta.  This way, some energy deposits 
to the ejecta producing a cocoon structure. 

In the work of \citep{Nagakura2014}, the jet opening angle was placed 
to be $15^{\circ}-45^{\circ}$, with an injected luminosity of 
$L \sim 10^{50} {\rm erg/s}$. As they pointed out, their results were 
similar to equivalent simulations in the context of the collapsar model 
\citep{Nagakura2011, Mizuta2013}. The opening angle at the base
of the jet is determined through the interaction of the jet and the 
surrounding disk. An important consequence of this study is to 
to find that irrespective to the initial opening angle, all 
jets succeed in breaking out and form what we would call
a structured jet with a cocoon.  Only for the model with an initial opening 
angle of $45^{\circ}$ this is not true and a choked jet with cocoon is 
formed. Due to the large cross section of the jet, 
it cannot go sideways into the cocoon and expands quasi-spherically.

In figure \ref{fig:nag} a model from \citep{Nagakura2014} is shown. 
The ejected mass is $10^{-3} \, M_{\odot}$ and the initial jet is injected 
with an opening angle of $15^{\circ}$. The density profile of the produced 
structure is shown for two snap shots, one at the 
time that the jet breaks out from the ejecta and the other at the end of 
the simulation. The average opening angle of the jet after break out, 
which has changed due to the interaction with the surrounding ejecta 
is $\theta_{jet}\sim 12.6^{\circ}$. 
Interestingly except the break out of the jet, a cocoon 
is formed and is clearly shown in the above mentioned figure \ref{fig:nag}. 
However, 
there does not exist in this study a detailed description of this 
component. The density profile for the ejecta used in this study has a steep profile 
$\rho \propto r^{-3.5}$ with a spherical shape.

In \citep{Murguia-Berthier2014} they  studied what is 
the influence of the neutrino driven wind to the 
expansion and propagation of the formed jet. 
They considered the post-merger production of 
neutrino fluxes that contribute to a wind density 
profile. They quantified this wind as:
\begin{equation}
\begin{aligned}
\dot{M}_w \sim 5 \times 10^{-4} \left( \frac{L_{\nu}}{10^{52{\rm erg s^{-1}}}}
\right) \, M_{\odot}s^{-1},
\end{aligned}
\label{eq:n-w}
\end{equation}
\cite{Qian1996,Rosswog2002b,Dessart2009}, 
which results in limiting the Lorentz factor of the jet:
\begin{equation}
\begin{aligned}
\Gamma_{\nu} \sim 10 \left( \frac{L_{jet}}{10^{52{\rm erg s^{-1}}}}
\right) \left( \frac{\dot{M}_w}{5 \times 10^{-4}M_{\odot}s^{-1}}
\right),
\end{aligned}
\label{eq:lfac_n-w}
\end{equation}
\begin{figure*}
  \begin{center}
    \includegraphics[width=0.24\columnwidth]{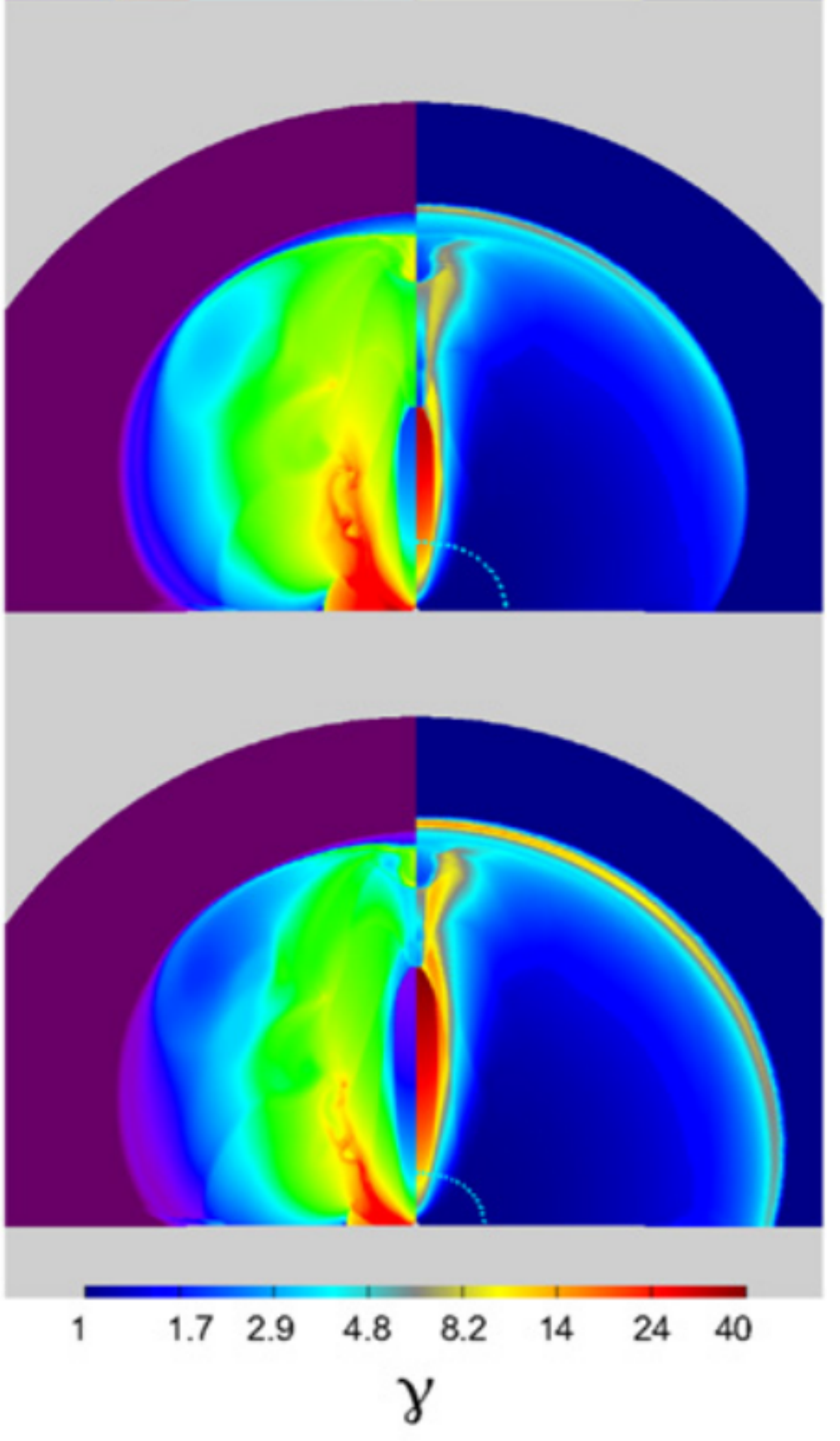}
  \end{center}
 \caption{A model from \citep{Duffell2015} where the shape of the 
 BNS ejecta is assumed to be oblate. 
In the upper and lower row  two different times are depicted $ct/R_0 = 15, \, 25$, 
 where $R_0 = 850 {\rm km}$ is the initial radius of the mass cloud. 
 The initial opening angle of both models is $60^{\circ}$. The left panel of each 
 plot shows the density and the right panel of each plot shows the Lorentz factor.
 The outer surface of the expanding mass ejecta is depicted with a dashed cyan curve.
 The model depicted in this figure with the oblate shape cloud 
 clearly produces a narrow relativistic outflow. In both cases the ratio between 
   the energy of the engine to the rest mass energy of the ejecta is 
   $E_{engine}/M_0c^2 = 0.024$, where $M_0=10^{-4}M_{\odot}$ is the mass of the 
   cloud ejecta.
   (Reprinted from \citep{Duffell2015}. 
 $\copyright$ AAS. Reproduced with permission.)}
  \label{fig:Duff}
\end{figure*}

Their wind profile depends on how long the neutrino 
driven wind was active. At the time that the wind stops, a jet is injected.
In figure \ref{fig:Murg_1} (left panel) they show a parameter
 study on whether the jet 
can break out or not from such a wind. The axes are the luminosity of 
the jet versus time, where $t_w$ depicts the time that the neutrino 
wind stops, supposedly when the merger remnant collapses to a BH. 
Matter is injected in the wind as $\dot{M}_w \sim 10^{-3} M_{\odot}s^{-1}$ 
with a velocity of $u \sim 0.3c$. The colored lines indicate a different 
termination time for the neutrino wind, where $t_w$ is the time 
that the neutrino wind stops. As a comparison 
the $T_{90}$ distribution (the duration distribution of short GRB from 
\cite{Gehrels2009,Nakar:2007yr}) is over plotted to show that when the neutrino 
wind operates for more than $t_w> 0.1 s$ then jet duration times that exceed the 
observed ones, are needed. Interestingly all jets with luminosity less 
than $10^{51} {\rm erg s^{-1}}$ are choked and never break out from the neutrino 
wind.  This can be regarded as the limiting value for 
a production of a structured jet with a cocoon or a choked jet with a cocoon.

However, this result strongly depends on the amount of mass that 
is ejected through this process. Thus the next thing to compare is jet luminosity 
with respect to the mass injection from 
the neutrino wind. The result is shown in figure \ref{fig:Murg_1} 
(right panel). 
The mass injection rate is plotted versus the jet luminosity 
and depict different regions in the parameter space. If the luminosity 
is low (on the left part of the figure) then the velocity of the head of 
the jet is not exceeding the velocity of the wind and consequently 
never breaks out resulting in a choked jet with a cocoon. Even for smaller 
luminosity, if the mass injection is less than $10^{-3}-10^{-4} M_{\odot}s^{-1}$
then a successful jet can be formed.  It is also known that in order 
to produce a successful jet, the jet injection time has to exceed the break 
out time through any medium. They further comment on the production of 
a cocoon as the jet advances through the ejecta and 	deposits some 
of its energy to form such a cocoon \citep{Ramirez-Ruiz2002}.

\textbf{Spherical versus oblate BNS ejecta}
In the above mentioned studies the shape of the density profile that 
is mimicking the BNS ejecta was spherical. So all results have to be 
interpreted as arising within a spherical expanding mass cloud. 
However, there is a possibility that this is not true \citep{Hotokezaka2013d}.
Recent simulations of BNS mergers show indeed that the merger 
ejecta and/or the post merger driven winds are not at all spherical 
\citep{palenzuela2015,Radice2016,Dietrich2016,Sekiguchi2016,Lehner2016,
Dietrich2017,Bovard2017,Fujibayashi2017b}. In \citep{Duffell2015} they 
consider the interaction of the jet with an oblate 
mass cloud mimicking the BNS ejecta, opposed to a spherical one. 
The earlier idea that the ejecta can provide the collimation of the jet
\citep{Rosswog2003c} is stronger in the case where the BNS ejecta 
have an elongated shape.  They inject a luminosity of 
\begin{equation}
\begin{aligned}
L = e M_{cloud} c^2/\tau
\end{aligned}
\label{eq:lum}
\end{equation}
where $M_{cloud}$ is the mass of the cloud, $e$ is the ratio of the energy 
deposited in the mass cloud and $\tau$ is the engine duration. 
Engines that act through an oblate cloud can collimate even wider 
initial angles. When the overall injected energy from the injected luminosity 
\eqref{eq:lum} is low, then the kinetic energy of the dynamical ejecta 
can be higher and this does not allow for collimation. In the other limit
where the injected energy  is large, then the mass of the ejecta  
cannot sufficiently provide any collimation to the outflow. In the latter 
case the outflow keeps the initial opening angle. 
For an initial opening angle of $29^{\circ}$, a mass cloud  
of $10^{-4}M_{\odot}$ and oblate shape, a jet with luminosity of 
$10^{48-49}$ is collimated significantly with a resulting 
opening angle fo $5^{\circ}-8^{\circ}$ when breaking out from the cloud.

\begin{figure*}
  \begin{center}
    \includegraphics[width=0.4\columnwidth]{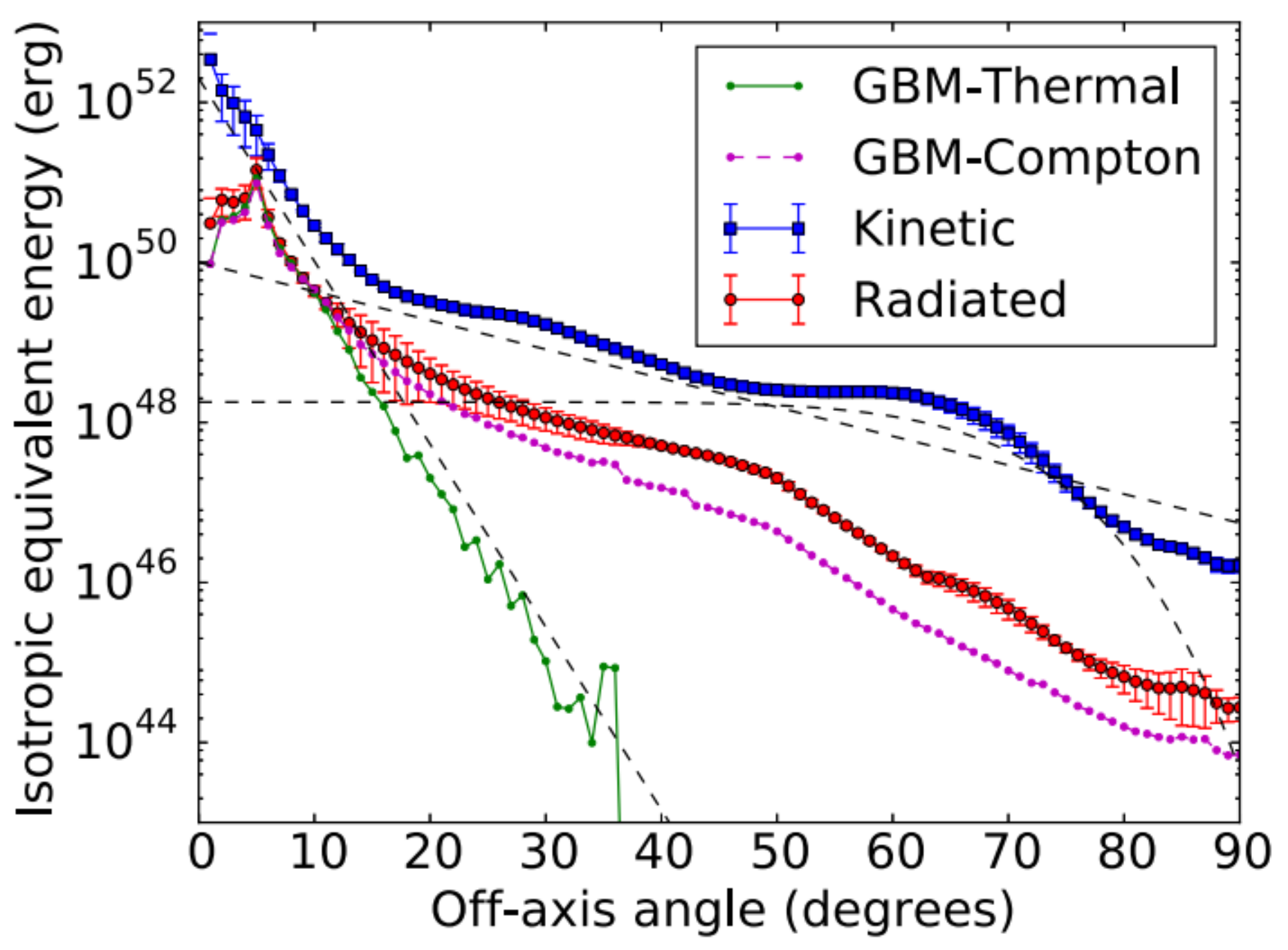}
    \includegraphics[width=0.4\columnwidth]{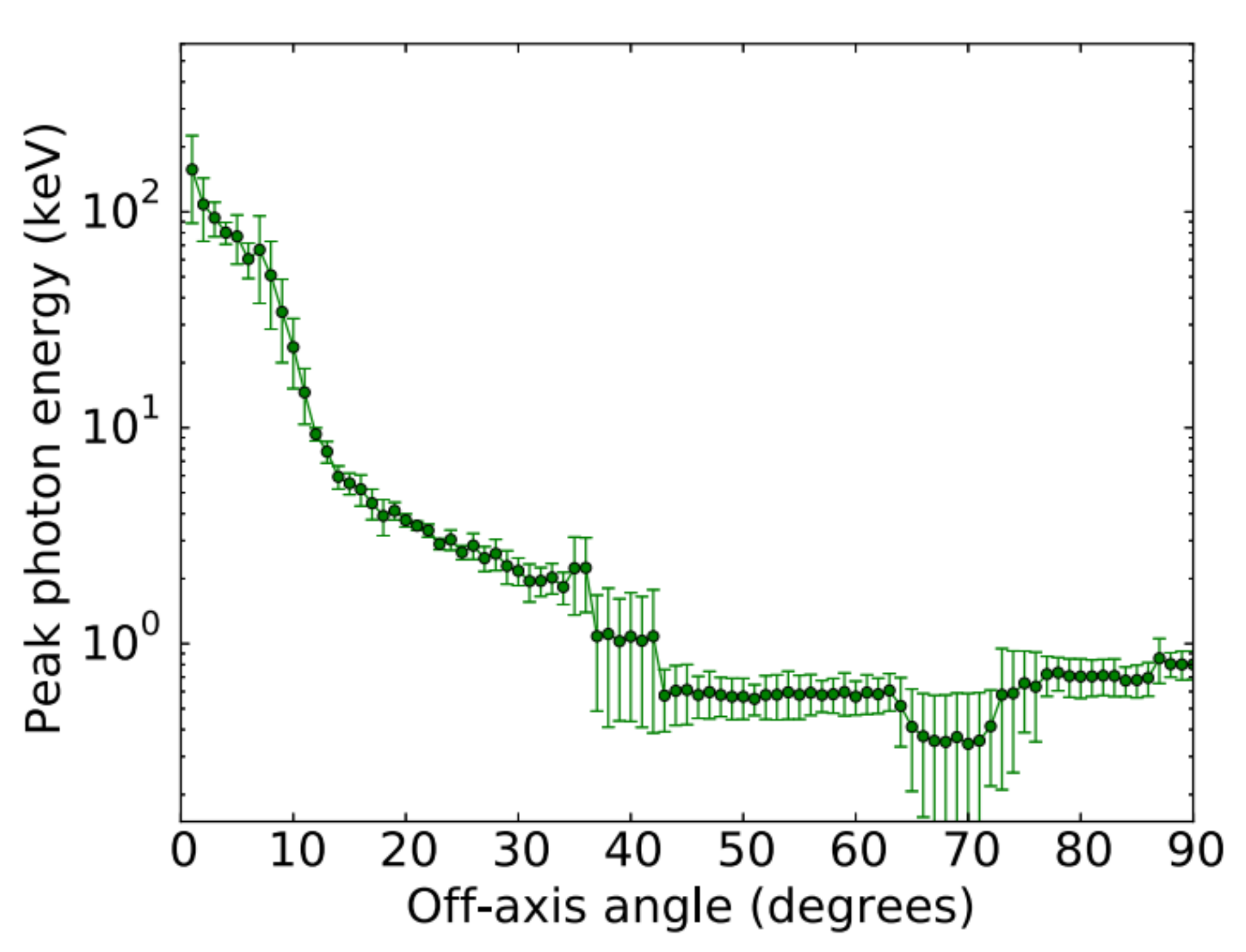}
  \end{center}
 \caption{Both figures are taken from \citep{Lazzati2017b}.
 \textit{Left panel:} Off-axis distribution of the isotropic equivalent energy. 
The error bars show the range of variation at each specific angle. 
The kinetic energy is shown in blue squares, while the bolometric energy 
is shown in red dots. The energy that Fermi (Gamma-ray Burst
Monitor (GBM) would  detect is shown as lines with dots, green solid
line is for a  thermal spectrum and magenta dashed line for 
 a Comptonized spectrum. The three components: the jet (exponential), the cocoon
  (exponential), and the shocked ambient medium (constant with sharp cutoff) 
   are overlaid on the kinetic energy profile as black dashed lines. 
 \textit{Right panel:} off-axis emission from the jet/cocoon  photosphere.
 The peak photon energy is depicted, while the symbols are as defined above.
 (Reprinted from \citep{Lazzati2017b}. 
 $\copyright$ AAS. Reproduced with permission.)}
  \label{fig:Lazz}
\end{figure*}
 In figure \ref{fig:Duff} a model is shown from \citep{Duffell2015}.
 In this model the mass cloud that mimics the BNS ejecta has an oblate shape..
  It is clearly seen that the interaction through the oblate 
  mass cloud produces a narrow outflow with high Lorentz factor.
We should also note here the possibility that jet formation may be accounted 
also to the production of a magentar after the BNS merger 
\citep{Bucciantini2012,Bromberg2018}

The next step was to use more realistic profiles taken from 
\citep{Perego2014,Siegel2014}, 
in order to continue a more detail study for the interaction of the jet with the 
neutrino driven and a magnetically driven wind  as studied in 
\citep{Murguia-Berthier2016}. 
They concluded that a jet with luminosity comparable to the observed ones 
from short GRBs can break out from such winds with the requirement of 
having an initial opening angle of less $\lesssim 20^{\circ}$.
They further used the observed duration of short GRBs to  set limits 
on the lifetime of the production of winds from a HMNS, which is 
determined by the time that the jet needs to break out.

\textbf{Observables from off-axis emission}
All of such simulations act as a first step towards understanding 
the jet and cocoon observables that follow a BNS merger. 
The next step was to see how these components would show up 
when observed off-axis. Furthermore, late radio counterparts 
from BNS mergers were long been proposed and expected \citep{Nakar2011}.
Wide angle signatures from jet and cocoon interactions 
were presented through semi-analytical calculations in \citep{Lazzati2017a}.
They  calculated the on-axis and off-axis emission of a short GRB. 
They included the prompt and afterglow
emission from a relativistic jet, as well as the prompt and
afterglow emission from the cocoon formed through the interaction 
of the jet and the surrounding ejected material. The energy of the cocoon 
was found to amount approximately $10 \%$  of the energy of the burst itself.
However, the cocoon energy strongly depends on the structure and size 
of the ejected material.

In the case of long GRBs the propagation of the jet 
through a baryon loaded region (such as the interior of a massive star)  
has been studied and shown a clear and robust observational picture. 
Nakar \& Piran \citep{Nakar2017} made a comprehensive  
(mostly analytical) study on the 
observable signatures of GRB cocoons. Their main focus was on the 
collapsar model for long GRBs, which envisions the propagation of 
a jet inside a massive star. While, their focus was cocoons emerging 
from long GRBs, short GRB cocoons should have an analogous signature 
(maybe weaker) as they indicated. 
All the formulas and equations reported in this study can give 
a quick in-depth description  of the characteristics of a cocoon 
and its emission. 
The analytical modeling by \citep{Bromberg2011}, calibrated 
by numerical results from \citep{Mizuta2013}, can be used to estimate 
the cocoon parameters through the jet break out time and the 
characteristics of the ejected matter.
\begin{figure*}
  \begin{center}
    \includegraphics[width=0.4\columnwidth]{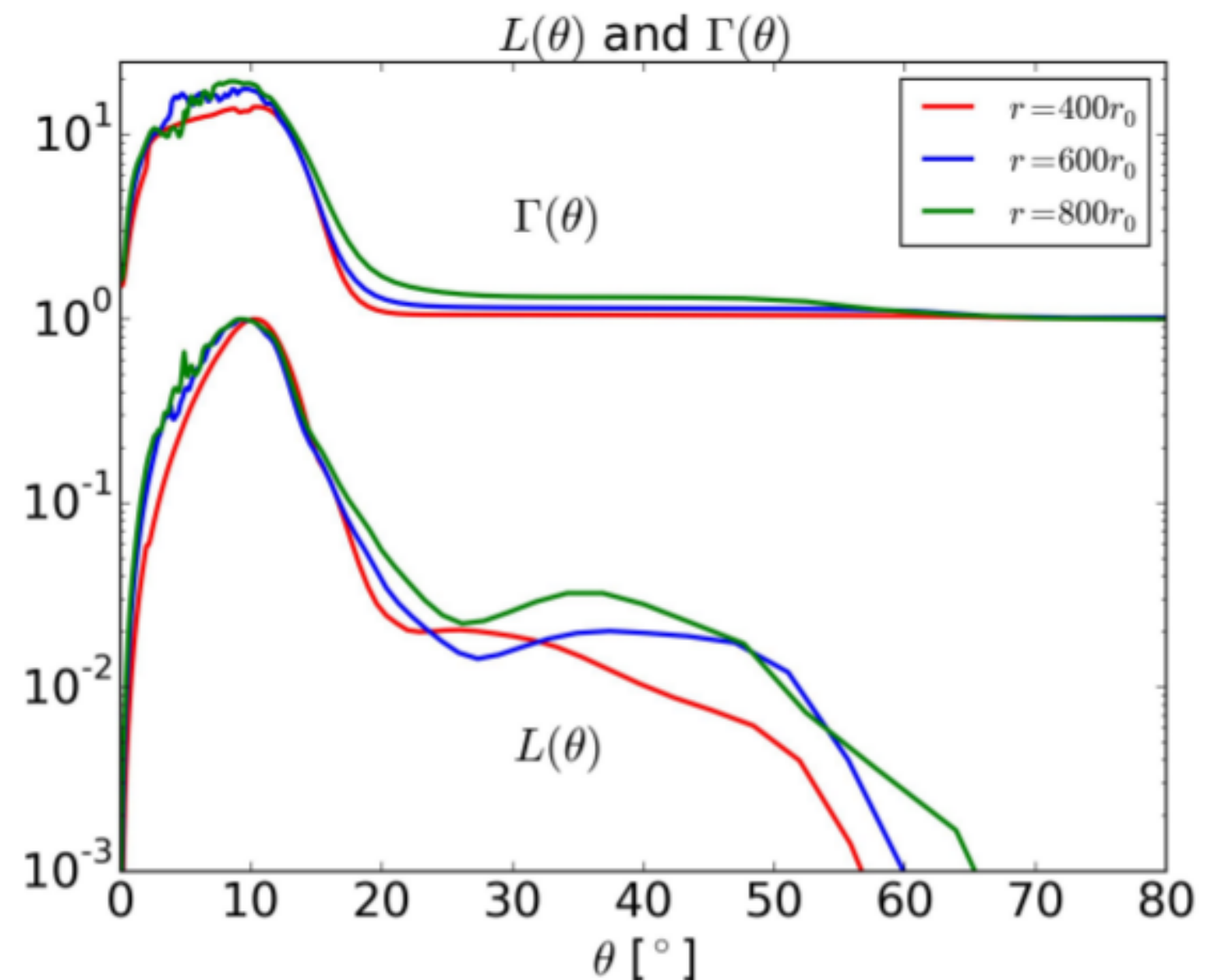}
  \end{center}
 \caption{the appearance of a jet model after break out from the BNS ejecta. 
 A model from \citep{Kathirgamaraju2018}. In the figure jet luminosity 
 $\Lambda (\theta)$ (in arbitrary units) and Lorentz factor as a function 
 of the observer's angle. Quantities are extracted at three different radii. 
 It is evident that even in angles greater than $20^{\circ}$, the luminosity is 
 reduced but still significant.
 (Reprinted from \citep{Kathirgamaraju2018}. 
 $\copyright$ Oxford University Press. Reproduced with permission.)}
  \label{fig:kath}
\end{figure*}
%
%
%

Simulations of short GRB cocoons can  provide more details 
on the production of the cocoon itself together with realistic characteristics 
for its shape and initial Lorentz factor which are key elements 
for a realistic description of any observables coming from it 
 \citep{Lazzati2017b, Gottlieb2018}. 
 The numerical setup from Lazzati et al. \citep{Lazzati2017b}  
 is an injected jet with luminosity 
 of $L_j = 10^{50} {\rm erg s^{-1}}$, an initial opening angle of 
 $\theta_j = 16^{\circ}$ and the duration of this engine was 
 defined to be $t_{enigne} = 1 {\rm s}$. 
 
 Through the isotropic equivalent energy three different components can be  
 identified. The core of the outflow, which is the initially injected jet 
 modified through the interaction with the ejecta and is the brightest part 
 confined in  $\theta_{jet}\sim 15^{\circ}$.
 The surrounding material of the jet that forms a hot bubble is the energized 
  cocoon which occupies a region between $15^{\circ}-45^{\circ}$. 
  The third component is a fairly isotropic 	wide angle structure 
  that stops at an angle of $65^{\circ}$. From the initial energy 
  of $10^{50} {\rm erg}$ that was injected, $5.5 \times 10^{49} {\rm erg}$ 
  stay in the confined jet, $3.8 \times 10^{48} {\rm erg}$ are given 
  to the surrounding cocoon and $7\times 10^{47}{\rm erg}$ are found in the  
  shocked ambient medium. The rest of the energy is stored in slow 
  moving material ($\Gamma < 1.1$). Figure \ref{fig:Lazz} shows  
  the results from \citep{Lazzati2017b},  
  in the left panel the isotropic equivalent energy where the three 
  components can be identified, in the right panel the peak photon energy 
  is plotted as seen from different angles. The cocoon emission 
  was studied in detail also by \citep{Gottlieb2018}. Their main focus was 
  the appearance of a kilonova following the radioactive heating 
  of the merger ejecta.
  
\textbf{Jet with core and sheath}
In a similar spirit  \citep{Kathirgamaraju2018} simulated the off-axis 
emission from a short GRB jet including magnetic field. 
They argued that for a realistic jet model, one whose Lorentz factor and 
luminosity vary smoothly with angle, detection can be achieved  for 
broader range of viewing angles. In figure \ref{fig:kath} the luminosity 
and Lorentz factor is shown from their model. It is clear that even 
in angles larger than $20^{\circ}$ the luminosity from the jet 
is significant. As the jet breaks out from the cocoon, 
the prompt emission is released \citep{Gottlieb2018b,Bromberg2018}. 
In figure \ref{fig:got}  the time that the shock breaks out 
is illustrated, from a simulation of \citep{Gottlieb2018b}. 
As the shock propagates through the expanding BNS ejecta it 
accumulates mass on top of the jet head. The wide parts of the jet 
are not collimated, they propagate conically inside the mass cloud. 
If the engine operates long enough, the shock breaks out and it is 
not choked inside the ejecta giving all its energy to them.  
The break out of this shock in the magentised case was studied 
by \citep{Bromberg2018}.

\begin{figure*}
  \begin{center}
    \includegraphics[width=0.8\columnwidth]{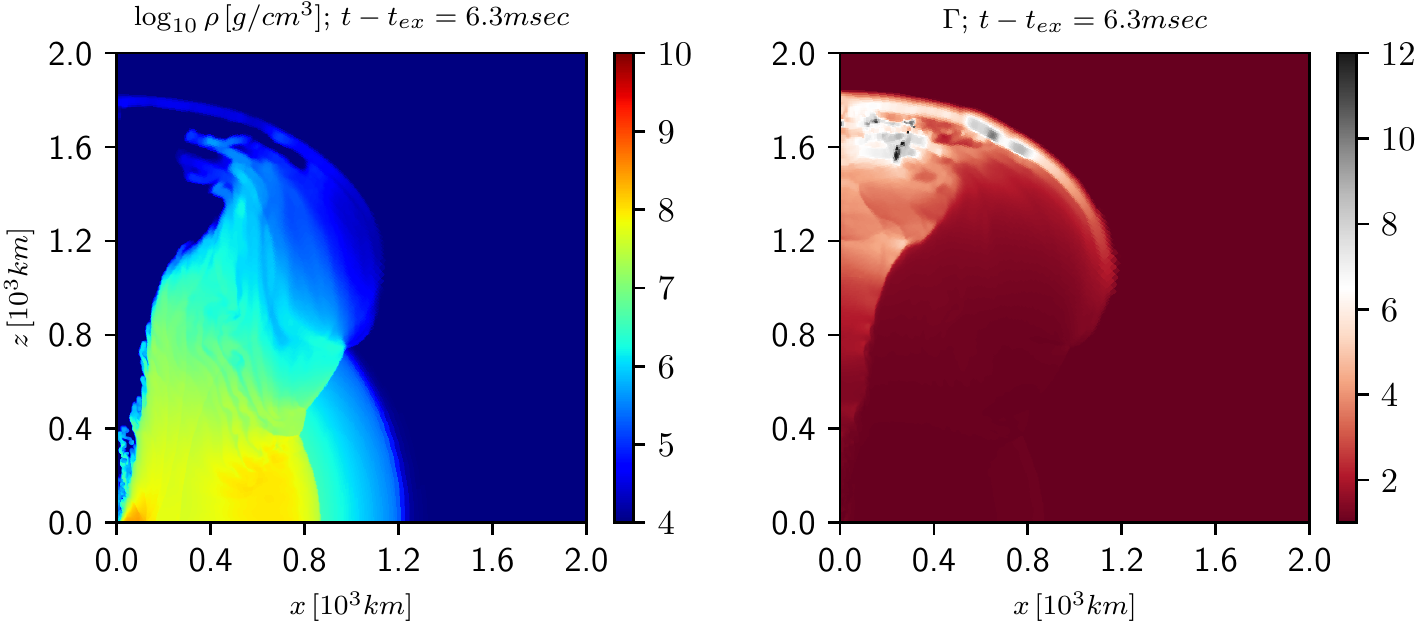}
  \end{center}
 \caption{A late magnetic explosion triggered by the collapse of the compact 
 remnant is shown, a model similar to the one from 
  \citep{Nathanail2018}. The amount of magnetic energy 
 released is on the order of $5\times 10^{51} {\rm erg}$. The left panel shows 
 the density and the right panel the Lorentz factor. The snapshot is taken 
 at the time that the shock enters the low density region and expands 
 sideways. It is interesting to note that there is no clear relativistic core  
  with a small opening angle, there are parts of the outflow in  bigger 
  angles that go slightly faster.}
  \label{fig:nath}
\end{figure*}

\textbf{Magnetic explosion}
It was argued in  the last part of section \ref{sec:BNS} that 
if the collapse of the compact remnant comes late (after a second), 
then the small amount of mass left at the torus cannot give 
a sufficient boundary for a magnetic jet to be launched. 
As such all the magnetic energy dissipates away producing 
an explosion. In figure \ref{fig:nath} such an explosion is depicted 
at the time that the outflow has entered a low density region.  
The main characteristics of a cocoon are still entering this picture. 
A main difference is that there does not exist an easily distinguishable 
relativistic core with a small opening angle. Faster moving 
material can be found in  bigger angles as can be seen in 
figure \ref{fig:nath}. This is a model from \citep{Nathanail2018b}

\textbf{Afterglow}
In late observations, following a BNS merger event, it is important to 
understand the signatures from different components and the differences in 
observations from different models. As the outflow, that was produced  
from the BNS merger, hits the ISM, a shock is produced 
were particles are energized and emit synchrotron  radiation. The outflow 
continues to  sweep up matter and begin to decelerate. This is the standard 
picture for the source of a GRB afterglow. In case of short GRBs from BNS 
mergers this has been discussed significantly before the detection 
of GW170817 \citep{Nakar2011,Piran2013,Hotokezaka2015MNRAS}.
Afterglow model predictions from numerical simulations have been 
studied in the context of long GRBs (e.g. \citep{vanEerten2011}), 
also as seen off-axis \citep{vanEerten2012}.

After the coincident detection of GW170817 together with GRB170817A and the following 
afterglow observations there 
is an enormous effort to analyze the data and fit them with realistic 
models in order to clarify what are the actual components  
that powered such emission. It would be unrealistic to review 
such ongoing effort. We restrict ourselves  to a brief overview 
of observations and the corresponding modeling of them. 

The prompt gamma-ray emission was reported in \citep{Goldstein2017,Savchenko2017}.
The first detection of X-rays from the event came nine days later 
\citep{Troja2017,Margutti2017}, whereas the first radio observations 
sixteen days after 
merger \citep{Hallinan2017}. The first interpretation acknowledged that 
we are observing something quite different than other short GRBs 
\citep{Kasliwal2017,Granot2017}.  

The ongoing effort for understanding the EM counterparts of GW170817 
include: afterglow modeling through hydrodynamic simulations of a jet 
propagating through the merger ejecta 
\citep{Lazzati2017c,Xie2018,Duffell2018},
radio imaging that could show the exact morphology of the outflow  
and polarization measurements that could help to distinguish different 
outflow structures 
\citep{Gill2018,Nakar2018b,Zrake2018,Granot2018}. 
Ideas that the merger event did not include  a jet have been 
proposed \citep{Salafia2017,Salafia2018,Tong2018} or models that 
follow the canonical 
picture with a short GRB jet \citep{Lamb2017,Ziaeepour2018}. 
Observation of GW170817 can give a deep understanding of short GRB modeling 
\citep{Veres2018,Nakar2018}.
It has been also proposed that the  afterglow may come from the interaction of the 
fast tail of the BNS ejecta with the ISM \citep{Hotokezaka2018}.
Also indicative may be what will be the appearance of the counter jet
 \citep{Yamazaki2018}.

Before finishing this section, we would like to gather some important points 
that should be kept in mind 
for the study of a relativistic outflow passing through the BNS ejecta. 

\bigskip

\textbf{Critical points}
\begin{itemize}
\item The amount of dynamically ejected matter strongly depend 
on the total mass of the binary and the mass ratio 
\cite{Rosswog1999,Aloy:2005,Dessart2009,Rezzolla:2010,Roberts2011,
Kyutoku2012,Rosswog2013a,Bauswein2013b,
Hotokezaka2013,Foucart2014,Siegel2014,
Wanajo2014,Sekiguchi2015,Radice2016,Sekiguchi2016,Lehner2016,
Siegel2017,Dietrich2017,Bovard2017,Fujibayashi2017,Fujibayashi2017b}.
\item The BNS ejecta are not spherical, but they rather have 
a unique structure for every different EOS used \citep{Bovard2017}.

\item The time that the engine begins to produce an outflow 
is extremely important, since this will depict how much mass has been 
ejected by neutrino and magnetic winds 
\citep{Perego2014,Siegel2014,Murguia-Berthier2016}.
\end{itemize}


\section{Conclusions}
\label{sec:conc}

In the years to come more detections of BNS mergers are expected from 
ground-based interferometers. It is important to analyze in detail 
observations of GW170817 and all its EM counterparts, starting with GRB170817A. 
It is equally important to reproduce realistic physics through numerical 
simulations in order to match and explain observations. 
This brief review can act as a quick introduction to BNS numerical relativity 
simulations for people interested in short GRB outflows through 
BNS ejecta. Or as a brief introduction  to short GRB jet simulations and 
setups used for people from BNS merger simulations. Overall, 
we want to point out the importance of combining knowledge from both 
paths in order for a consistent picture to be drawn at the end. 

In section \ref{sec:BNS} we went through studies from numerical relativity 
for magnetized BNS mergers. We highlight important aspects of this physical 
process as given in the literature. Points, such as the magnetic field amplification, 
the difficulty of launching a relativistic jet and certainly 
the mass ejection during merger and all possible  winds produced 
after merger can become clear through detailed studies.  At the end 
of the section we state several important points (importance is a 
subjective criterion).

Next step is to take  these ingredients from BNS simulations 
and study any outflow emerging after merger. A relativistic outflow 
has been observed from a BNS merger  \citep{Alexander2017,Haggard2017}.
Thus, we need to understand how it was launched and what is the 
initial structure of this outflow. Furthermore, how it will evolve through 
its interaction with the BNS ejecta. In section \ref{sec:shortGRB} we 
briefly go through previous works on these aspects. This is 
a rapidly evolving sub-field, especially after the detection. 
Now, any model and idea can be simulated and be exposed to the data
that followed GW170817. However, we should keep in mind that  
a BNS can have a different evolution even with a slightly difference 
in mass. At the end, modeling and studying outflows of such events 
should be inspired by GW170817.



\vspace{6pt} 

\section*{Acknowledgements}

It is a pleasure to thank D. Giannios, R. Gold, E. Most and 
V. Paschalidis for reading the manuscript
and giving valuable comments.   
The author is supported by an  Alexander von Humboldt Fellowship.


\reftitle{References}





\begin{thebibliography}{999}

\providecommand{\natexlab}[1]{#1}

\bibitem[{The LIGO Scientific Collaboration} and {The Virgo
  Collaboration}(2017)]{Abbott2017}
{The LIGO Scientific Collaboration}.; {The Virgo Collaboration}.
\newblock GW170817: Observation of Gravitational Waves from a Binary Neutron
  Star Inspiral.
\newblock {\em Phys. Rev. Lett.} {\bf 2017}, {\em 119},~161101.
\newblock
  doi:{\changeurlcolor{black}\href{https://doi.org/10.1103/PhysRevLett.119.161101}{\detokenize{10.1103/PhysRevLett.119.161101}}}.

\bibitem[{The LIGO Scientific Collaboration} \em{et~al.}(2017){The LIGO
  Scientific Collaboration}, {the Virgo Collaboration}, {Abbott}, {Abbott},
  {Abbott}, {Acernese}, {Ackley}, a{Adams}, {Adams}, {Addesso}, and
  et~al.]{Abbott2017b}
{The LIGO Scientific Collaboration}.; {the Virgo Collaboration}.; {Abbott},
  B.P.; {Abbott}, R.; {Abbott}, T.D.; {Acernese}, F.; {Ackley}, K.; a{Adams},
  C.; {Adams}, T.; {Addesso}, P.; et~al..
\newblock Multi-messenger Observations of a Binary Neutron Star Merger.
\newblock {\em Astrophys. J. Lett.} {\bf 2017}, {\em 848},~L12.

\bibitem[{Goldstein} \em{et~al.}(2017){Goldstein}, {Veres}, {Burns}, {Briggs},
  {Hamburg}, {Kocevski}, {Wilson-Hodge}, {Preece}, {Poolakkil}, {Roberts},
  {Hui}, {Connaughton}, {Racusin}, {von Kienlin}, {Dal Canton}, {Christensen},
  {Littenberg}, {Siellez}, {Blackburn}, {Broida}, {Bissaldi}, {Cleveland},
  {Gibby}, {Giles}, {Kippen}, {McBreen}, {McEnery}, {Meegan}, {Paciesas}, and
  {Stanbro}]{Goldstein2017}
{Goldstein}, A.; {Veres}, P.; {Burns}, E.; {Briggs}, M.S.; {Hamburg}, R.;
  {Kocevski}, D.; {Wilson-Hodge}, C.A.; {Preece}, R.D.; {Poolakkil}, S.;
  {Roberts}, O.J.; {Hui}, C.M.; {Connaughton}, V.; {Racusin}, J.; {von
  Kienlin}, A.; {Dal Canton}, T.; {Christensen}, N.; {Littenberg}, T.;
  {Siellez}, K.; {Blackburn}, L.; {Broida}, J.; {Bissaldi}, E.; {Cleveland},
  W.H.; {Gibby}, M.H.; {Giles}, M.M.; {Kippen}, R.M.; {McBreen}, S.; {McEnery},
  J.; {Meegan}, C.A.; {Paciesas}, W.S.; {Stanbro}, M.
\newblock {An Ordinary Short Gamma-Ray Burst with Extraordinary Implications:
  Fermi-GBM Detection of GRB 170817A}.
\newblock {\em Astrophys. J. Letters} {\bf 2017}, {\em 848},~L14,
  \href{http://xxx.lanl.gov/abs/1710.05446}{{\normalfont
  [arXiv:astro-ph.HE/1710.05446]}}.
\newblock
  doi:{\changeurlcolor{black}\href{https://doi.org/10.3847/2041-8213/aa8f41}{\detokenize{10.3847/2041-8213/aa8f41}}}.

\bibitem[{Savchenko} \em{et~al.}(2017){Savchenko}, {Ferrigno}, {Kuulkers},
  {Bazzano}, {Bozzo}, {Brandt}, {Chenevez}, {Courvoisier}, {Diehl}, {Domingo},
  {Hanlon}, {Jourdain}, {von Kienlin}, {Laurent}, {Lebrun}, {Lutovinov},
  {Martin-Carrillo}, {Mereghetti}, {Natalucci}, {Rodi}, {Roques}, {Sunyaev},
  and {Ubertini}]{Savchenko2017}
{Savchenko}, V.; {Ferrigno}, C.; {Kuulkers}, E.; {Bazzano}, A.; {Bozzo}, E.;
  {Brandt}, S.; {Chenevez}, J.; {Courvoisier}, T.J.L.; {Diehl}, R.; {Domingo},
  A.; {Hanlon}, L.; {Jourdain}, E.; {von Kienlin}, A.; {Laurent}, P.; {Lebrun},
  F.; {Lutovinov}, A.; {Martin-Carrillo}, A.; {Mereghetti}, S.; {Natalucci},
  L.; {Rodi}, J.; {Roques}, J.P.; {Sunyaev}, R.; {Ubertini}, P.
\newblock {INTEGRAL Detection of the First Prompt Gamma-Ray Signal Coincident
  with the Gravitational-wave Event GW170817}.
\newblock {\em Astrophys. J. Letters} {\bf 2017}, {\em 848},~L15,
  \href{http://xxx.lanl.gov/abs/1710.05449}{{\normalfont
  [arXiv:astro-ph.HE/1710.05449]}}.
\newblock
  doi:{\changeurlcolor{black}\href{https://doi.org/10.3847/2041-8213/aa8f94}{\detokenize{10.3847/2041-8213/aa8f94}}}.

\bibitem[{Coulter} \em{et~al.}(2017){Coulter}, {Foley}, {Kilpatrick}, {Drout},
  {Piro}, {Shappee}, {Siebert}, {Simon}, {Ulloa}, {Kasen}, {Madore},
  {Murguia-Berthier}, {Pan}, {Prochaska}, {Ramirez-Ruiz}, {Rest}, and
  {Rojas-Bravo}]{Coulter2017}
{Coulter}, D.A.; {Foley}, R.J.; {Kilpatrick}, C.D.; {Drout}, M.R.; {Piro},
  A.L.; {Shappee}, B.J.; {Siebert}, M.R.; {Simon}, J.D.; {Ulloa}, N.; {Kasen},
  D.; {Madore}, B.F.; {Murguia-Berthier}, A.; {Pan}, Y.C.; {Prochaska}, J.X.;
  {Ramirez-Ruiz}, E.; {Rest}, A.; {Rojas-Bravo}, C.
\newblock {Swope Supernova Survey 2017a (SSS17a), the optical counterpart to a
  gravitational wave source}.
\newblock {\em Science} {\bf 2017}, {\em 358},~1556--1558,
  \href{http://xxx.lanl.gov/abs/1710.05452}{{\normalfont
  [arXiv:astro-ph.HE/1710.05452]}}.
\newblock
  doi:{\changeurlcolor{black}\href{https://doi.org/10.1126/science.aap9811}{\detokenize{10.1126/science.aap9811}}}.

\bibitem[{Soares-Santos} \em{et~al.}(2017){Soares-Santos}, {Holz}, {Annis},
  {Chornock}, {Herner}, {Berger}, {Brout}, {Chen}, {Kessler}, {Sako}, {Allam},
  {Tucker}, {Butler}, {Palmese}, {Doctor}, {Diehl}, {Frieman}, {Yanny}, {Lin},
  {Scolnic}, {Cowperthwaite}, {Neilsen}, {Marriner}, {Dark Energy Survey}, and
  {Dark Energy Camera GW-EM Collaboration}]{Soares-Santos2017}
{Soares-Santos}, M.; {Holz}, D.E.; {Annis}, J.; {Chornock}, R.; {Herner}, K.;
  {Berger}, E.; {Brout}, D.; {Chen}, H.Y.; {Kessler}, R.; {Sako}, M.; {Allam},
  S.; {Tucker}, D.L.; {Butler}, R.E.; {Palmese}, A.; {Doctor}, Z.; {Diehl},
  H.T.; {Frieman}, J.; {Yanny}, B.; {Lin}, H.; {Scolnic}, D.; {Cowperthwaite},
  P.; {Neilsen}, E.; {Marriner}, J.; {Dark Energy Survey}.; {Dark Energy Camera
  GW-EM Collaboration}.
\newblock {The Electromagnetic Counterpart of the Binary Neutron Star Merger
  LIGO/Virgo GW170817. I. Discovery of the Optical Counterpart Using the Dark
  Energy Camera}.
\newblock {\em Astrophys. J. Letters} {\bf 2017}, {\em 848},~L16,
  \href{http://xxx.lanl.gov/abs/1710.05459}{{\normalfont
  [arXiv:astro-ph.HE/1710.05459]}}.
\newblock
  doi:{\changeurlcolor{black}\href{https://doi.org/10.3847/2041-8213/aa9059}{\detokenize{10.3847/2041-8213/aa9059}}}.

\bibitem[{Arcavi} \em{et~al.}(2017){Arcavi}, {Hosseinzadeh}, {Howell},
  {McCully}, {Poznanski}, {Kasen}, {Barnes}, {Zaltzman}, {Vasylyev}, {Maoz},
  and {Valenti}]{Arcavi2017}
{Arcavi}, I.; {Hosseinzadeh}, G.; {Howell}, D.A.; {McCully}, C.; {Poznanski},
  D.; {Kasen}, D.; {Barnes}, J.; {Zaltzman}, M.; {Vasylyev}, S.; {Maoz}, D.;
  {Valenti}, S.
\newblock {Optical emission from a kilonova following a
  gravitational-wave-detected neutron-star merger}.
\newblock {\em Nature} {\bf 2017}, {\em 551},~64--66,
  \href{http://xxx.lanl.gov/abs/1710.05843}{{\normalfont
  [arXiv:astro-ph.HE/1710.05843]}}.
\newblock
  doi:{\changeurlcolor{black}\href{https://doi.org/10.1038/nature24291}{\detokenize{10.1038/nature24291}}}.

\bibitem[{Nicholl} \em{et~al.}(2017){Nicholl}, {Berger}, {Kasen}, {Metzger},
  {Elias}, {Brice{\~n}o}, {Alexander}, {Blanchard}, {Chornock},
  {Cowperthwaite}, {Eftekhari}, {Fong}, {Margutti}, {Villar}, {Williams},
  {Brown}, {Annis}, {Bahramian}, {Brout}, {Brown}, {Chen}, {Clemens},
  {Dennihy}, {Dunlap}, {Holz}, {Marchesini}, {Massaro}, {Moskowitz},
  {Pelisoli}, {Rest}, {Ricci}, {Sako}, {Soares-Santos}, and
  {Strader}]{Nicholl2017}
{Nicholl}, M.; {Berger}, E.; {Kasen}, D.; {Metzger}, B.D.; {Elias}, J.;
  {Brice{\~n}o}, C.; {Alexander}, K.D.; {Blanchard}, P.K.; {Chornock}, R.;
  {Cowperthwaite}, P.S.; {Eftekhari}, T.; {Fong}, W.; {Margutti}, R.; {Villar},
  V.A.; {Williams}, P.K.G.; {Brown}, W.; {Annis}, J.; {Bahramian}, A.; {Brout},
  D.; {Brown}, D.A.; {Chen}, H.Y.; {Clemens}, J.C.; {Dennihy}, E.; {Dunlap},
  B.; {Holz}, D.E.; {Marchesini}, E.; {Massaro}, F.; {Moskowitz}, N.;
  {Pelisoli}, I.; {Rest}, A.; {Ricci}, F.; {Sako}, M.; {Soares-Santos}, M.;
  {Strader}, J.
\newblock {The Electromagnetic Counterpart of the Binary Neutron Star Merger
  LIGO/Virgo GW170817. III. Optical and UV Spectra of a Blue Kilonova from Fast
  Polar Ejecta}.
\newblock {\em Astrophys. J. Letters} {\bf 2017}, {\em 848},~L18,
  \href{http://xxx.lanl.gov/abs/1710.05456}{{\normalfont
  [arXiv:astro-ph.HE/1710.05456]}}.
\newblock
  doi:{\changeurlcolor{black}\href{https://doi.org/10.3847/2041-8213/aa9029}{\detokenize{10.3847/2041-8213/aa9029}}}.

\bibitem[{Pian} \em{et~al.}(2017){Pian}, {D'Avanzo}, {Benetti}, {Branchesi},
  {Brocato}, {Campana}, {Cappellaro}, {Covino}, {D'Elia}, {Fynbo}, {Getman},
  {Ghirlanda}, {Ghisellini}, {Grado}, {Greco}, {Hjorth}, {Kouveliotou},
  {Levan}, {Limatola}, {Malesani}, {Mazzali}, {Melandri}, {M{\o}ller},
  {Vergani}, and {Vergani}]{Pian2017}
{Pian}, E.; {D'Avanzo}, P.; {Benetti}, S.; {Branchesi}, M.; {Brocato}, E.;
  {Campana}, S.; {Cappellaro}, E.; {Covino}, S.; {D'Elia}, V.; {Fynbo}, J.P.U.;
  {Getman}, F.; {Ghirlanda}, G.; {Ghisellini}, G.; {Grado}, A.; {Greco}, G.;
  {Hjorth}, J.; {Kouveliotou}, C.; {Levan}, A.; {Limatola}, L.; {Malesani}, D.;
  {Mazzali}, P.A.; {Melandri}, A.; {M{\o}ller}, P.; {Vergani}, S.D.; {Vergani},
  D.
\newblock {Spectroscopic identification of r-process nucleosynthesis in a
  double neutron-star merger}.
\newblock {\em Nature} {\bf 2017}, {\em 551},~67--70,
  \href{http://xxx.lanl.gov/abs/1710.05858}{{\normalfont
  [arXiv:astro-ph.HE/1710.05858]}}.
\newblock
  doi:{\changeurlcolor{black}\href{https://doi.org/10.1038/nature24298}{\detokenize{10.1038/nature24298}}}.

\bibitem[{Smartt} and {Chen}(2017)]{Smartt2017}
{Smartt}, S.; {Chen}, T.e.a.
\newblock A kilonova as the electromagnetic counterpart to a gravitational-wave
  source.
\newblock {\em Nature} {\bf 2017}, {\em 551},~75--79,
  \href{http://xxx.lanl.gov/abs/1710.05841}{{\normalfont [1710.05841]}}.
\newblock
  doi:{\changeurlcolor{black}\href{https://doi.org/10.1038/nature24303}{\detokenize{10.1038/nature24303}}}.

\bibitem[{Tanvir} \em{et~al.}(2017){Tanvir}, {Levan},
  {Gonz{\'a}lez-Fern{\'a}ndez}, {Korobkin}, {Mandel}, {Rosswog}, {Hjorth},
  {D'Avanzo}, {Fruchter}, {Fryer}, {Kangas}, {Milvang-Jensen}, {Rosetti},
  {Steeghs}, {Wollaeger}, {Cano}, {Copperwheat}, and {Wijers}]{Tanvir2017}
{Tanvir}, N.R.; {Levan}, A.J.; {Gonz{\'a}lez-Fern{\'a}ndez}, C.; {Korobkin},
  O.; {Mandel}, I.; {Rosswog}, S.; {Hjorth}, J.; {D'Avanzo}, P.; {Fruchter},
  A.S.; {Fryer}, C.L.; {Kangas}, T.; {Milvang-Jensen}, B.; {Rosetti}, S.;
  {Steeghs}, D.; {Wollaeger}, R.T.; {Cano}, Z.; {Copperwheat}, C.M.; {Wijers},
  R.A.M.J.
\newblock {The Emergence of a Lanthanide-rich Kilonova Following the Merger of
  Two Neutron Stars}.
\newblock {\em Astrophys. J. Letters} {\bf 2017}, {\em 848},~L27,
  \href{http://xxx.lanl.gov/abs/1710.05455}{{\normalfont
  [arXiv:astro-ph.HE/1710.05455]}}.
\newblock
  doi:{\changeurlcolor{black}\href{https://doi.org/10.3847/2041-8213/aa90b6}{\detokenize{10.3847/2041-8213/aa90b6}}}.

\bibitem[{Utsumi} \em{et~al.}(2017){Utsumi}, {Tanaka}, {Tominaga}, {Yoshida},
  {Barway}, {Nagayama}, {Zenko}, {Aoki}, {Fujiyoshi}, {Furusawa}, {Kawabata},
  {Koshida}, {Lee}, {Morokuma}, {Motohara}, {Nakata}, {Ohsawa}, {Ohta},
  {Okita}, {Tajitsu}, {Tanaka}, {Terai}, {Yasuda}, {Abe}, {Asakura}, {Bond},
  and {Miyazaki}]{Ustumi2017}
{Utsumi}, Y.; {Tanaka}, M.; {Tominaga}, N.; {Yoshida}, M.; {Barway}, S.;
  {Nagayama}, T.; {Zenko}, T.; {Aoki}, K.; {Fujiyoshi}, T.; {Furusawa}, H.;
  {Kawabata}, K.S.; {Koshida}, S.; {Lee}, C.H.; {Morokuma}, T.; {Motohara}, K.;
  {Nakata}, F.; {Ohsawa}, R.; {Ohta}, K.; {Okita}, H.; {Tajitsu}, A.; {Tanaka},
  I.; {Terai}, T.; {Yasuda}, N.; {Abe}, F.; {Asakura}, Y.; {Bond}, I.A.;
  {Miyazaki}, S.a.
\newblock {J-GEM observations of an electromagnetic counterpart to the neutron
  star merger GW170817}.
\newblock {\em Publ. of the Astr. Soc. of Japan} {\bf 2017}, {\em 69},~101,
  \href{http://xxx.lanl.gov/abs/1710.05848}{{\normalfont
  [arXiv:astro-ph.HE/1710.05848]}}.
\newblock
  doi:{\changeurlcolor{black}\href{https://doi.org/10.1093/pasj/psx118}{\detokenize{10.1093/pasj/psx118}}}.

\bibitem[{Kilpatrick} \em{et~al.}(2017){Kilpatrick}, {Foley}, {Kasen},
  {Murguia-Berthier}, {Ramirez-Ruiz}, {Coulter}, {Drout}, {Piro}, {Shappee},
  {Boutsia}, {Contreras}, {Di Mille}, {Madore}, {Morrell}, {Pan}, {Prochaska},
  {Rest}, {Rojas-Bravo}, {Siebert}, {Simon}, and {Ulloa}]{Kilpatrick2017}
{Kilpatrick}, C.D.; {Foley}, R.J.; {Kasen}, D.; {Murguia-Berthier}, A.;
  {Ramirez-Ruiz}, E.; {Coulter}, D.A.; {Drout}, M.R.; {Piro}, A.L.; {Shappee},
  B.J.; {Boutsia}, K.; {Contreras}, C.; {Di Mille}, F.; {Madore}, B.F.;
  {Morrell}, N.; {Pan}, Y.C.; {Prochaska}, J.X.; {Rest}, A.; {Rojas-Bravo}, C.;
  {Siebert}, M.R.; {Simon}, J.D.; {Ulloa}, N.
\newblock {Electromagnetic evidence that SSS17a is the result of a binary
  neutron star merger}.
\newblock {\em Science} {\bf 2017}, {\em 358},~1583--1587,
  \href{http://xxx.lanl.gov/abs/1710.05434}{{\normalfont
  [arXiv:astro-ph.HE/1710.05434]}}.
\newblock
  doi:{\changeurlcolor{black}\href{https://doi.org/10.1126/science.aaq0073}{\detokenize{10.1126/science.aaq0073}}}.

\bibitem[{Kasliwal} \em{et~al.}(2017){Kasliwal}, {Nakar}, {Singer}, {Kaplan},
  {Cook}, {Van Sistine}, {Lau}, {Fremling}, {Gottlieb}, {Jencson}, {Adams},
  {Feindt}, {Hotokezaka}, {Ghosh}, {Perley}, {Yu}, {Piran}, {Allison},
  {Anupama}, {Balasubramanian}, {Bannister}, {Bally}, {Barnes}, and
  {Barway}]{Kasliwal2017}
{Kasliwal}, M.M.; {Nakar}, E.; {Singer}, L.P.; {Kaplan}, D.L.; {Cook}, D.O.;
  {Van Sistine}, A.; {Lau}, R.M.; {Fremling}, C.; {Gottlieb}, O.; {Jencson},
  J.E.; {Adams}, S.M.; {Feindt}, U.; {Hotokezaka}, K.; {Ghosh}, S.; {Perley},
  D.A.; {Yu}, P.C.; {Piran}, T.; {Allison}, J.R.; {Anupama}, G.C.;
  {Balasubramanian}, A.; {Bannister}, K.W.; {Bally}, J.; {Barnes}, J.;
  {Barway}, S.a.
\newblock {Illuminating gravitational waves: A concordant picture of photons
  from a neutron star merger}.
\newblock {\em Science} {\bf 2017}, {\em 358},~1559--1565,
  \href{http://xxx.lanl.gov/abs/1710.05436}{{\normalfont
  [arXiv:astro-ph.HE/1710.05436]}}.
\newblock
  doi:{\changeurlcolor{black}\href{https://doi.org/10.1126/science.aap9455}{\detokenize{10.1126/science.aap9455}}}.

\bibitem[{Covino} \em{et~al.}(2017){Covino}, {Wiersema}, {Fan}, {Toma},
  {Higgins}, {Melandri}, {D'Avanzo}, {Mundell}, and {Wijers}]{Covino2017}
{Covino}, S.; {Wiersema}, K.; {Fan}, Y.Z.; {Toma}, K.; {Higgins}, A.B.;
  {Melandri}, A.; {D'Avanzo}, P.; {Mundell}, C.G.; {Wijers}, R.A.M.J.
\newblock {The unpolarized macronova associated with the gravitational wave
  event GW 170817}.
\newblock {\em Nature Astronomy} {\bf 2017}, {\em 1},~791--794,
  \href{http://xxx.lanl.gov/abs/1710.05849}{{\normalfont
  [arXiv:astro-ph.HE/1710.05849]}}.
\newblock
  doi:{\changeurlcolor{black}\href{https://doi.org/10.1038/s41550-017-0285-z}{\detokenize{10.1038/s41550-017-0285-z}}}.

\bibitem[{Cowperthwaite} \em{et~al.}(2017){Cowperthwaite}, {Berger}, {Villar},
  {Metzger}, {Nicholl}, {Chornock}, {Blanchard}, {Fong}, {Margutti},
  {Soares-Santos}, {Alexander}, {Allam}, {Annis}, {Brout}, {Brown}, {Butler},
  {Chen}, {Diehl}, {Doctor}, {Drout}, {Eftekhari}, {Farr}, {Finley}, {Foley},
  {Frieman}, {Fryer}, {Garc{\'{\i}}a-Bellido}, {Gill}, {Guillochon}, {Herner},
  {Holz}, {Kasen}, {Kessler}, {Marriner}, {Matheson}, {Neilsen}, {Quataert},
  {Palmese}, {Rest}, {Sako}, {Scolnic}, {Smith}, {Tucker}, {Williams},
  {Balbinot}, {Carlin}, {Cook}, {Durret}, {Li}, {Lopes}, {Louren{\c c}o},
  {Marshall}, {Medina}, {Muir}, {Mu{\~n}oz}, {Sauseda}, {Schlegel}, {Secco},
  {Vivas}, {Wester}, {Zenteno}, {Zhang}, {Abbott}, {Banerji}, {Bechtol},
  {Benoit-L{\'e}vy}, {Bertin}, {Buckley-Geer}, {Burke}, {Capozzi}, {Carnero
  Rosell}, {Carrasco Kind}, {Castander}, {Crocce}, {Cunha}, {D'Andrea}, {da
  Costa}, {Davis}, {DePoy}, {Desai}, {Dietrich}, {Drlica-Wagner}, {Eifler},
  {Evrard}, {Fernandez}, {Flaugher}, {Fosalba}, {Gaztanaga}, {Gerdes},
  {Giannantonio}, {Goldstein}, {Gruen}, {Gruendl}, {Gutierrez}, {Honscheid},
  {Jain}, {James}, {Jeltema}, {Johnson}, {Johnson}, {Kent}, {Krause}, {Kron},
  {Kuehn}, {Nuropatkin}, {Lahav}, {Lima}, {Lin}, {Maia}, {March}, {Martini},
  {McMahon}, {Menanteau}, {Miller}, {Miquel}, {Mohr}, {Neilsen}, {Nichol},
  {Ogando}, {Plazas}, {Roe}, {Romer}, {Roodman}, {Rykoff}, {Sanchez},
  {Scarpine}, {Schindler}, {Schubnell}, {Sevilla-Noarbe}, {Smith}, {Smith},
  {Sobreira}, {Suchyta}, {Swanson}, {Tarle}, {Thomas}, {Thomas}, {Troxel},
  {Vikram}, {Walker}, {Wechsler}, {Weller}, {Yanny}, and
  {Zuntz}]{Cowperthwaite2017}
{Cowperthwaite}, P.S.; {Berger}, E.; {Villar}, V.A.; {Metzger}, B.D.;
  {Nicholl}, M.; {Chornock}, R.; {Blanchard}, P.K.; {Fong}, W.; {Margutti}, R.;
  {Soares-Santos}, M.; {Alexander}, K.D.; {Allam}, S.; {Annis}, J.; {Brout},
  D.; {Brown}, D.A.; {Butler}, R.E.; {Chen}, H.Y.; {Diehl}, H.T.; {Doctor}, Z.;
  {Drout}, M.R.; {Eftekhari}, T.; {Farr}, B.; {Finley}, D.A.; {Foley}, R.J.;
  {Frieman}, J.A.; {Fryer}, C.L.; {Garc{\'{\i}}a-Bellido}, J.; {Gill}, M.S.S.;
  {Guillochon}, J.; {Herner}, K.; {Holz}, D.E.; {Kasen}, D.; {Kessler}, R.;
  {Marriner}, J.; {Matheson}, T.; {Neilsen}, Jr., E.H.; {Quataert}, E.;
  {Palmese}, A.; {Rest}, A.; {Sako}, M.; {Scolnic}, D.M.; {Smith}, N.;
  {Tucker}, D.L.; {Williams}, P.K.G.; {Balbinot}, E.; {Carlin}, J.L.; {Cook},
  E.R.; {Durret}, F.; {Li}, T.S.; {Lopes}, P.A.A.; {Louren{\c c}o}, A.C.C.;
  {Marshall}, J.L.; {Medina}, G.E.; {Muir}, J.; {Mu{\~n}oz}, R.R.; {Sauseda},
  M.; {Schlegel}, D.J.; {Secco}, L.F.; {Vivas}, A.K.; {Wester}, W.; {Zenteno},
  A.; {Zhang}, Y.; {Abbott}, T.M.C.; {Banerji}, M.; {Bechtol}, K.;
  {Benoit-L{\'e}vy}, A.; {Bertin}, E.; {Buckley-Geer}, E.; {Burke}, D.L.;
  {Capozzi}, D.; {Carnero Rosell}, A.; {Carrasco Kind}, M.; {Castander}, F.J.;
  {Crocce}, M.; {Cunha}, C.E.; {D'Andrea}, C.B.; {da Costa}, L.N.; {Davis}, C.;
  {DePoy}, D.L.; {Desai}, S.; {Dietrich}, J.P.; {Drlica-Wagner}, A.; {Eifler},
  T.F.; {Evrard}, A.E.; {Fernandez}, E.; {Flaugher}, B.; {Fosalba}, P.;
  {Gaztanaga}, E.; {Gerdes}, D.W.; {Giannantonio}, T.; {Goldstein}, D.A.;
  {Gruen}, D.; {Gruendl}, R.A.; {Gutierrez}, G.; {Honscheid}, K.; {Jain}, B.;
  {James}, D.J.; {Jeltema}, T.; {Johnson}, M.W.G.; {Johnson}, M.D.; {Kent}, S.;
  {Krause}, E.; {Kron}, R.; {Kuehn}, K.; {Nuropatkin}, N.; {Lahav}, O.; {Lima},
  M.; {Lin}, H.; {Maia}, M.A.G.; {March}, M.; {Martini}, P.; {McMahon}, R.G.;
  {Menanteau}, F.; {Miller}, C.J.; {Miquel}, R.; {Mohr}, J.J.; {Neilsen}, E.;
  {Nichol}, R.C.; {Ogando}, R.L.C.; {Plazas}, A.A.; {Roe}, N.; {Romer}, A.K.;
  {Roodman}, A.; {Rykoff}, E.S.; {Sanchez}, E.; {Scarpine}, V.; {Schindler},
  R.; {Schubnell}, M.; {Sevilla-Noarbe}, I.; {Smith}, M.; {Smith}, R.C.;
  {Sobreira}, F.; {Suchyta}, E.; {Swanson}, M.E.C.; {Tarle}, G.; {Thomas}, D.;
  {Thomas}, R.C.; {Troxel}, M.A.; {Vikram}, V.; {Walker}, A.R.; {Wechsler},
  R.H.; {Weller}, J.; {Yanny}, B.; {Zuntz}, J.
\newblock {The Electromagnetic Counterpart of the Binary Neutron Star Merger
  LIGO/Virgo GW170817. II. UV, Optical, and Near-infrared Light Curves and
  Comparison to Kilonova Models}.
\newblock {\em Astrophys. J. Lett.} {\bf 2017}, {\em 848},~L17,
  \href{http://xxx.lanl.gov/abs/1710.05840}{{\normalfont
  [arXiv:astro-ph.HE/1710.05840]}}.
\newblock
  doi:{\changeurlcolor{black}\href{https://doi.org/10.3847/2041-8213/aa8fc7}{\detokenize{10.3847/2041-8213/aa8fc7}}}.

\bibitem[{Buckley} \em{et~al.}(2018){Buckley}, {Andreoni}, {Barway}, {Cooke},
  {Crawford}, {Gorbovskoy}, {Gromadzki}, {Lipunov}, {Mao}, {Potter},
  {Pretorius}, {Pritchard}, {Romero-Colmenero}, {Shara}, {V{\"a}is{\"a}nen},
  and {Williams}]{Buckley2018}
{Buckley}, D.A.H.; {Andreoni}, I.; {Barway}, S.; {Cooke}, J.; {Crawford}, S.M.;
  {Gorbovskoy}, E.; {Gromadzki}, M.; {Lipunov}, V.; {Mao}, J.; {Potter}, S.B.;
  {Pretorius}, M.L.; {Pritchard}, T.A.; {Romero-Colmenero}, E.; {Shara}, M.M.;
  {V{\"a}is{\"a}nen}, P.; {Williams}, T.B.
\newblock {A comparison between SALT/SAAO observations and kilonova models for
  AT 2017gfo: the first electromagnetic counterpart of a gravitational wave
  transient - GW170817}.
\newblock {\em Mon. Not. R. Astron. Soc.} {\bf 2018}, {\em 474},~L71--L75,
  \href{http://xxx.lanl.gov/abs/1710.05855}{{\normalfont
  [arXiv:astro-ph.HE/1710.05855]}}.
\newblock
  doi:{\changeurlcolor{black}\href{https://doi.org/10.1093/mnrasl/slx196}{\detokenize{10.1093/mnrasl/slx196}}}.

\bibitem[{Drout} \em{et~al.}(2017){Drout}, {Piro}, {Shappee}, {Kilpatrick},
  {Simon}, {Contreras}, {Coulter}, {Foley}, {Siebert}, {Morrell}, {Boutsia},
  {Di Mille}, {Holoien}, {Kasen}, {Kollmeier}, {Madore}, {Monson},
  {Murguia-Berthier}, {Pan}, {Prochaska}, {Ramirez-Ruiz}, {Rest}, {Adams},
  {Alatalo}, {Ba{\~n}ados}, {Baughman}, {Beers}, {Bernstein}, {Bitsakis},
  {Campillay}, {Hansen}, {Higgs}, {Ji}, {Maravelias}, {Marshall}, {Moni Bidin},
  {Prieto}, {Rasmussen}, {Rojas-Bravo}, {Strom}, {Ulloa},
  {Vargas-Gonz{\'a}lez}, {Wan}, and {Whitten}]{Drout2017}
{Drout}, M.R.; {Piro}, A.L.; {Shappee}, B.J.; {Kilpatrick}, C.D.; {Simon},
  J.D.; {Contreras}, C.; {Coulter}, D.A.; {Foley}, R.J.; {Siebert}, M.R.;
  {Morrell}, N.; {Boutsia}, K.; {Di Mille}, F.; {Holoien}, T.W.S.; {Kasen}, D.;
  {Kollmeier}, J.A.; {Madore}, B.F.; {Monson}, A.J.; {Murguia-Berthier}, A.;
  {Pan}, Y.C.; {Prochaska}, J.X.; {Ramirez-Ruiz}, E.; {Rest}, A.; {Adams}, C.;
  {Alatalo}, K.; {Ba{\~n}ados}, E.; {Baughman}, J.; {Beers}, T.C.; {Bernstein},
  R.A.; {Bitsakis}, T.; {Campillay}, A.; {Hansen}, T.T.; {Higgs}, C.R.; {Ji},
  A.P.; {Maravelias}, G.; {Marshall}, J.L.; {Moni Bidin}, C.; {Prieto}, J.L.;
  {Rasmussen}, K.C.; {Rojas-Bravo}, C.; {Strom}, A.L.; {Ulloa}, N.;
  {Vargas-Gonz{\'a}lez}, J.; {Wan}, Z.; {Whitten}, D.D.
\newblock {Light Curves of the Neutron Star Merger GW170817/SSS17a:
  Implications for R-Process Nucleosynthesis}.
\newblock {\em Science, in press} {\bf 2017},
  \href{http://xxx.lanl.gov/abs/1710.05443}{{\normalfont
  [arXiv:astro-ph.HE/1710.05443]}}.

\bibitem[{Evans} \em{et~al.}(2017){Evans}, {Cenko}, {Kennea}, {Emery}, {Kuin},
  {Korobkin}, {Wollaeger}, {Tagliaferri}, {Tanvir}, and
  {Tohuvavohu}]{Evans2017}
{Evans}, P.A.; {Cenko}, S.B.; {Kennea}, J.A.; {Emery}, S.W.K.; {Kuin}, N.P.M.;
  {Korobkin}, O.; {Wollaeger}, R.T.; {Tagliaferri}, G.; {Tanvir}, N.R.;
  {Tohuvavohu}, A.
\newblock {Swift and NuSTAR observations of GW170817: Detection of a blue
  kilonova}.
\newblock {\em Science} {\bf 2017}, {\em 358},~1565--1570,
  \href{http://xxx.lanl.gov/abs/1710.05437}{{\normalfont
  [arXiv:astro-ph.HE/1710.05437]}}.
\newblock
  doi:{\changeurlcolor{black}\href{https://doi.org/10.1126/science.aap9580}{\detokenize{10.1126/science.aap9580}}}.

\bibitem[{Arcavi}(2018)]{Arcavi2018}
{Arcavi}, I.
\newblock {The First Hours of the GW170817 Kilonova and the Importance of Early
  Optical and Ultraviolet Observations for Constraining Emission Models}.
\newblock {\em Astrophys. J. Lett.} {\bf 2018}, {\em 855},~L23,
  \href{http://xxx.lanl.gov/abs/1802.02164}{{\normalfont
  [arXiv:astro-ph.HE/1802.02164]}}.
\newblock
  doi:{\changeurlcolor{black}\href{https://doi.org/10.3847/2041-8213/aab267}{\detokenize{10.3847/2041-8213/aab267}}}.

\bibitem[{Eichler} \em{et~al.}(1989){Eichler}, {Livio}, {Piran}, and
  {Schramm}]{Eichler89}
{Eichler}, D.; {Livio}, M.; {Piran}, T.; {Schramm}, D.N.
\newblock {Nucleosynthesis, neutrino bursts and gamma-rays from coalescing
  neutron stars}.
\newblock {\em Nature} {\bf 1989}, {\em 340},~126--128.
\newblock
  doi:{\changeurlcolor{black}\href{https://doi.org/10.1038/340126a0}{\detokenize{10.1038/340126a0}}}.

\bibitem[{Narayan} \em{et~al.}(1992){Narayan}, {Paczynski}, and
  {Piran}]{Narayan92}
{Narayan}, R.; {Paczynski}, B.; {Piran}, T.
\newblock {Gamma-ray bursts as the death throes of massive binary stars}.
\newblock {\em Astrophys. J. Lett.} {\bf 1992}, {\em 395},~L83--L86,
  \href{http://xxx.lanl.gov/abs/astro-ph/9204001}{{\normalfont
  [astro-ph/9204001]}}.
\newblock
  doi:{\changeurlcolor{black}\href{https://doi.org/10.1086/186493}{\detokenize{10.1086/186493}}}.

\bibitem[Mochkovitch \em{et~al.}(1993)Mochkovitch, Hernanz, Isern, and
  Martin]{Mochkovitch93}
Mochkovitch, R.; Hernanz, M.; Isern, J.; Martin, X.
\newblock {\em Nature} {\bf 1993}, {\em 361},~236.

\bibitem[{Annala} \em{et~al.}(2018){Annala}, {Gorda}, {Kurkela}, and
  {Vuorinen}]{Annala2017}
{Annala}, E.; {Gorda}, T.; {Kurkela}, A.; {Vuorinen}, A.
\newblock {Gravitational-Wave Constraints on the Neutron-Star-Matter Equation
  of State}.
\newblock {\em Phys. Rev. Lett.} {\bf 2018}, {\em 120},~172703,
  \href{http://xxx.lanl.gov/abs/1711.02644}{{\normalfont
  [arXiv:astro-ph.HE/1711.02644]}}.
\newblock
  doi:{\changeurlcolor{black}\href{https://doi.org/10.1103/PhysRevLett.120.172703}{\detokenize{10.1103/PhysRevLett.120.172703}}}.

\bibitem[{Bauswein} \em{et~al.}(2017){Bauswein}, {Just}, {Janka}, and
  {Stergioulas}]{Bauswein2017b}
{Bauswein}, A.; {Just}, O.; {Janka}, H.T.; {Stergioulas}, N.
\newblock {Neutron-star Radius Constraints from GW170817 and Future
  Detections}.
\newblock {\em Astrophys. J. Lett.} {\bf 2017}, {\em 850},~L34,
  \href{http://xxx.lanl.gov/abs/1710.06843}{{\normalfont
  [arXiv:astro-ph.HE/1710.06843]}}.
\newblock
  doi:{\changeurlcolor{black}\href{https://doi.org/10.3847/2041-8213/aa9994}{\detokenize{10.3847/2041-8213/aa9994}}}.

\bibitem[{Margalit} and {Metzger}(2017)]{Margalit2017}
{Margalit}, B.; {Metzger}, B.D.
\newblock {Constraining the Maximum Mass of Neutron Stars from Multi-messenger
  Observations of GW170817}.
\newblock {\em Astrophys. J. Lett.} {\bf 2017}, {\em 850},~L19,
  \href{http://xxx.lanl.gov/abs/1710.05938}{{\normalfont
  [arXiv:astro-ph.HE/1710.05938]}}.
\newblock
  doi:{\changeurlcolor{black}\href{https://doi.org/10.3847/2041-8213/aa991c}{\detokenize{10.3847/2041-8213/aa991c}}}.

\bibitem[{Radice} \em{et~al.}(2018){Radice}, {Perego}, {Zappa}, and
  {Bernuzzi}]{Radice2017b}
{Radice}, D.; {Perego}, A.; {Zappa}, F.; {Bernuzzi}, S.
\newblock {GW170817: Joint Constraint on the Neutron Star Equation of State
  from Multimessenger Observations}.
\newblock {\em Astrophys. J. Lett.} {\bf 2018}, {\em 852},~L29,
  \href{http://xxx.lanl.gov/abs/1711.03647}{{\normalfont
  [arXiv:astro-ph.HE/1711.03647]}}.
\newblock
  doi:{\changeurlcolor{black}\href{https://doi.org/10.3847/2041-8213/aaa402}{\detokenize{10.3847/2041-8213/aaa402}}}.

\bibitem[{Rezzolla} \em{et~al.}(2018){Rezzolla}, {Most}, and
  {Weih}]{Rezzolla2017}
{Rezzolla}, L.; {Most}, E.R.; {Weih}, L.R.
\newblock {Using Gravitational-wave Observations and Quasi-universal Relations
  to Constrain the Maximum Mass of Neutron Stars}.
\newblock {\em Astrophys. J. Lett.} {\bf 2018}, {\em 852},~L25,
  \href{http://xxx.lanl.gov/abs/1711.00314}{{\normalfont
  [arXiv:astro-ph.HE/1711.00314]}}.
\newblock
  doi:{\changeurlcolor{black}\href{https://doi.org/10.3847/2041-8213/aaa401}{\detokenize{10.3847/2041-8213/aaa401}}}.

\bibitem[{Ruiz} \em{et~al.}(2018){Ruiz}, {Shapiro}, and {Tsokaros}]{Ruiz2017}
{Ruiz}, M.; {Shapiro}, S.L.; {Tsokaros}, A.
\newblock {GW170817, general relativistic magnetohydrodynamic simulations, and
  the neutron star maximum mass}.
\newblock {\em Phys. Rev. D} {\bf 2018}, {\em 97},~021501,
  \href{http://xxx.lanl.gov/abs/1711.00473}{{\normalfont
  [arXiv:astro-ph.HE/1711.00473]}}.
\newblock
  doi:{\changeurlcolor{black}\href{https://doi.org/10.1103/PhysRevD.97.021501}{\detokenize{10.1103/PhysRevD.97.021501}}}.

\bibitem[{Shibata} \em{et~al.}(2017){Shibata}, {Fujibayashi}, {Hotokezaka},
  {Kiuchi}, {Kyutoku}, {Sekiguchi}, and {Tanaka}]{Shibata2017c}
{Shibata}, M.; {Fujibayashi}, S.; {Hotokezaka}, K.; {Kiuchi}, K.; {Kyutoku},
  K.; {Sekiguchi}, Y.; {Tanaka}, M.
\newblock {Modeling GW170817 based on numerical relativity and its
  implications}.
\newblock {\em Phys. Rev. D} {\bf 2017}, {\em 96},~123012,
  \href{http://xxx.lanl.gov/abs/1710.07579}{{\normalfont
  [arXiv:astro-ph.HE/1710.07579]}}.
\newblock
  doi:{\changeurlcolor{black}\href{https://doi.org/10.1103/PhysRevD.96.123012}{\detokenize{10.1103/PhysRevD.96.123012}}}.

\bibitem[{De} \em{et~al.}(2018){De}, {Finstad}, {Lattimer}, {Brown}, {Berger},
  and {Biwer}]{De2018}
{De}, S.; {Finstad}, D.; {Lattimer}, J.M.; {Brown}, D.A.; {Berger}, E.;
  {Biwer}, C.M.
\newblock {Constraining the nuclear equation of state with GW170817}.
\newblock {\em ArXiv e-prints} {\bf 2018},
  \href{http://xxx.lanl.gov/abs/1804.08583}{{\normalfont
  [arXiv:astro-ph.HE/1804.08583]}}.

\bibitem[{Tews} \em{et~al.}(2018{\natexlab{a}}){Tews}, {Carlson}, {Gandolfi},
  and {Reddy}]{Tews2018a}
{Tews}, I.; {Carlson}, J.; {Gandolfi}, S.; {Reddy}, S.
\newblock {Constraining the speed of sound inside neutron stars with chiral
  effective field theory interactions and observations}.
\newblock {\em ArXiv e-prints} {\bf 2018},
  \href{http://xxx.lanl.gov/abs/1801.01923}{{\normalfont
  [arXiv:nucl-th/1801.01923]}}.

\bibitem[{Tews} \em{et~al.}(2018{\natexlab{b}}){Tews}, {Margueron}, and
  {Reddy}]{Tews2018}
{Tews}, I.; {Margueron}, J.; {Reddy}, S.
\newblock {How well does GW170817 constrain the equation of state of dense
  matter?}
\newblock {\em ArXiv e-prints} {\bf 2018},
  \href{http://xxx.lanl.gov/abs/1804.02783}{{\normalfont
  [arXiv:nucl-th/1804.02783]}}.

\bibitem[{Alsing} \em{et~al.}(2017){Alsing}, {Silva}, and {Berti}]{Alsing2017}
{Alsing}, J.; {Silva}, H.O.; {Berti}, E.
\newblock {Evidence for a maximum mass cut-off in the neutron star mass
  distribution and constraints on the equation of state}.
\newblock {\em ArXiv e-prints} {\bf 2017},
  \href{http://xxx.lanl.gov/abs/1709.07889}{{\normalfont
  [arXiv:astro-ph.HE/1709.07889]}}.

\bibitem[{Burgio} \em{et~al.}(2018){Burgio}, {Drago}, {Pagliara}, {Schulze},
  and {Wei}]{Burgio2018}
{Burgio}, G.F.; {Drago}, A.; {Pagliara}, G.; {Schulze}, H.J.; {Wei}, J.B.
\newblock {Has deconfined quark matter been detected during
  GW170817/AT2017gfo?}
\newblock {\em ArXiv e-prints} {\bf 2018},
  \href{http://xxx.lanl.gov/abs/1803.09696}{{\normalfont
  [arXiv:astro-ph.HE/1803.09696]}}.

\bibitem[{Raithel} \em{et~al.}(2018){Raithel}, {{\"O}zel}, and
  {Psaltis}]{Raithel2018}
{Raithel}, C.; {{\"O}zel}, F.; {Psaltis}, D.
\newblock {Tidal deformability from GW170817 as a direct probe of the neutron
  star radius}.
\newblock {\em ArXiv e-prints} {\bf 2018},
  \href{http://xxx.lanl.gov/abs/1803.07687}{{\normalfont
  [arXiv:astro-ph.HE/1803.07687]}}.

\bibitem[{Paschalidis} \em{et~al.}(2017){Paschalidis}, {Yagi},
  {Alvarez-Castillo}, {Blaschke}, and {Sedrakian}]{Paschalidis2017}
{Paschalidis}, V.; {Yagi}, K.; {Alvarez-Castillo}, D.; {Blaschke}, D.B.;
  {Sedrakian}, A.
\newblock {Implications from GW170817 and I-Love-Q relations for relativistic
  hybrid stars}.
\newblock {\em arXiv:1712.00451} {\bf 2017},
  \href{http://xxx.lanl.gov/abs/1712.00451}{{\normalfont
  [arXiv:astro-ph.HE/1712.00451]}}.

\bibitem[{Lattimer} and {Schramm}(1974)]{Lattimer74}
{Lattimer}, J.M.; {Schramm}, D.N.
\newblock {Black-hole-neutron-star collisions}.
\newblock {\em Astrophys. J. Lett.} {\bf 1974}, {\em 192},~L145--L147.
\newblock
  doi:{\changeurlcolor{black}\href{https://doi.org/10.1086/181612}{\detokenize{10.1086/181612}}}.

\bibitem[Li and Paczynski(1998)]{Li:1998}
Li, L.X.; Paczynski, B.
\newblock {Transient events from neutron star mergers}.
\newblock {\em Astrophys. J.} {\bf 1998}, {\em 507},~L59,
  \href{http://xxx.lanl.gov/abs/astro-ph/9807272}{{\normalfont
  [arXiv:astro-ph/astro-ph/9807272]}}.
\newblock
  doi:{\changeurlcolor{black}\href{https://doi.org/10.1086/311680}{\detokenize{10.1086/311680}}}.

\bibitem[{Kasen} \em{et~al.}(2017){Kasen}, {Metzger}, {Barnes}, {Quataert}, and
  {Ramirez-Ruiz}]{Kasen2017}
{Kasen}, D.; {Metzger}, B.; {Barnes}, J.; {Quataert}, E.; {Ramirez-Ruiz}, E.
\newblock {Origin of the heavy elements in binary neutron-star mergers from a
  gravitational-wave event}.
\newblock {\em Nature} {\bf 2017}, {\em 551},~80--84,
  \href{http://xxx.lanl.gov/abs/1710.05463}{{\normalfont
  [arXiv:astro-ph.HE/1710.05463]}}.
\newblock
  doi:{\changeurlcolor{black}\href{https://doi.org/10.1038/nature24453}{\detokenize{10.1038/nature24453}}}.

\bibitem[{Tanaka} \em{et~al.}(2017){Tanaka}, {Utsumi}, {Mazzali}, {Tominaga},
  {Yoshida}, {Sekiguchi}, {Morokuma}, {Motohara}, {Ohta}, {Kawabata}, {Abe},
  {Aoki}, {Asakura}, {Baar}, {Barway}, {Bond}, {Doi}, {Fujiyoshi}, {Furusawa},
  {Honda}, {Itoh}, {Kawabata}, {Kawai}, {Kim}, {Lee}, {Miyazaki}, {Morihana},
  {Nagashima}, {Nagayama}, {Nakaoka}, {Nakata}, {Ohsawa}, {Ohshima}, {Okita},
  {Saito}, {Sumi}, {Tajitsu}, {Takahashi}, {Takayama}, {Tamura}, {Tanaka},
  {Terai}, {Tristram}, {Yasuda}, and {Zenko}]{Tanaka2017}
{Tanaka}, M.; {Utsumi}, Y.; {Mazzali}, P.A.; {Tominaga}, N.; {Yoshida}, M.;
  {Sekiguchi}, Y.; {Morokuma}, T.; {Motohara}, K.; {Ohta}, K.; {Kawabata},
  K.S.; {Abe}, F.; {Aoki}, K.; {Asakura}, Y.; {Baar}, S.; {Barway}, S.; {Bond},
  I.A.; {Doi}, M.; {Fujiyoshi}, T.; {Furusawa}, H.; {Honda}, S.; {Itoh}, Y.;
  {Kawabata}, M.; {Kawai}, N.; {Kim}, J.H.; {Lee}, C.H.; {Miyazaki}, S.;
  {Morihana}, K.; {Nagashima}, H.; {Nagayama}, T.; {Nakaoka}, T.; {Nakata}, F.;
  {Ohsawa}, R.; {Ohshima}, T.; {Okita}, H.; {Saito}, T.; {Sumi}, T.; {Tajitsu},
  A.; {Takahashi}, J.; {Takayama}, M.; {Tamura}, Y.; {Tanaka}, I.; {Terai}, T.;
  {Tristram}, P.J.; {Yasuda}, N.; {Zenko}, T.
\newblock {Kilonova from post-merger ejecta as an optical and near-Infrared
  counterpart of GW170817}.
\newblock {\em Public. Astron. Soc. of Japan} {\bf 2017}, {\em 69},~102,
  \href{http://xxx.lanl.gov/abs/1710.05850}{{\normalfont
  [arXiv:astro-ph.HE/1710.05850]}}.
\newblock
  doi:{\changeurlcolor{black}\href{https://doi.org/10.1093/pasj/psx121}{\detokenize{10.1093/pasj/psx121}}}.

\bibitem[{Murguia-Berthier} \em{et~al.}(2017){Murguia-Berthier},
  {Ramirez-Ruiz}, {Kilpatrick}, {Foley}, {Kasen}, {Lee}, {Piro}, {Coulter},
  {Drout}, {Madore}, {Shappee}, {Pan}, {Prochaska}, {Rest}, {Rojas-Bravo},
  {Siebert}, and {Simon}]{Murguia-Berthier2017}
{Murguia-Berthier}, A.; {Ramirez-Ruiz}, E.; {Kilpatrick}, C.D.; {Foley}, R.J.;
  {Kasen}, D.; {Lee}, W.H.; {Piro}, A.L.; {Coulter}, D.A.; {Drout}, M.R.;
  {Madore}, B.F.; {Shappee}, B.J.; {Pan}, Y.C.; {Prochaska}, J.X.; {Rest}, A.;
  {Rojas-Bravo}, C.; {Siebert}, M.R.; {Simon}, J.D.
\newblock {A Neutron Star Binary Merger Model for GW170817/GRB 170817A/SSS17a}.
\newblock {\em Astrophys. J. Lett.} {\bf 2017}, {\em 848},~L34,
  \href{http://xxx.lanl.gov/abs/1710.05453}{{\normalfont
  [arXiv:astro-ph.HE/1710.05453]}}.
\newblock
  doi:{\changeurlcolor{black}\href{https://doi.org/10.3847/2041-8213/aa91b3}{\detokenize{10.3847/2041-8213/aa91b3}}}.

\bibitem[{Waxman} \em{et~al.}(2017){Waxman}, {Ofek}, {Kushnir}, and
  {Gal-Yam}]{Waxman2017}
{Waxman}, E.; {Ofek}, E.; {Kushnir}, D.; {Gal-Yam}, A.
\newblock {Constraints on the ejecta of the GW170817 neutron-star merger from
  its electromagnetic emission}.
\newblock {\em ArXiv e-prints} {\bf 2017},
  \href{http://xxx.lanl.gov/abs/1711.09638}{{\normalfont
  [arXiv:astro-ph.HE/1711.09638]}}.

\bibitem[{Villar} \em{et~al.}(2017){Villar}, {Guillochon}, {Berger}, {Metzger},
  {Cowperthwaite}, {Nicholl}, {Alexander}, {Blanchard}, {Chornock},
  {Eftekhari}, {Fong}, {Margutti}, and {Williams}]{Villar2017}
{Villar}, V.A.; {Guillochon}, J.; {Berger}, E.; {Metzger}, B.D.;
  {Cowperthwaite}, P.S.; {Nicholl}, M.; {Alexander}, K.D.; {Blanchard}, P.K.;
  {Chornock}, R.; {Eftekhari}, T.; {Fong}, W.; {Margutti}, R.; {Williams},
  P.K.G.
\newblock {The Combined Ultraviolet, Optical, and Near-infrared Light Curves of
  the Kilonova Associated with the Binary Neutron Star Merger GW170817: Unified
  Data Set, Analytic Models, and Physical Implications}.
\newblock {\em Astrophys. J. Letters} {\bf 2017}, {\em 851},~L21,
  \href{http://xxx.lanl.gov/abs/1710.11576}{{\normalfont
  [arXiv:astro-ph.HE/1710.11576]}}.
\newblock
  doi:{\changeurlcolor{black}\href{https://doi.org/10.3847/2041-8213/aa9c84}{\detokenize{10.3847/2041-8213/aa9c84}}}.

\bibitem[{Utsumi} \em{et~al.}(2017){Utsumi}, {Tanaka}, {Tominaga}, {Yoshida},
  {Barway}, {Nagayama}, {Zenko}, {Aoki}, {Fujiyoshi}, {Furusawa}, {Kawabata},
  {Koshida}, {Lee}, {Morokuma}, {Motohara}, {Nakata}, {Ohsawa}, {Ohta},
  {Okita}, {Tajitsu}, {Tanaka}, {Terai}, {Yasuda}, {Abe}, {Asakura}, {Bond},
  {Miyazaki}, {Sumi}, {Tristram}, {Honda}, {Itoh}, {Itoh}, {Kawabata},
  {Morihana}, {Nagashima}, {Nakaoka}, {Ohshima}, {Takahashi}, {Takayama},
  {Aoki}, {Baar}, {Doi}, {Finet}, {Kanda}, {Kawai}, {Kim}, {Kuroda}, {Liu},
  {Matsubayashi}, {Murata}, {Nagai}, {Saito}, {Saito}, {Sako}, {Sekiguchi},
  {Tamura}, {Tanaka}, {Uemura}, and {Yamaguchi}]{Utsumi2017}
{Utsumi}, Y.; {Tanaka}, M.; {Tominaga}, N.; {Yoshida}, M.; {Barway}, S.;
  {Nagayama}, T.; {Zenko}, T.; {Aoki}, K.; {Fujiyoshi}, T.; {Furusawa}, H.;
  {Kawabata}, K.S.; {Koshida}, S.; {Lee}, C.H.; {Morokuma}, T.; {Motohara}, K.;
  {Nakata}, F.; {Ohsawa}, R.; {Ohta}, K.; {Okita}, H.; {Tajitsu}, A.; {Tanaka},
  I.; {Terai}, T.; {Yasuda}, N.; {Abe}, F.; {Asakura}, Y.; {Bond}, I.A.;
  {Miyazaki}, S.; {Sumi}, T.; {Tristram}, P.J.; {Honda}, S.; {Itoh}, R.;
  {Itoh}, Y.; {Kawabata}, M.; {Morihana}, K.; {Nagashima}, H.; {Nakaoka}, T.;
  {Ohshima}, T.; {Takahashi}, J.; {Takayama}, M.; {Aoki}, W.; {Baar}, S.;
  {Doi}, M.; {Finet}, F.; {Kanda}, N.; {Kawai}, N.; {Kim}, J.H.; {Kuroda}, D.;
  {Liu}, W.; {Matsubayashi}, K.; {Murata}, K.L.; {Nagai}, H.; {Saito}, T.;
  {Saito}, Y.; {Sako}, S.; {Sekiguchi}, Y.; {Tamura}, Y.; {Tanaka}, M.;
  {Uemura}, M.; {Yamaguchi}, M.S.
\newblock {J-GEM observations of an electromagnetic counterpart to the neutron
  star merger GW170817}.
\newblock {\em Public. Astron. Soc. of Japan} {\bf 2017}, {\em 69},~101,
  \href{http://xxx.lanl.gov/abs/1710.05848}{{\normalfont
  [arXiv:astro-ph.HE/1710.05848]}}.
\newblock
  doi:{\changeurlcolor{black}\href{https://doi.org/10.1093/pasj/psx118}{\detokenize{10.1093/pasj/psx118}}}.

\bibitem[{Perego} \em{et~al.}(2017){Perego}, {Radice}, and
  {Bernuzzi}]{Perego2017}
{Perego}, A.; {Radice}, D.; {Bernuzzi}, S.
\newblock {AT 2017gfo: An Anisotropic and Three-component Kilonova Counterpart
  of GW170817}.
\newblock {\em Astrophys. J. Lett.} {\bf 2017}, {\em 850},~L37,
  \href{http://xxx.lanl.gov/abs/1711.03982}{{\normalfont
  [arXiv:astro-ph.HE/1711.03982]}}.
\newblock
  doi:{\changeurlcolor{black}\href{https://doi.org/10.3847/2041-8213/aa9ab9}{\detokenize{10.3847/2041-8213/aa9ab9}}}.

\bibitem[{Metzger} \em{et~al.}(2018){Metzger}, {Thompson}, and
  {Quataert}]{Metzger2018}
{Metzger}, B.D.; {Thompson}, T.A.; {Quataert}, E.
\newblock {A Magnetar Origin for the Kilonova Ejecta in GW170817}.
\newblock {\em Astrophys. J.} {\bf 2018}, {\em 856},~101,
  \href{http://xxx.lanl.gov/abs/1801.04286}{{\normalfont
  [arXiv:astro-ph.HE/1801.04286]}}.
\newblock
  doi:{\changeurlcolor{black}\href{https://doi.org/10.3847/1538-4357/aab095}{\detokenize{10.3847/1538-4357/aab095}}}.

\bibitem[{Troja} \em{et~al.}(2017){Troja}, {Piro}, {van Eerten}, {Wollaeger},
  {Im}, {Fox}, {Butler}, {Cenko}, {Sakamoto}, {Fryer}, {Ricci}, {Lien},
  {Veilleux}, {Wieringa}, and {Yoon}]{Troja2017}
{Troja}, E.; {Piro}, L.; {van Eerten}, H.; {Wollaeger}, R.T.; {Im}, M.; {Fox},
  O.D.; {Butler}, N.R.; {Cenko}, S.B.; {Sakamoto}, T.; {Fryer}, C.L.; {Ricci},
  R.; {Lien}, A.; {Veilleux}, S.; {Wieringa}, M.H.; {Yoon}, Y.
\newblock {The X-ray counterpart to the gravitational-wave event GW170817}.
\newblock {\em Nature} {\bf 2017}, {\em 551},~71--74,
  \href{http://xxx.lanl.gov/abs/1710.05433}{{\normalfont
  [arXiv:astro-ph.HE/1710.05433]}}.
\newblock
  doi:{\changeurlcolor{black}\href{https://doi.org/10.1038/nature24290}{\detokenize{10.1038/nature24290}}}.

\bibitem[{Margutti} \em{et~al.}(2017){Margutti}, {Berger}, {Fong}, {Guidorzi},
  {Alexander}, {Metzger}, {Blanchard}, {Cowperthwaite}, {Chornock},
  {Eftekhari}, {Nicholl}, {Villar}, {Williams}, {Annis}, {Brown}, {Chen},
  {Doctor}, {Frieman}, {Holz}, {Sako}, and {Soares-Santos}]{Margutti2017}
{Margutti}, R.; {Berger}, E.; {Fong}, W.; {Guidorzi}, C.; {Alexander}, K.D.;
  {Metzger}, B.D.; {Blanchard}, P.K.; {Cowperthwaite}, P.S.; {Chornock}, R.;
  {Eftekhari}, T.; {Nicholl}, M.; {Villar}, V.A.; {Williams}, P.K.G.; {Annis},
  J.; {Brown}, D.A.; {Chen}, H.; {Doctor}, Z.; {Frieman}, J.A.; {Holz}, D.E.;
  {Sako}, M.; {Soares-Santos}, M.
\newblock {The Electromagnetic Counterpart of the Binary Neutron Star Merger
  LIGO/Virgo GW170817. V. Rising X-Ray Emission from an Off-axis Jet}.
\newblock {\em Astrophys. J. Letters} {\bf 2017}, {\em 848},~L20,
  \href{http://xxx.lanl.gov/abs/1710.05431}{{\normalfont
  [arXiv:astro-ph.HE/1710.05431]}}.
\newblock
  doi:{\changeurlcolor{black}\href{https://doi.org/10.3847/2041-8213/aa9057}{\detokenize{10.3847/2041-8213/aa9057}}}.

\bibitem[{Hallinan} \em{et~al.}(2017){Hallinan}, {Corsi}, {Mooley},
  {Hotokezaka}, {Nakar}, {Kasliwal}, {Kaplan}, {Frail}, {Myers}, {Murphy},
  {De}, {Dobie}, {Allison}, {Bannister}, {Bhalerao}, {Chandra}, {Clarke},
  {Giacintucci}, {Ho}, {Horesh}, {Kassim}, {Kulkarni}, {Lenc}, {Lockman},
  {Lynch}, {Nichols}, {Nissanke}, {Palliyaguru}, {Peters}, {Piran}, {Rana},
  {Sadler}, and {Singer}]{Hallinan2017}
{Hallinan}, G.; {Corsi}, A.; {Mooley}, K.P.; {Hotokezaka}, K.; {Nakar}, E.;
  {Kasliwal}, M.M.; {Kaplan}, D.L.; {Frail}, D.A.; {Myers}, S.T.; {Murphy}, T.;
  {De}, K.; {Dobie}, D.; {Allison}, J.R.; {Bannister}, K.W.; {Bhalerao}, V.;
  {Chandra}, P.; {Clarke}, T.E.; {Giacintucci}, S.; {Ho}, A.Y.Q.; {Horesh}, A.;
  {Kassim}, N.E.; {Kulkarni}, S.R.; {Lenc}, E.; {Lockman}, F.J.; {Lynch}, C.;
  {Nichols}, D.; {Nissanke}, S.; {Palliyaguru}, N.; {Peters}, W.M.; {Piran},
  T.; {Rana}, J.; {Sadler}, E.M.; {Singer}, L.P.
\newblock {A radio counterpart to a neutron star merger}.
\newblock {\em Science} {\bf 2017}, {\em 358},~1579--1583,
  \href{http://xxx.lanl.gov/abs/1710.05435}{{\normalfont
  [arXiv:astro-ph.HE/1710.05435]}}.
\newblock
  doi:{\changeurlcolor{black}\href{https://doi.org/10.1126/science.aap9855}{\detokenize{10.1126/science.aap9855}}}.

\bibitem[{Alexander} \em{et~al.}(2017){Alexander}, {Berger}, {Fong},
  {Williams}, {Guidorzi}, {Margutti}, {Metzger}, {Annis}, {Blanchard}, {Brout},
  {Brown}, {Chen}, {Chornock}, {Cowperthwaite}, {Drout}, {Eftekhari},
  {Frieman}, {Holz}, {Nicholl}, {Rest}, {Sako}, {Soares-Santos}, and
  {Villar}]{Alexander2017}
{Alexander}, K.D.; {Berger}, E.; {Fong}, W.; {Williams}, P.K.G.; {Guidorzi},
  C.; {Margutti}, R.; {Metzger}, B.D.; {Annis}, J.; {Blanchard}, P.K.; {Brout},
  D.; {Brown}, D.A.; {Chen}, H.Y.; {Chornock}, R.; {Cowperthwaite}, P.S.;
  {Drout}, M.; {Eftekhari}, T.; {Frieman}, J.; {Holz}, D.E.; {Nicholl}, M.;
  {Rest}, A.; {Sako}, M.; {Soares-Santos}, M.; {Villar}, V.A.
\newblock {The Electromagnetic Counterpart of the Binary Neutron Star Merger
  LIGO/Virgo GW170817. VI. Radio Constraints on a Relativistic Jet and
  Predictions for Late-time Emission from the Kilonova Ejecta}.
\newblock {\em Astrophys. J. Letters} {\bf 2017}, {\em 848},~L21,
  \href{http://xxx.lanl.gov/abs/1710.05457}{{\normalfont
  [arXiv:astro-ph.HE/1710.05457]}}.
\newblock
  doi:{\changeurlcolor{black}\href{https://doi.org/10.3847/2041-8213/aa905d}{\detokenize{10.3847/2041-8213/aa905d}}}.

\bibitem[{Haggard} \em{et~al.}(2017){Haggard}, {Nynka}, {Ruan}, {Kalogera},
  {Cenko}, {Evans}, and {Kennea}]{Haggard2017}
{Haggard}, D.; {Nynka}, M.; {Ruan}, J.J.; {Kalogera}, V.; {Cenko}, S.B.;
  {Evans}, P.; {Kennea}, J.A.
\newblock {A Deep Chandra X-Ray Study of Neutron Star Coalescence GW170817}.
\newblock {\em Astrophys. J. Letters} {\bf 2017}, {\em 848},~L25,
  \href{http://xxx.lanl.gov/abs/1710.05852}{{\normalfont
  [arXiv:astro-ph.HE/1710.05852]}}.
\newblock
  doi:{\changeurlcolor{black}\href{https://doi.org/10.3847/2041-8213/aa8ede}{\detokenize{10.3847/2041-8213/aa8ede}}}.

\bibitem[{Granot} \em{et~al.}(2017){Granot}, {Guetta}, and {Gill}]{Granot2017}
{Granot}, J.; {Guetta}, D.; {Gill}, R.
\newblock {Lessons from the Short GRB 170817A: The First Gravitational-wave
  Detection of a Binary Neutron Star Merger}.
\newblock {\em Astrophys. J. Letters} {\bf 2017}, {\em 850},~L24,
  \href{http://xxx.lanl.gov/abs/1710.06407}{{\normalfont
  [arXiv:astro-ph.HE/1710.06407]}}.
\newblock
  doi:{\changeurlcolor{black}\href{https://doi.org/10.3847/2041-8213/aa991d}{\detokenize{10.3847/2041-8213/aa991d}}}.

\bibitem[{Mooley} \em{et~al.}(2018){Mooley}, {Nakar}, {Hotokezaka}, {Hallinan},
  {Corsi}, {Frail}, {Horesh}, {Murphy}, {Lenc}, {Kaplan}, {de}, {Dobie},
  {Chandra}, {Deller}, {Gottlieb}, {Kasliwal}, {Kulkarni}, {Myers}, {Nissanke},
  {Piran}, {Lynch}, {Bhalerao}, {Bourke}, {Bannister}, and
  {Singer}]{Mooley2018}
{Mooley}, K.P.; {Nakar}, E.; {Hotokezaka}, K.; {Hallinan}, G.; {Corsi}, A.;
  {Frail}, D.A.; {Horesh}, A.; {Murphy}, T.; {Lenc}, E.; {Kaplan}, D.L.; {de},
  K.; {Dobie}, D.; {Chandra}, P.; {Deller}, A.; {Gottlieb}, O.; {Kasliwal},
  M.M.; {Kulkarni}, S.R.; {Myers}, S.T.; {Nissanke}, S.; {Piran}, T.; {Lynch},
  C.; {Bhalerao}, V.; {Bourke}, S.; {Bannister}, K.W.; {Singer}, L.P.
\newblock {A mildly relativistic wide-angle outflow in the neutron-star merger
  event GW170817}.
\newblock {\em Nature} {\bf 2018}, {\em 554},~207--210,
  \href{http://xxx.lanl.gov/abs/1711.11573}{{\normalfont
  [arXiv:astro-ph.HE/1711.11573]}}.
\newblock
  doi:{\changeurlcolor{black}\href{https://doi.org/10.1038/nature25452}{\detokenize{10.1038/nature25452}}}.

\bibitem[{Ruan} \em{et~al.}(2018){Ruan}, {Nynka}, {Haggard}, {Kalogera}, and
  {Evans}]{Ruan2018}
{Ruan}, J.J.; {Nynka}, M.; {Haggard}, D.; {Kalogera}, V.; {Evans}, P.
\newblock {Brightening X-Ray Emission from GW170817/GRB 170817A: Further
  Evidence for an Outflow}.
\newblock {\em Astrophys. J. Letters} {\bf 2018}, {\em 853},~L4,
  \href{http://xxx.lanl.gov/abs/1712.02809}{{\normalfont
  [arXiv:astro-ph.HE/1712.02809]}}.
\newblock
  doi:{\changeurlcolor{black}\href{https://doi.org/10.3847/2041-8213/aaa4f3}{\detokenize{10.3847/2041-8213/aaa4f3}}}.

\bibitem[{Pooley} \em{et~al.}(2018){Pooley}, {Kumar}, {Wheeler}, and
  {Grossan}]{Pooley2018}
{Pooley}, D.; {Kumar}, P.; {Wheeler}, J.C.; {Grossan}, B.
\newblock {GW170817 Most Likely Made a Black Hole}.
\newblock {\em Astrophys. J. Letters} {\bf 2018}, {\em 859},~L23,
  \href{http://xxx.lanl.gov/abs/1712.03240}{{\normalfont
  [arXiv:astro-ph.HE/1712.03240]}}.
\newblock
  doi:{\changeurlcolor{black}\href{https://doi.org/10.3847/2041-8213/aac3d6}{\detokenize{10.3847/2041-8213/aac3d6}}}.

\bibitem[{Margutti} \em{et~al.}(2018){Margutti}, {Alexander}, {Xie}, {Sironi},
  {Metzger}, {Kathirgamaraju}, {Fong}, {Blanchard}, {Berger}, {MacFadyen},
  {Giannios}, {Guidorzi}, {Hajela}, {Chornock}, {Cowperthwaite}, {Eftekhari},
  {Nicholl}, {Villar}, {Williams}, and {Zrake}]{Margutti2018}
{Margutti}, R.; {Alexander}, K.D.; {Xie}, X.; {Sironi}, L.; {Metzger}, B.D.;
  {Kathirgamaraju}, A.; {Fong}, W.; {Blanchard}, P.K.; {Berger}, E.;
  {MacFadyen}, A.; {Giannios}, D.; {Guidorzi}, C.; {Hajela}, A.; {Chornock},
  R.; {Cowperthwaite}, P.S.; {Eftekhari}, T.; {Nicholl}, M.; {Villar}, V.A.;
  {Williams}, P.K.G.; {Zrake}, J.
\newblock {The Binary Neutron Star Event LIGO/Virgo GW170817 160 Days after
  Merger: Synchrotron Emission across the Electromagnetic Spectrum}.
\newblock {\em Astrophys. J. Letters} {\bf 2018}, {\em 856},~L18,
  \href{http://xxx.lanl.gov/abs/1801.03531}{{\normalfont
  [arXiv:astro-ph.HE/1801.03531]}}.
\newblock
  doi:{\changeurlcolor{black}\href{https://doi.org/10.3847/2041-8213/aab2ad}{\detokenize{10.3847/2041-8213/aab2ad}}}.

\bibitem[{Lyman} \em{et~al.}(2018){Lyman}, {Lamb}, {Levan}, {Mandel}, {Tanvir},
  {Kobayashi}, {Gompertz}, {Hjorth}, {Fruchter}, {Kangas}, {Steeghs}, {Steele},
  {Cano}, {Copperwheat}, {Evans}, {Fynbo}, {Gall}, {Im}, {Izzo}, {Jakobsson},
  {Milvang-Jensen}, {O'Brien}, {Osborne}, {Palazzi}, {Perley}, {Pian},
  {Rosswog}, {Rowlinson}, {Schulze}, {Stanway}, {Sutton}, {Th{\"o}ne}, {de
  Ugarte Postigo}, {Watson}, {Wiersema}, and {Wijers}]{Lyman2018}
{Lyman}, J.D.; {Lamb}, G.P.; {Levan}, A.J.; {Mandel}, I.; {Tanvir}, N.R.;
  {Kobayashi}, S.; {Gompertz}, B.; {Hjorth}, J.; {Fruchter}, A.S.; {Kangas},
  T.; {Steeghs}, D.; {Steele}, I.A.; {Cano}, Z.; {Copperwheat}, C.; {Evans},
  P.A.; {Fynbo}, J.P.U.; {Gall}, C.; {Im}, M.; {Izzo}, L.; {Jakobsson}, P.;
  {Milvang-Jensen}, B.; {O'Brien}, P.; {Osborne}, J.P.; {Palazzi}, E.;
  {Perley}, D.A.; {Pian}, E.; {Rosswog}, S.; {Rowlinson}, A.; {Schulze}, S.;
  {Stanway}, E.R.; {Sutton}, P.; {Th{\"o}ne}, C.C.; {de Ugarte Postigo}, A.;
  {Watson}, D.J.; {Wiersema}, K.; {Wijers}, R.A.M.J.
\newblock {The optical afterglow of the short gamma-ray burst associated with
  GW170817}.
\newblock {\em Nature Astronomy} {\bf 2018},
  \href{http://xxx.lanl.gov/abs/1801.02669}{{\normalfont
  [arXiv:astro-ph.HE/1801.02669]}}.
\newblock
  doi:{\changeurlcolor{black}\href{https://doi.org/10.1038/s41550-018-0511-3}{\detokenize{10.1038/s41550-018-0511-3}}}.

\bibitem[{Li} \em{et~al.}(2018){Li}, {Li}, {Huang}, {Geng}, {Yu}, and
  {Song}]{Li2018}
{Li}, B.; {Li}, L.B.; {Huang}, Y.F.; {Geng}, J.J.; {Yu}, Y.B.; {Song}, L.M.
\newblock {Continued Brightening of the Afterglow of GW170817/GRB 170817A as
  Being Due to a Delayed Energy Injection}.
\newblock {\em Astrophys. J. Letters} {\bf 2018}, {\em 859},~L3,
  \href{http://xxx.lanl.gov/abs/1802.10397}{{\normalfont
  [arXiv:astro-ph.HE/1802.10397]}}.
\newblock
  doi:{\changeurlcolor{black}\href{https://doi.org/10.3847/2041-8213/aac2c5}{\detokenize{10.3847/2041-8213/aac2c5}}}.

\bibitem[{Dobie} \em{et~al.}(2018){Dobie}, {Kaplan}, {Murphy}, {Lenc},
  {Mooley}, {Lynch}, {Corsi}, {Frail}, {Kasliwal}, and {Hallinan}]{Dobie2018}
{Dobie}, D.; {Kaplan}, D.L.; {Murphy}, T.; {Lenc}, E.; {Mooley}, K.P.; {Lynch},
  C.; {Corsi}, A.; {Frail}, D.; {Kasliwal}, M.; {Hallinan}, G.
\newblock {A Turnover in the Radio Light Curve of GW170817}.
\newblock {\em Astrophys. J. Letters} {\bf 2018}, {\em 858},~L15,
  \href{http://xxx.lanl.gov/abs/1803.06853}{{\normalfont
  [arXiv:astro-ph.HE/1803.06853]}}.
\newblock
  doi:{\changeurlcolor{black}\href{https://doi.org/10.3847/2041-8213/aac105}{\detokenize{10.3847/2041-8213/aac105}}}.

\bibitem[{Alexander} \em{et~al.}(2018){Alexander}, {Margutti}, {Blanchard},
  {Fong}, {Berger}, {Hajela}, {Eftekhari}, {Chornock}, {Cowperthwaite},
  {Giannios}, {Guidorzi}, {Kathirgamaraju}, {MacFadyen}, {Metzger}, {Nicholl},
  {Sironi}, {Villar}, {Williams}, {Xie}, and {Zrake}]{Alexander2018}
{Alexander}, K.D.; {Margutti}, R.; {Blanchard}, P.K.; {Fong}, W.; {Berger}, E.;
  {Hajela}, A.; {Eftekhari}, T.; {Chornock}, R.; {Cowperthwaite}, P.S.;
  {Giannios}, D.; {Guidorzi}, C.; {Kathirgamaraju}, A.; {MacFadyen}, A.;
  {Metzger}, B.D.; {Nicholl}, M.; {Sironi}, L.; {Villar}, V.A.; {Williams},
  P.K.G.; {Xie}, X.; {Zrake}, J.
\newblock {A Decline in the X-ray through Radio Emission from GW170817
  Continues to Support an Off-Axis Structured Jet}.
\newblock {\em ArXiv e-prints} {\bf 2018},
  \href{http://xxx.lanl.gov/abs/1805.02870}{{\normalfont
  [arXiv:astro-ph.HE/1805.02870]}}.

\bibitem[{Nynka} \em{et~al.}(2018){Nynka}, {Ruan}, {Haggard}, and
  {Evans}]{Nynka2018}
{Nynka}, M.; {Ruan}, J.J.; {Haggard}, D.; {Evans}, P.A.
\newblock {Fading of the X-ray Afterglow of Neutron Star Merger
  GW170817/GRB170817A at 260 days}.
\newblock {\em ArXiv e-prints} {\bf 2018},
  \href{http://xxx.lanl.gov/abs/1805.04093}{{\normalfont
  [arXiv:astro-ph.HE/1805.04093]}}.

\bibitem[Faber and Rasio(2012)]{Faber2012:lrr}
Faber, J.A.; Rasio, F.A.
\newblock Binary Neutron Star Mergers.
\newblock {\em Living Rev. Relativity} {\bf 2012}, {\em 15}.

\bibitem[Baiotti and Rezzolla(2017)]{Baiotti2016}
Baiotti, L.; Rezzolla, L.
\newblock {Binary neutron-star mergers: a review of Einstein's richest
  laboratory}.
\newblock {\em Rept. Prog. Phys.} {\bf 2017}, {\em 80},~096901,
  \href{http://xxx.lanl.gov/abs/1607.03540}{{\normalfont
  [arXiv:gr-qc/1607.03540]}}.
\newblock
  doi:{\changeurlcolor{black}\href{https://doi.org/10.1088/1361-6633/aa67bb}{\detokenize{10.1088/1361-6633/aa67bb}}}.

\bibitem[{Shibata} \em{et~al.}(2006){Shibata}, {Liu}, {Shapiro}, and
  {Stephens}]{Shibata06c}
{Shibata}, M.; {Liu}, Y.T.; {Shapiro}, S.L.; {Stephens}, B.C.
\newblock {Magnetorotational collapse of massive stellar cores to neutron
  stars: Simulations in full general relativity}.
\newblock {\em Phys. Rev. D} {\bf 2006}, {\em 74},~104026,
  \href{http://xxx.lanl.gov/abs/arXiv:astro-ph/0610840}{{\normalfont
  [arXiv:astro-ph/0610840]}}.
\newblock
  doi:{\changeurlcolor{black}\href{https://doi.org/10.1103/PhysRevD.74.104026}{\detokenize{10.1103/PhysRevD.74.104026}}}.

\bibitem[{Paschalidis}(2017)]{Paschalidis2016}
{Paschalidis}, V.
\newblock {General relativistic simulations of compact binary mergers as
  engines for short gamma-ray bursts}.
\newblock {\em Classical and Quantum Gravity} {\bf 2017}, {\em 34},~084002,
  \href{http://xxx.lanl.gov/abs/1611.01519}{{\normalfont
  [arXiv:astro-ph.HE/1611.01519]}}.
\newblock
  doi:{\changeurlcolor{black}\href{https://doi.org/10.1088/1361-6382/aa61ce}{\detokenize{10.1088/1361-6382/aa61ce}}}.

\bibitem[{Berger}(2014)]{Berger2013b}
{Berger}, E.
\newblock {Short-Duration Gamma-Ray Bursts}.
\newblock {\em Annual Review of Astron. and Astrophys.} {\bf 2014}, {\em
  52},~43--105,  \href{http://xxx.lanl.gov/abs/1311.2603}{{\normalfont
  [arXiv:astro-ph.HE/1311.2603]}}.
\newblock
  doi:{\changeurlcolor{black}\href{https://doi.org/10.1146/annurev-astro-081913-035926}{\detokenize{10.1146/annurev-astro-081913-035926}}}.

\bibitem[{Fong} \em{et~al.}(2015){Fong}, {Berger}, {Margutti}, and
  {Zauderer}]{Fong2015}
{Fong}, W.; {Berger}, E.; {Margutti}, R.; {Zauderer}, B.A.
\newblock {A Decade of Short-duration Gamma-Ray Burst Broadband Afterglows:
  Energetics, Circumburst Densities, and Jet Opening Angles}.
\newblock {\em Astrophys. J.} {\bf 2015}, {\em 815},~102,
  \href{http://xxx.lanl.gov/abs/1509.02922}{{\normalfont
  [arXiv:astro-ph.HE/1509.02922]}}.
\newblock
  doi:{\changeurlcolor{black}\href{https://doi.org/10.1088/0004-637X/815/2/102}{\detokenize{10.1088/0004-637X/815/2/102}}}.

\bibitem[{Rosswog}(2015)]{Rosswog2015}
{Rosswog}, S.
\newblock {The multi-messenger picture of compact binary mergers}.
\newblock {\em International Journal of Modern Physics D} {\bf 2015}, {\em
  24},~1530012--52,  \href{http://xxx.lanl.gov/abs/1501.02081}{{\normalfont
  [arXiv:astro-ph.HE/1501.02081]}}.
\newblock
  doi:{\changeurlcolor{black}\href{https://doi.org/10.1142/S0218271815300128}{\detokenize{10.1142/S0218271815300128}}}.

\bibitem[{Fern{\'a}ndez} \em{et~al.}(2015){Fern{\'a}ndez}, {Quataert},
  {Schwab}, {Kasen}, and {Rosswog}]{Fernandez2015b}
{Fern{\'a}ndez}, R.; {Quataert}, E.; {Schwab}, J.; {Kasen}, D.; {Rosswog}, S.
\newblock {The interplay of disc wind and dynamical ejecta in the aftermath of
  neutron star-black hole mergers}.
\newblock {\em Mon. Not. R. Astron. Soc.} {\bf 2015}, {\em 449},~390--402,
  \href{http://xxx.lanl.gov/abs/1412.5588}{{\normalfont
  [arXiv:astro-ph.HE/1412.5588]}}.
\newblock
  doi:{\changeurlcolor{black}\href{https://doi.org/10.1093/mnras/stv238}{\detokenize{10.1093/mnras/stv238}}}.

\bibitem[{Thielemann} \em{et~al.}(2017){Thielemann}, {Eichler}, {Panov}, and
  {Wehmeyer}]{Thielemann2017b}
{Thielemann}, F.K.; {Eichler}, M.; {Panov}, I.V.; {Wehmeyer}, B.
\newblock {Neutron Star Mergers and Nucleosynthesis of Heavy Elements}.
\newblock {\em Annual Review of Nuclear and Particle Science} {\bf 2017}, {\em
  67},~253--274,  \href{http://xxx.lanl.gov/abs/1710.02142}{{\normalfont
  [arXiv:astro-ph.HE/1710.02142]}}.
\newblock
  doi:{\changeurlcolor{black}\href{https://doi.org/10.1146/annurev-nucl-101916-123246}{\detokenize{10.1146/annurev-nucl-101916-123246}}}.

\bibitem[{Metzger}(2017)]{Metzger2017}
{Metzger}, B.D.
\newblock {Kilonovae}.
\newblock {\em Living Reviews in Relativity} {\bf 2017}, {\em 20},~3,
  \href{http://xxx.lanl.gov/abs/1610.09381}{{\normalfont
  [arXiv:astro-ph.HE/1610.09381]}}.
\newblock
  doi:{\changeurlcolor{black}\href{https://doi.org/10.1007/s41114-017-0006-z}{\detokenize{10.1007/s41114-017-0006-z}}}.

\bibitem[{Paschalidis} and {Stergioulas}(2017)]{Paschalidis2017b}
{Paschalidis}, V.; {Stergioulas}, N.
\newblock {Rotating stars in relativity}.
\newblock {\em Living Reviews in Relativity} {\bf 2017}, {\em 20},~7,
  \href{http://xxx.lanl.gov/abs/1612.03050}{{\normalfont
  [arXiv:astro-ph.HE/1612.03050]}}.
\newblock
  doi:{\changeurlcolor{black}\href{https://doi.org/10.1007/s41114-017-0008-x}{\detokenize{10.1007/s41114-017-0008-x}}}.

\bibitem[Nakar(2007)]{Nakar:2007yr}
Nakar, E.
\newblock {Short-hard gamma-ray bursts}.
\newblock {\em Phys. Rep.} {\bf 2007}, {\em 442},~166--236,
  \href{http://xxx.lanl.gov/abs/astro-ph/0701748}{{\normalfont
  [astro-ph/0701748]}}.
\newblock
  doi:{\changeurlcolor{black}\href{https://doi.org/10.1016/j.physrep.2007.02.005}{\detokenize{10.1016/j.physrep.2007.02.005}}}.

\bibitem[Lee and Ramirez-Ruiz(2007)]{Lee:2007js}
Lee, W.H.; Ramirez-Ruiz, E.
\newblock {The Progenitors of Short Gamma-Ray Bursts}.
\newblock {\em New J. Phys.} {\bf 2007}, {\em 9},~17,
  \href{http://xxx.lanl.gov/abs/astro-ph/0701874}{{\normalfont
  [astro-ph/0701874]}}.
\newblock
  doi:{\changeurlcolor{black}\href{https://doi.org/10.1088/1367-2630/9/1/017}{\detokenize{10.1088/1367-2630/9/1/017}}}.

\bibitem[{Piran}(2005)]{Piran:2004ba}
{Piran}, T.
\newblock {The physics of gamma-ray bursts}.
\newblock {\em Reviews of Modern Physics} {\bf 2005}, {\em 76},~1143--1210,
  \href{http://xxx.lanl.gov/abs/astro-ph/0405503}{{\normalfont
  [astro-ph/0405503]}}.
\newblock
  doi:{\changeurlcolor{black}\href{https://doi.org/10.1103/RevModPhys.76.1143}{\detokenize{10.1103/RevModPhys.76.1143}}}.

\bibitem[Meszaros(2006)]{Meszaros:2006rc}
Meszaros, P.
\newblock {Gamma-Ray Bursts}.
\newblock {\em Rep. Prog. Phys.} {\bf 2006}, {\em 69},~2259--2322,
  \href{http://xxx.lanl.gov/abs/astro-ph/0605208}{{\normalfont
  [astro-ph/0605208]}}.

\bibitem[{Kumar} and {Smoot}(2014)]{Kumar2014}
{Kumar}, P.; {Smoot}, G.F.
\newblock {Some implications of inverse-Compton scattering of hot cocoon
  radiation by relativistic jets in gamma-ray bursts}.
\newblock {\em Mon. Not. R. Astron. Soc.} {\bf 2014}, {\em 445},~528--543,
  \href{http://xxx.lanl.gov/abs/1402.2656}{{\normalfont
  [arXiv:astro-ph.HE/1402.2656]}}.
\newblock
  doi:{\changeurlcolor{black}\href{https://doi.org/10.1093/mnras/stu1638}{\detokenize{10.1093/mnras/stu1638}}}.

\bibitem[{Tauris} \em{et~al.}(2017){Tauris}, {Kramer}, {Freire}, {Wex},
  {Janka}, {Langer}, {Podsiadlowski}, {Bozzo}, {Chaty}, {Kruckow}, {van den
  Heuvel}, {Antoniadis}, {Breton}, and {Champion}]{Tauris2017}
{Tauris}, T.M.; {Kramer}, M.; {Freire}, P.C.C.; {Wex}, N.; {Janka}, H.T.;
  {Langer}, N.; {Podsiadlowski}, P.; {Bozzo}, E.; {Chaty}, S.; {Kruckow}, M.U.;
  {van den Heuvel}, E.P.J.; {Antoniadis}, J.; {Breton}, R.P.; {Champion}, D.J.
\newblock {Formation of Double Neutron Star Systems}.
\newblock {\em Astrophys. J.} {\bf 2017}, {\em 846},~170,
  \href{http://xxx.lanl.gov/abs/1706.09438}{{\normalfont
  [arXiv:astro-ph.HE/1706.09438]}}.
\newblock
  doi:{\changeurlcolor{black}\href{https://doi.org/10.3847/1538-4357/aa7e89}{\detokenize{10.3847/1538-4357/aa7e89}}}.

\bibitem[{Baumgarte} \em{et~al.}(2000){Baumgarte}, {Shapiro}, and
  {Shibata}]{Baumgarte00bb}
{Baumgarte}, T.W.; {Shapiro}, S.L.; {Shibata}, M.
\newblock {On the Maximum Mass of Differentially Rotating Neutron Stars}.
\newblock {\em Astrophys. J. Lett.} {\bf 2000}, {\em 528},~L29--L32,
  \href{http://xxx.lanl.gov/abs/arXiv:astro-ph/9910565}{{\normalfont
  [arXiv:astro-ph/9910565]}}.
\newblock
  doi:{\changeurlcolor{black}\href{https://doi.org/10.1086/312425}{\detokenize{10.1086/312425}}}.

\bibitem[{Sekiguchi} \em{et~al.}(2011){Sekiguchi}, {Kiuchi}, {Kyutoku}, and
  {Shibata}]{Sekiguchi2011}
{Sekiguchi}, Y.; {Kiuchi}, K.; {Kyutoku}, K.; {Shibata}, M.
\newblock {Gravitational Waves and Neutrino Emission from the Merger of Binary
  Neutron Stars}.
\newblock {\em Phys. Rev. Lett.} {\bf 2011}, {\em 107},~051102,
  \href{http://xxx.lanl.gov/abs/1105.2125}{{\normalfont
  [arXiv:gr-qc/1105.2125]}}.
\newblock
  doi:{\changeurlcolor{black}\href{https://doi.org/10.1103/PhysRevLett.107.051102}{\detokenize{10.1103/PhysRevLett.107.051102}}}.

\bibitem[{Paschalidis} \em{et~al.}(2012){Paschalidis}, {Etienne}, and
  {Shapiro}]{Paschalidis2012}
{Paschalidis}, V.; {Etienne}, Z.B.; {Shapiro}, S.L.
\newblock {Importance of cooling in triggering the collapse of hypermassive
  neutron stars}.
\newblock {\em Phys. Rev. D} {\bf 2012}, {\em 86},~064032,
  \href{http://xxx.lanl.gov/abs/1208.5487}{{\normalfont
  [arXiv:astro-ph.HE/1208.5487]}}.
\newblock
  doi:{\changeurlcolor{black}\href{https://doi.org/10.1103/PhysRevD.86.064032}{\detokenize{10.1103/PhysRevD.86.064032}}}.

\bibitem[{Kaplan}(2014)]{Kaplan2014}
{Kaplan}, J.D.
\newblock {Where Tori Fear to Tread: Hypermassive Neutron Star Remnants and
  Absolute Event Horizons or Topics in Computational General Relativity}.
\newblock PhD thesis, California Institute of Technology,  2014.

\bibitem[{Rosswog} \em{et~al.}(1999){Rosswog}, {Liebend{\"o}rfer},
  {Thielemann}, {Davies}, {Benz}, and {Piran}]{Rosswog1999}
{Rosswog}, S.; {Liebend{\"o}rfer}, M.; {Thielemann}, F.K.; {Davies}, M.B.;
  {Benz}, W.; {Piran}, T.
\newblock {Mass ejection in neutron star mergers}.
\newblock {\em Astron. Astrophys.} {\bf 1999}, {\em 341},~499--526,
  \href{http://xxx.lanl.gov/abs/astro-ph/9811367}{{\normalfont
  [astro-ph/9811367]}}.

\bibitem[{Aloy} \em{et~al.}(2005){Aloy}, {Janka}, and {M{\"u}ller}]{Aloy:2005}
{Aloy}, M.A.; {Janka}, H.; {M{\"u}ller}, E.
\newblock {Relativistic outflows from remnants of compact object mergers and
  their viability for short gamma-ray bursts}.
\newblock {\em Astron. Astrophys.} {\bf 2005}, {\em 436},~273--311,
  \href{http://xxx.lanl.gov/abs/arXiv:astro-ph/0408291}{{\normalfont
  [arXiv:astro-ph/0408291]}}.
\newblock
  doi:{\changeurlcolor{black}\href{https://doi.org/10.1051/0004-6361:20041865}{\detokenize{10.1051/0004-6361:20041865}}}.

\bibitem[{Dessart} \em{et~al.}(2009){Dessart}, {Ott}, {Burrows}, {Rosswog}, and
  {Livne}]{Dessart2009}
{Dessart}, L.; {Ott}, C.D.; {Burrows}, A.; {Rosswog}, S.; {Livne}, E.
\newblock {Neutrino Signatures and the Neutrino-Driven Wind in Binary Neutron
  Star Mergers}.
\newblock {\em Astrophys. J.} {\bf 2009}, {\em 690},~1681--1705,
  \href{http://xxx.lanl.gov/abs/0806.4380}{{\normalfont [0806.4380]}}.
\newblock
  doi:{\changeurlcolor{black}\href{https://doi.org/10.1088/0004-637X/690/2/1681}{\detokenize{10.1088/0004-637X/690/2/1681}}}.

\bibitem[{Rezzolla} \em{et~al.}(2010){Rezzolla}, {Baiotti}, {Giacomazzo},
  {Link}, and {Font}]{Rezzolla:2010}
{Rezzolla}, L.; {Baiotti}, L.; {Giacomazzo}, B.; {Link}, D.; {Font}, J.A.
\newblock {Accurate evolutions of unequal-mass neutron-star binaries:
  properties of the torus and short GRB engines}.
\newblock {\em Class. Quantum Grav.} {\bf 2010}, {\em 27},~114105,
  \href{http://xxx.lanl.gov/abs/1001.3074}{{\normalfont
  [arXiv:gr-qc/1001.3074]}}.
\newblock
  doi:{\changeurlcolor{black}\href{https://doi.org/10.1088/0264-9381/27/11/114105}{\detokenize{10.1088/0264-9381/27/11/114105}}}.

\bibitem[{Roberts} \em{et~al.}(2011){Roberts}, {Kasen}, {Lee}, and
  {Ramirez-Ruiz}]{Roberts2011}
{Roberts}, L.F.; {Kasen}, D.; {Lee}, W.H.; {Ramirez-Ruiz}, E.
\newblock {Electromagnetic Transients Powered by Nuclear Decay in the Tidal
  Tails of Coalescing Compact Binaries}.
\newblock {\em Astrophys. J. Lett.} {\bf 2011}, {\em 736},~L21,
  \href{http://xxx.lanl.gov/abs/1104.5504}{{\normalfont
  [arXiv:astro-ph.HE/1104.5504]}}.
\newblock
  doi:{\changeurlcolor{black}\href{https://doi.org/10.1088/2041-8205/736/1/L21}{\detokenize{10.1088/2041-8205/736/1/L21}}}.

\bibitem[{Kyutoku} \em{et~al.}(2014){Kyutoku}, {Ioka}, and
  {Shibata}]{Kyutoku2012}
{Kyutoku}, K.; {Ioka}, K.; {Shibata}, M.
\newblock {Ultrarelativistic electromagnetic counterpart to binary neutron star
  mergers}.
\newblock {\em Mon. Not. R. Astron.Soc.} {\bf 2014}, {\em 437},~L6--L10,
  \href{http://xxx.lanl.gov/abs/1209.5747}{{\normalfont
  [arXiv:astro-ph.HE/1209.5747]}}.
\newblock
  doi:{\changeurlcolor{black}\href{https://doi.org/10.1093/mnrasl/slt128}{\detokenize{10.1093/mnrasl/slt128}}}.

\bibitem[{Rosswog}(2013)]{Rosswog2013a}
{Rosswog}, S.
\newblock {The dynamic ejecta of compact object mergers and eccentric
  collisions}.
\newblock {\em Royal Society of London Philosophical Transactions Series A}
  {\bf 2013}, {\em 371},~20272,
  \href{http://xxx.lanl.gov/abs/1210.6549}{{\normalfont
  [arXiv:astro-ph.HE/1210.6549]}}.
\newblock
  doi:{\changeurlcolor{black}\href{https://doi.org/10.1098/rsta.2012.0272}{\detokenize{10.1098/rsta.2012.0272}}}.

\bibitem[{Bauswein} \em{et~al.}(2013){Bauswein}, {Goriely}, and
  {Janka}]{Bauswein2013b}
{Bauswein}, A.; {Goriely}, S.; {Janka}, H.T.
\newblock {Systematics of Dynamical Mass Ejection, Nucleosynthesis, and
  Radioactively Powered Electromagnetic Signals from Neutron-star Mergers}.
\newblock {\em Astrophys. J.} {\bf 2013}, {\em 773},~78,
  \href{http://xxx.lanl.gov/abs/1302.6530}{{\normalfont
  [arXiv:astro-ph.SR/1302.6530]}}.
\newblock
  doi:{\changeurlcolor{black}\href{https://doi.org/10.1088/0004-637X/773/1/78}{\detokenize{10.1088/0004-637X/773/1/78}}}.

\bibitem[{Hotokezaka} \em{et~al.}(2013){Hotokezaka}, {Kiuchi}, {Kyutoku},
  {Okawa}, {Sekiguchi}, {Shibata}, and {Taniguchi}]{Hotokezaka2013}
{Hotokezaka}, K.; {Kiuchi}, K.; {Kyutoku}, K.; {Okawa}, H.; {Sekiguchi}, Y.i.;
  {Shibata}, M.; {Taniguchi}, K.
\newblock {Mass ejection from the merger of binary neutron stars}.
\newblock {\em Phys. Rev. D} {\bf 2013}, {\em 87},~024001,
  \href{http://xxx.lanl.gov/abs/1212.0905}{{\normalfont
  [arXiv:astro-ph.HE/1212.0905]}}.
\newblock
  doi:{\changeurlcolor{black}\href{https://doi.org/10.1103/PhysRevD.87.024001}{\detokenize{10.1103/PhysRevD.87.024001}}}.

\bibitem[{Foucart} \em{et~al.}(2014){Foucart}, {Deaton}, {Duez}, {O'Connor},
  {Ott}, {Haas}, {Kidder}, {Pfeiffer}, {Scheel}, and {Szilagyi}]{Foucart2014}
{Foucart}, F.; {Deaton}, M.B.; {Duez}, M.D.; {O'Connor}, E.; {Ott}, C.D.;
  {Haas}, R.; {Kidder}, L.E.; {Pfeiffer}, H.P.; {Scheel}, M.A.; {Szilagyi}, B.
\newblock {Neutron star-black hole mergers with a nuclear equation of state and
  neutrino cooling: Dependence in the binary parameters}.
\newblock {\em Phys. Rev. D} {\bf 2014}, {\em 90},~024026,
  \href{http://xxx.lanl.gov/abs/1405.1121}{{\normalfont
  [arXiv:astro-ph.HE/1405.1121]}}.
\newblock
  doi:{\changeurlcolor{black}\href{https://doi.org/10.1103/PhysRevD.90.024026}{\detokenize{10.1103/PhysRevD.90.024026}}}.

\bibitem[{Siegel} \em{et~al.}(2014){Siegel}, {Ciolfi}, and
  {Rezzolla}]{Siegel2014}
{Siegel}, D.M.; {Ciolfi}, R.; {Rezzolla}, L.
\newblock {Magnetically Driven Winds from Differentially Rotating Neutron Stars
  and X-Ray Afterglows of Short Gamma-Ray Bursts}.
\newblock {\em Astrophys. J.} {\bf 2014}, {\em 785},~L6,
  \href{http://xxx.lanl.gov/abs/1401.4544}{{\normalfont
  [arXiv:astro-ph.HE/1401.4544]}}.
\newblock
  doi:{\changeurlcolor{black}\href{https://doi.org/10.1088/2041-8205/785/1/L6}{\detokenize{10.1088/2041-8205/785/1/L6}}}.

\bibitem[{Wanajo} \em{et~al.}(2014){Wanajo}, {Sekiguchi}, {Nishimura},
  {Kiuchi}, {Kyutoku}, and {Shibata}]{Wanajo2014}
{Wanajo}, S.; {Sekiguchi}, Y.; {Nishimura}, N.; {Kiuchi}, K.; {Kyutoku}, K.;
  {Shibata}, M.
\newblock {Production of All the r-process Nuclides in the Dynamical Ejecta of
  Neutron Star Mergers}.
\newblock {\em Astrophys. J.} {\bf 2014}, {\em 789},~L39,
  \href{http://xxx.lanl.gov/abs/1402.7317}{{\normalfont
  [arXiv:astro-ph.SR/1402.7317]}}.
\newblock
  doi:{\changeurlcolor{black}\href{https://doi.org/10.1088/2041-8205/789/2/L39}{\detokenize{10.1088/2041-8205/789/2/L39}}}.

\bibitem[{Sekiguchi} \em{et~al.}(2015){Sekiguchi}, {Kiuchi}, {Kyutoku}, and
  {Shibata}]{Sekiguchi2015}
{Sekiguchi}, Y.; {Kiuchi}, K.; {Kyutoku}, K.; {Shibata}, M.
\newblock {Dynamical mass ejection from binary neutron star mergers:
  Radiation-hydrodynamics study in general relativity}.
\newblock {\em Phys. Rev. D} {\bf 2015}, {\em 91},~064059,
  \href{http://xxx.lanl.gov/abs/1502.06660}{{\normalfont
  [arXiv:astro-ph.HE/1502.06660]}}.
\newblock
  doi:{\changeurlcolor{black}\href{https://doi.org/10.1103/PhysRevD.91.064059}{\detokenize{10.1103/PhysRevD.91.064059}}}.

\bibitem[{Radice} \em{et~al.}(2016){Radice}, {Galeazzi}, {Lippuner}, {Roberts},
  {Ott}, and {Rezzolla}]{Radice2016}
{Radice}, D.; {Galeazzi}, F.; {Lippuner}, J.; {Roberts}, L.F.; {Ott}, C.D.;
  {Rezzolla}, L.
\newblock {Dynamical Mass Ejection from Binary Neutron Star Mergers}.
\newblock {\em Mon. Not. R. Astron. Soc.} {\bf 2016}, {\em 460},~3255--3271,
  \href{http://xxx.lanl.gov/abs/1601.02426}{{\normalfont
  [arXiv:astro-ph.HE/1601.02426]}}.
\newblock
  doi:{\changeurlcolor{black}\href{https://doi.org/10.1093/mnras/stw1227}{\detokenize{10.1093/mnras/stw1227}}}.

\bibitem[{Sekiguchi} \em{et~al.}(2016){Sekiguchi}, {Kiuchi}, {Kyutoku},
  {Shibata}, and {Taniguchi}]{Sekiguchi2016}
{Sekiguchi}, Y.; {Kiuchi}, K.; {Kyutoku}, K.; {Shibata}, M.; {Taniguchi}, K.
\newblock {Dynamical mass ejection from the merger of asymmetric binary neutron
  stars: Radiation-hydrodynamics study in general relativity}.
\newblock {\em Phys. Rev. D} {\bf 2016}, {\em 93},~124046,
  \href{http://xxx.lanl.gov/abs/1603.01918}{{\normalfont
  [arXiv:astro-ph.HE/1603.01918]}}.
\newblock
  doi:{\changeurlcolor{black}\href{https://doi.org/10.1103/PhysRevD.93.124046}{\detokenize{10.1103/PhysRevD.93.124046}}}.

\bibitem[{Lehner} \em{et~al.}(2016){Lehner}, {Liebling}, {Palenzuela},
  {Caballero}, {O'Connor}, {Anderson}, and {Neilsen}]{Lehner2016}
{Lehner}, L.; {Liebling}, S.L.; {Palenzuela}, C.; {Caballero}, O.L.;
  {O'Connor}, E.; {Anderson}, M.; {Neilsen}, D.
\newblock {Unequal mass binary neutron star mergers and multimessenger
  signals}.
\newblock {\em Classical and Quantum Gravity} {\bf 2016}, {\em 33},~184002,
  \href{http://xxx.lanl.gov/abs/1603.00501}{{\normalfont
  [arXiv:gr-qc/1603.00501]}}.
\newblock
  doi:{\changeurlcolor{black}\href{https://doi.org/10.1088/0264-9381/33/18/184002}{\detokenize{10.1088/0264-9381/33/18/184002}}}.

\bibitem[{Siegel} and {Metzger}(2017)]{Siegel2017}
{Siegel}, D.M.; {Metzger}, B.D.
\newblock {Three-Dimensional General-Relativistic Magnetohydrodynamic
  Simulations of Remnant Accretion Disks from Neutron Star Mergers: Outflows
  and r -Process Nucleosynthesis}.
\newblock {\em Physical Review Letters} {\bf 2017}, {\em 119},~231102,
  \href{http://xxx.lanl.gov/abs/1705.05473}{{\normalfont
  [arXiv:astro-ph.HE/1705.05473]}}.
\newblock
  doi:{\changeurlcolor{black}\href{https://doi.org/10.1103/PhysRevLett.119.231102}{\detokenize{10.1103/PhysRevLett.119.231102}}}.

\bibitem[{Dietrich} \em{et~al.}(2017){Dietrich}, {Ujevic}, {Tichy}, {Bernuzzi},
  and {Br{\"u}gmann}]{Dietrich2017}
{Dietrich}, T.; {Ujevic}, M.; {Tichy}, W.; {Bernuzzi}, S.; {Br{\"u}gmann}, B.
\newblock {Gravitational waves and mass ejecta from binary neutron star
  mergers: Effect of the mass ratio}.
\newblock {\em Phys. Rev. D} {\bf 2017}, {\em 95},~024029,
  \href{http://xxx.lanl.gov/abs/1607.06636}{{\normalfont
  [arXiv:gr-qc/1607.06636]}}.
\newblock
  doi:{\changeurlcolor{black}\href{https://doi.org/10.1103/PhysRevD.95.024029}{\detokenize{10.1103/PhysRevD.95.024029}}}.

\bibitem[{Bovard} \em{et~al.}(2017){Bovard}, {Martin}, {Guercilena}, {Arcones},
  {Rezzolla}, and {Korobkin}]{Bovard2017}
{Bovard}, L.; {Martin}, D.; {Guercilena}, F.; {Arcones}, A.; {Rezzolla}, L.;
  {Korobkin}, O.
\newblock {On r-process nucleosynthesis from matter ejected in binary neutron
  star mergers}.
\newblock {\em Phys. Rev. D} {\bf 2017}, {\em 96},~124005,
  \href{http://xxx.lanl.gov/abs/1709.09630}{{\normalfont
  [arXiv:gr-qc/1709.09630]}}.

\bibitem[{Fujibayashi} \em{et~al.}(2017){Fujibayashi}, {Sekiguchi}, {Kiuchi},
  and {Shibata}]{Fujibayashi2017}
{Fujibayashi}, S.; {Sekiguchi}, Y.; {Kiuchi}, K.; {Shibata}, M.
\newblock {Properties of Neutrino-driven Ejecta from the Remnant of a Binary
  Neutron Star Merger: Pure Radiation Hydrodynamics Case}.
\newblock {\em Astrophys. J.} {\bf 2017}, {\em 846},~114.
\newblock
  doi:{\changeurlcolor{black}\href{https://doi.org/10.3847/1538-4357/aa8039}{\detokenize{10.3847/1538-4357/aa8039}}}.

\bibitem[{Fujibayashi} \em{et~al.}(2018){Fujibayashi}, {Kiuchi}, {Nishimura},
  {Sekiguchi}, and {Shibata}]{Fujibayashi2017b}
{Fujibayashi}, S.; {Kiuchi}, K.; {Nishimura}, N.; {Sekiguchi}, Y.; {Shibata},
  M.
\newblock {Mass Ejection from the Remnant of a Binary Neutron Star Merger:
  Viscous-radiation Hydrodynamics Study}.
\newblock {\em Astrophys. J.} {\bf 2018}, {\em 860},~64,
  \href{http://xxx.lanl.gov/abs/1711.02093}{{\normalfont
  [arXiv:astro-ph.HE/1711.02093]}}.
\newblock
  doi:{\changeurlcolor{black}\href{https://doi.org/10.3847/1538-4357/aabafd}{\detokenize{10.3847/1538-4357/aabafd}}}.

\bibitem[{Kastaun} and {Galeazzi}(2015)]{Kastaun2014}
{Kastaun}, W.; {Galeazzi}, F.
\newblock {Properties of hypermassive neutron stars formed in mergers of
  spinning binaries}.
\newblock {\em Phys. Rev. D} {\bf 2015}, {\em 91},~064027,
  \href{http://xxx.lanl.gov/abs/1411.7975}{{\normalfont
  [arXiv:gr-qc/1411.7975]}}.
\newblock
  doi:{\changeurlcolor{black}\href{https://doi.org/10.1103/PhysRevD.91.064027}{\detokenize{10.1103/PhysRevD.91.064027}}}.

\bibitem[{Kastaun} \em{et~al.}(2017){Kastaun}, {Ciolfi}, {Endrizzi}, and
  {Giacomazzo}]{Kastaun2017}
{Kastaun}, W.; {Ciolfi}, R.; {Endrizzi}, A.; {Giacomazzo}, B.
\newblock {Structure of stable binary neutron star merger remnants: Role of
  initial spin}.
\newblock {\em Phys. Rev. D} {\bf 2017}, {\em 96},~043019,
  \href{http://xxx.lanl.gov/abs/1612.03671}{{\normalfont
  [arXiv:astro-ph.HE/1612.03671]}}.
\newblock
  doi:{\changeurlcolor{black}\href{https://doi.org/10.1103/PhysRevD.96.043019}{\detokenize{10.1103/PhysRevD.96.043019}}}.

\bibitem[{Hanauske} \em{et~al.}(2017){Hanauske}, {Takami}, {Bovard},
  {Rezzolla}, {Font}, {Galeazzi}, and {St{\"o}cker}]{Hanauske2016}
{Hanauske}, M.; {Takami}, K.; {Bovard}, L.; {Rezzolla}, L.; {Font}, J.A.;
  {Galeazzi}, F.; {St{\"o}cker}, H.
\newblock {Rotational properties of hypermassive neutron stars from binary
  mergers}.
\newblock {\em Phys. Rev. D} {\bf 2017}, {\em 96},~043004,
  \href{http://xxx.lanl.gov/abs/1611.07152}{{\normalfont
  [arXiv:gr-qc/1611.07152]}}.
\newblock
  doi:{\changeurlcolor{black}\href{https://doi.org/10.1103/PhysRevD.96.043004}{\detokenize{10.1103/PhysRevD.96.043004}}}.

\bibitem[{Obergaulinger} \em{et~al.}(2010){Obergaulinger}, {Aloy}, and
  {M{\"u}ller}]{Obergaulinger10}
{Obergaulinger}, M.; {Aloy}, M.A.; {M{\"u}ller}, E.
\newblock {Local simulations of the magnetized Kelvin-Helmholtz instability in
  neutron-star mergers}.
\newblock {\em Astron. Astrophys.} {\bf 2010}, {\em 515},~A30,
  \href{http://xxx.lanl.gov/abs/1003.6031}{{\normalfont
  [arXiv:astro-ph.SR/1003.6031]}}.
\newblock
  doi:{\changeurlcolor{black}\href{https://doi.org/10.1051/0004-6361/200913386}{\detokenize{10.1051/0004-6361/200913386}}}.

\bibitem[{Rasio} and {Shapiro}(1999)]{Rasio99}
{Rasio}, F.; {Shapiro}, S.
\newblock {TOPICAL REVIEW: Coalescing binary neutron stars}.
\newblock {\em Class. Quantum Grav.} {\bf 1999}, {\em 16},~R1--R29,
  \href{http://xxx.lanl.gov/abs/gr-qc/9902019}{{\normalfont [gr-qc/9902019]}}.
\newblock
  doi:{\changeurlcolor{black}\href{https://doi.org/10.1088/0264-9381/16/6/201}{\detokenize{10.1088/0264-9381/16/6/201}}}.

\bibitem[{Rosswog} \em{et~al.}(2003){Rosswog}, {Ramirez-Ruiz}, and
  {Davies}]{Rosswog02}
{Rosswog}, S.; {Ramirez-Ruiz}, E.; {Davies}, M.B.
\newblock {High-resolution calculations of merging neutron stars - III.
  Gamma-ray bursts}.
\newblock {\em Mon. Not. R. Astron. Soc.} {\bf 2003}, {\em 345},~1077--1090,
  \href{http://xxx.lanl.gov/abs/astro-ph/0306418}{{\normalfont
  [astro-ph/0306418]}}.
\newblock
  doi:{\changeurlcolor{black}\href{https://doi.org/10.1046/j.1365-2966.2003.07032.x}{\detokenize{10.1046/j.1365-2966.2003.07032.x}}}.

\bibitem[{Price} and {Rosswog}(2006)]{Price06}
{Price}, D.J.; {Rosswog}, S.
\newblock {Producing Ultrastrong Magnetic Fields in Neutron Star Mergers}.
\newblock {\em Science} {\bf 2006}, {\em 312},~719--722,
  \href{http://xxx.lanl.gov/abs/astro-ph/0603845}{{\normalfont
  [astro-ph/0603845]}}.
\newblock
  doi:{\changeurlcolor{black}\href{https://doi.org/10.1126/science.1125201}{\detokenize{10.1126/science.1125201}}}.

\bibitem[Liu \em{et~al.}(2008)Liu, Shapiro, Etienne, and Taniguchi]{Liu:2008xy}
Liu, Y.T.; Shapiro, S.L.; Etienne, Z.B.; Taniguchi, K.
\newblock {General relativistic simulations of magnetized binary neutron star
  mergers}.
\newblock {\em Phys. Rev. D} {\bf 2008}, {\em 78},~024012,
  \href{http://xxx.lanl.gov/abs/0803.4193}{{\normalfont
  [arXiv:astro-ph/0803.4193]}}.
\newblock
  doi:{\changeurlcolor{black}\href{https://doi.org/10.1103/PhysRevD.78.024012}{\detokenize{10.1103/PhysRevD.78.024012}}}.

\bibitem[{Anderson} \em{et~al.}(2008){Anderson}, {Hirschmann}, {Lehner},
  {Liebling}, {Motl}, {Neilsen}, {Palenzuela}, and {Tohline}]{Anderson2008}
{Anderson}, M.; {Hirschmann}, E.W.; {Lehner}, L.; {Liebling}, S.L.; {Motl},
  P.M.; {Neilsen}, D.; {Palenzuela}, C.; {Tohline}, J.E.
\newblock {Magnetized Neutron-Star Mergers and Gravitational-Wave Signals}.
\newblock {\em Phys. Rev. Lett.} {\bf 2008}, {\em 100},~191101,
  \href{http://xxx.lanl.gov/abs/0801.4387}{{\normalfont
  [arXiv:gr-qc/0801.4387]}}.
\newblock
  doi:{\changeurlcolor{black}\href{https://doi.org/10.1103/PhysRevLett.100.191101}{\detokenize{10.1103/PhysRevLett.100.191101}}}.

\bibitem[{Giacomazzo} \em{et~al.}(2011){Giacomazzo}, {Rezzolla}, and
  {Baiotti}]{Giacomazzo2011b}
{Giacomazzo}, B.; {Rezzolla}, L.; {Baiotti}, L.
\newblock {Accurate evolutions of inspiralling and magnetized neutron stars:
  Equal-mass binaries}.
\newblock {\em Phys. Rev. D} {\bf 2011}, {\em 83},~044014,
  \href{http://xxx.lanl.gov/abs/1009.2468}{{\normalfont
  [arXiv:gr-qc/1009.2468]}}.
\newblock
  doi:{\changeurlcolor{black}\href{https://doi.org/10.1103/PhysRevD.83.044014}{\detokenize{10.1103/PhysRevD.83.044014}}}.

\bibitem[{Giacomazzo} \em{et~al.}(2015){Giacomazzo}, {Zrake}, {Duffell},
  {MacFadyen}, and {Perna}]{Giacomazzo:2014b}
{Giacomazzo}, B.; {Zrake}, J.; {Duffell}, P.C.; {MacFadyen}, A.I.; {Perna}, R.
\newblock {Producing Magnetar Magnetic Fields in the Merger of Binary Neutron
  Stars}.
\newblock {\em Astrophys. J.} {\bf 2015}, {\em 809},~39,
  \href{http://xxx.lanl.gov/abs/1410.0013}{{\normalfont
  [arXiv:astro-ph.HE/1410.0013]}}.
\newblock
  doi:{\changeurlcolor{black}\href{https://doi.org/10.1088/0004-637X/809/1/39}{\detokenize{10.1088/0004-637X/809/1/39}}}.

\bibitem[{Dionysopoulou} \em{et~al.}(2015){Dionysopoulou}, {Alic}, and
  {Rezzolla}]{Dionysopoulou2015}
{Dionysopoulou}, K.; {Alic}, D.; {Rezzolla}, L.
\newblock {General-relativistic resistive-magnetohydrodynamic simulations of
  binary neutron stars}.
\newblock {\em Phys. Rev. D} {\bf 2015}, {\em 92},~084064,
  \href{http://xxx.lanl.gov/abs/1502.02021}{{\normalfont
  [arXiv:gr-qc/1502.02021]}}.
\newblock
  doi:{\changeurlcolor{black}\href{https://doi.org/10.1103/PhysRevD.92.084064}{\detokenize{10.1103/PhysRevD.92.084064}}}.

\bibitem[{Palenzuela} \em{et~al.}(2015){Palenzuela}, {Liebling}, {Neilsen},
  {Lehner}, {Caballero}, {O'Connor}, and {Anderson}]{Palenzuela2015}
{Palenzuela}, C.; {Liebling}, S.L.; {Neilsen}, D.; {Lehner}, L.; {Caballero},
  O.L.; {O'Connor}, E.; {Anderson}, M.
\newblock {Effects of the microphysical equation of state in the mergers of
  magnetized neutron stars with neutrino cooling}.
\newblock {\em Phys. Rev. D} {\bf 2015}, {\em 92},~044045,
  \href{http://xxx.lanl.gov/abs/1505.01607}{{\normalfont
  [arXiv:gr-qc/1505.01607]}}.
\newblock
  doi:{\changeurlcolor{black}\href{https://doi.org/10.1103/PhysRevD.92.044045}{\detokenize{10.1103/PhysRevD.92.044045}}}.

\bibitem[{Kiuchi} \em{et~al.}(2014){Kiuchi}, {Kyutoku}, {Sekiguchi}, {Shibata},
  and {Wada}]{Kiuchi2014}
{Kiuchi}, K.; {Kyutoku}, K.; {Sekiguchi}, Y.; {Shibata}, M.; {Wada}, T.
\newblock {High resolution numerical relativity simulations for the merger of
  binary magnetized neutron stars}.
\newblock {\em Phys. Rev. D} {\bf 2014}, {\em 90},~041502,
  \href{http://xxx.lanl.gov/abs/1407.2660}{{\normalfont
  [arXiv:astro-ph.HE/1407.2660]}}.
\newblock
  doi:{\changeurlcolor{black}\href{https://doi.org/10.1103/PhysRevD.90.041502}{\detokenize{10.1103/PhysRevD.90.041502}}}.

\bibitem[{Zrake} and {MacFadyen}(2013)]{Zrake2013b}
{Zrake}, J.; {MacFadyen}, A.I.
\newblock {Magnetic Energy Production by Turbulence in Binary Neutron Star
  Mergers}.
\newblock {\em Astrophys. J.} {\bf 2013}, {\em 769},~L29,
  \href{http://xxx.lanl.gov/abs/1303.1450}{{\normalfont
  [arXiv:astro-ph.HE/1303.1450]}}.
\newblock
  doi:{\changeurlcolor{black}\href{https://doi.org/10.1088/2041-8205/769/2/L29}{\detokenize{10.1088/2041-8205/769/2/L29}}}.

\bibitem[{Kiuchi} \em{et~al.}(2018){Kiuchi}, {Kyutoku}, {Sekiguchi}, and
  {Shibata}]{Kiuchi2017}
{Kiuchi}, K.; {Kyutoku}, K.; {Sekiguchi}, Y.; {Shibata}, M.
\newblock {Global simulations of strongly magnetized remnant massive neutron
  stars formed in binary neutron star mergers}.
\newblock {\em Phys. Rev. D} {\bf 2018}, {\em 97},~124039,
  \href{http://xxx.lanl.gov/abs/1710.01311}{{\normalfont
  [arXiv:astro-ph.HE/1710.01311]}}.
\newblock
  doi:{\changeurlcolor{black}\href{https://doi.org/10.1103/PhysRevD.97.124039}{\detokenize{10.1103/PhysRevD.97.124039}}}.

\bibitem[{Kiuchi} \em{et~al.}(2015){Kiuchi}, {Cerd{\'a}-Dur{\'a}n}, {Kyutoku},
  {Sekiguchi}, and {Shibata}]{Kiuchi2015a}
{Kiuchi}, K.; {Cerd{\'a}-Dur{\'a}n}, P.; {Kyutoku}, K.; {Sekiguchi}, Y.;
  {Shibata}, M.
\newblock {Efficient magnetic-field amplification due to the Kelvin-Helmholtz
  instability in binary neutron star mergers}.
\newblock {\em Phys. Rev. D} {\bf 2015}, {\em 92},~124034,
  \href{http://xxx.lanl.gov/abs/1509.09205}{{\normalfont
  [arXiv:astro-ph.HE/1509.09205]}}.
\newblock
  doi:{\changeurlcolor{black}\href{https://doi.org/10.1103/PhysRevD.92.124034}{\detokenize{10.1103/PhysRevD.92.124034}}}.

\bibitem[{Harutyunyan} \em{et~al.}(2018){Harutyunyan}, {Nathanail}, {Rezzolla},
  and {Sedrakian}]{Harutyunyan2018}
{Harutyunyan}, A.; {Nathanail}, A.; {Rezzolla}, L.; {Sedrakian}, A.
\newblock {Electrical Resistivity and Hall Effect in Binary Neutron-Star
  Mergers}.
\newblock {\em ArXiv e-prints} {\bf 2018},
  \href{http://xxx.lanl.gov/abs/1803.09215}{{\normalfont
  [arXiv:astro-ph.HE/1803.09215]}}.

\bibitem[{Moesta} \em{et~al.}(2013){Moesta}, {Mundim}, {Faber}, {Noble},
  {Bode}, {Haas}, {Loeffler}, {Ott}, {Reisswig}, and {Schnetter}]{Moesta2013}
{Moesta}, P.; {Mundim}, B.; {Faber}, J.; {Noble}, S.; {Bode}, T.; {Haas}, R.;
  {Loeffler}, F.; {Ott}, C.; {Reisswig}, C.; {Schnetter}, E.
\newblock {General relativistic magneto-hydrodynamics with the Einstein
  Toolkit}.
\newblock  APS Meeting Abstracts,  2013, p. 10001.

\bibitem[{Rembiasz} \em{et~al.}(2016){Rembiasz}, {Guilet}, {Obergaulinger},
  {Cerd{\'a}-Dur{\'a}n}, {Aloy}, and {M{\"u}ller}]{Rembiasz2016}
{Rembiasz}, T.; {Guilet}, J.; {Obergaulinger}, M.; {Cerd{\'a}-Dur{\'a}n}, P.;
  {Aloy}, M.A.; {M{\"u}ller}, E.
\newblock {On the maximum magnetic field amplification by the magnetorotational
  instability in core-collapse supernovae}.
\newblock {\em Mon. Not. R. Astron. Soc.} {\bf 2016}, {\em 460},~3316--3334,
  \href{http://xxx.lanl.gov/abs/1603.00466}{{\normalfont
  [arXiv:astro-ph.SR/1603.00466]}}.
\newblock
  doi:{\changeurlcolor{black}\href{https://doi.org/10.1093/mnras/stw1201}{\detokenize{10.1093/mnras/stw1201}}}.

\bibitem[{Zrake} and {MacFadyen}(2013)]{Zrake2013}
{Zrake}, J.; {MacFadyen}, A.I.
\newblock {Spectral and Intermittency Properties of Relativistic Turbulence}.
\newblock {\em Astrophys. J.} {\bf 2013}, {\em 763},~L12,
  \href{http://xxx.lanl.gov/abs/1210.4066}{{\normalfont
  [arXiv:astro-ph.HE/1210.4066]}}.
\newblock
  doi:{\changeurlcolor{black}\href{https://doi.org/10.1088/2041-8205/763/1/L12}{\detokenize{10.1088/2041-8205/763/1/L12}}}.

\bibitem[{Kawamura} \em{et~al.}(2016){Kawamura}, {Giacomazzo}, {Kastaun},
  {Ciolfi}, {Endrizzi}, {Baiotti}, and {Perna}]{Kawamura2016}
{Kawamura}, T.; {Giacomazzo}, B.; {Kastaun}, W.; {Ciolfi}, R.; {Endrizzi}, A.;
  {Baiotti}, L.; {Perna}, R.
\newblock {Binary neutron star mergers and short gamma-ray bursts: Effects of
  magnetic field orientation, equation of state, and mass ratio}.
\newblock {\em Phys. Rev. D} {\bf 2016}, {\em 94},~064012,
  \href{http://xxx.lanl.gov/abs/1607.01791}{{\normalfont
  [arXiv:astro-ph.HE/1607.01791]}}.
\newblock
  doi:{\changeurlcolor{black}\href{https://doi.org/10.1103/PhysRevD.94.064012}{\detokenize{10.1103/PhysRevD.94.064012}}}.

\bibitem[{Rezzolla} \em{et~al.}(2011){Rezzolla}, {Giacomazzo}, {Baiotti},
  {Granot}, {Kouveliotou}, and {Aloy}]{Rezzolla:2011}
{Rezzolla}, L.; {Giacomazzo}, B.; {Baiotti}, L.; {Granot}, J.; {Kouveliotou},
  C.; {Aloy}, M.A.
\newblock {The Missing Link: Merging Neutron Stars Naturally Produce Jet-like
  Structures and Can Power Short Gamma-ray Bursts}.
\newblock {\em Astrophys. J. Letters} {\bf 2011}, {\em 732},~L6,
  \href{http://xxx.lanl.gov/abs/1101.4298}{{\normalfont
  [arXiv:astro-ph.HE/1101.4298]}}.
\newblock
  doi:{\changeurlcolor{black}\href{https://doi.org/10.1088/2041-8205/732/1/L6}{\detokenize{10.1088/2041-8205/732/1/L6}}}.

\bibitem[{Blandford} and {Znajek}(1977)]{Blandford1977}
{Blandford}, R.D.; {Znajek}, R.L.
\newblock {Electromagnetic extraction of energy from Kerr black holes}.
\newblock {\em Mon. Not. R. Astron. Soc.} {\bf 1977}, {\em 179},~433--456.

\bibitem[{Contopoulos}(1995)]{Contopoulos1995}
{Contopoulos}, J.
\newblock {A Simple Type of Magnetically Driven Jets: an Astrophysical Plasma
  Gun}.
\newblock {\em Astrophys. J.} {\bf 1995}, {\em 450},~616.
\newblock
  doi:{\changeurlcolor{black}\href{https://doi.org/10.1086/176170}{\detokenize{10.1086/176170}}}.

\bibitem[{Ruiz} \em{et~al.}(2016){Ruiz}, {Lang}, {Paschalidis}, and
  {Shapiro}]{Ruiz2016}
{Ruiz}, M.; {Lang}, R.N.; {Paschalidis}, V.; {Shapiro}, S.L.
\newblock {Binary Neutron Star Mergers: A Jet Engine for Short Gamma-Ray
  Bursts}.
\newblock {\em Astrophys. J. Lett.} {\bf 2016}, {\em 824},~L6,
  \href{http://xxx.lanl.gov/abs/1604.02455}{{\normalfont
  [arXiv:astro-ph.HE/1604.02455]}}.
\newblock
  doi:{\changeurlcolor{black}\href{https://doi.org/10.3847/2041-8205/824/1/L6}{\detokenize{10.3847/2041-8205/824/1/L6}}}.

\bibitem[{Endrizzi} \em{et~al.}(2016){Endrizzi}, {Ciolfi}, {Giacomazzo},
  {Kastaun}, and {Kawamura}]{Endrizzi2016}
{Endrizzi}, A.; {Ciolfi}, R.; {Giacomazzo}, B.; {Kastaun}, W.; {Kawamura}, T.
\newblock {General relativistic magnetohydrodynamic simulations of binary
  neutron star mergers with the APR4 equation of state}.
\newblock {\em Classical and Quantum Gravity} {\bf 2016}, {\em 33},~164001,
  \href{http://xxx.lanl.gov/abs/1604.03445}{{\normalfont
  [arXiv:astro-ph.HE/1604.03445]}}.
\newblock
  doi:{\changeurlcolor{black}\href{https://doi.org/10.1088/0264-9381/33/16/164001}{\detokenize{10.1088/0264-9381/33/16/164001}}}.

\bibitem[{Ruffert} and {Janka}(1999)]{Ruffert99b}
{Ruffert}, M.; {Janka}, H.T.
\newblock {Gamma-ray bursts from accreting black holes in neutron star
  mergers}.
\newblock {\em Astron. Astrophys.} {\bf 1999}, {\em 344},~573--606,
  \href{http://xxx.lanl.gov/abs/astro-ph/9809280}{{\normalfont
  [astro-ph/9809280]}}.

\bibitem[{Just} \em{et~al.}(2016){Just}, {Obergaulinger}, {Janka}, {Bauswein},
  and {Schwarz}]{Just2016}
{Just}, O.; {Obergaulinger}, M.; {Janka}, H.T.; {Bauswein}, A.; {Schwarz}, N.
\newblock {Neutron-star Merger Ejecta as Obstacles to Neutrino-powered Jets of
  Gamma-Ray Bursts}.
\newblock {\em Astrophys. J. Lett.} {\bf 2016}, {\em 816},~L30,
  \href{http://xxx.lanl.gov/abs/1510.04288}{{\normalfont
  [arXiv:astro-ph.HE/1510.04288]}}.
\newblock
  doi:{\changeurlcolor{black}\href{https://doi.org/10.3847/2041-8205/816/2/L30}{\detokenize{10.3847/2041-8205/816/2/L30}}}.

\bibitem[{Komissarov}(2001)]{Komissarov2001}
{Komissarov}, S.S.
\newblock {Direct numerical simulations of the Blandford-Znajek effect}.
\newblock {\em Mon. Not. R. Astron. Soc.} {\bf 2001}, {\em 326},~L41--L44.
\newblock
  doi:{\changeurlcolor{black}\href{https://doi.org/10.1046/j.1365-8711.2001.04863.x}{\detokenize{10.1046/j.1365-8711.2001.04863.x}}}.

\bibitem[{Komissarov} \em{et~al.}(2007){Komissarov}, {Barkov}, and
  {Lyutikov}]{Komissarov2007c}
{Komissarov}, S.S.; {Barkov}, M.; {Lyutikov}, M.
\newblock {Tearing instability in relativistic magnetically dominated plasmas}.
\newblock {\em Mon. Not. R. Astron. Soc.} {\bf 2007}, {\em 374},~415--426,
  \href{http://xxx.lanl.gov/abs/astro-ph/0606375}{{\normalfont
  [astro-ph/0606375]}}.
\newblock
  doi:{\changeurlcolor{black}\href{https://doi.org/10.1111/j.1365-2966.2006.11152.x}{\detokenize{10.1111/j.1365-2966.2006.11152.x}}}.

\bibitem[{Nathanail} and {Contopoulos}(2014)]{Nathanail2014}
{Nathanail}, A.; {Contopoulos}, I.
\newblock {Black Hole Magnetospheres}.
\newblock {\em Astrophys. J.} {\bf 2014}, {\em 788},~186,
  \href{http://xxx.lanl.gov/abs/1404.0549}{{\normalfont
  [arXiv:astro-ph.HE/1404.0549]}}.
\newblock
  doi:{\changeurlcolor{black}\href{https://doi.org/10.1088/0004-637X/788/2/186}{\detokenize{10.1088/0004-637X/788/2/186}}}.

\bibitem[{Gralla} \em{et~al.}(2016){Gralla}, {Lupsasca}, and
  {Rodriguez}]{Gralla2016}
{Gralla}, S.E.; {Lupsasca}, A.; {Rodriguez}, M.J.
\newblock {Electromagnetic jets from stars and black holes}.
\newblock {\em Phys. Rev. D} {\bf 2016}, {\em 93},~044038,
  \href{http://xxx.lanl.gov/abs/1504.02113}{{\normalfont
  [arXiv:gr-qc/1504.02113]}}.
\newblock
  doi:{\changeurlcolor{black}\href{https://doi.org/10.1103/PhysRevD.93.044038}{\detokenize{10.1103/PhysRevD.93.044038}}}.

\bibitem[{Shapiro}(2017)]{Shapiro2017}
{Shapiro}, S.L.
\newblock {Black holes, disks, and jets following binary mergers and stellar
  collapse: The narrow range of electromagnetic luminosities and accretion
  rates}.
\newblock {\em Phys. Rev. D} {\bf 2017}, {\em 95},~101303,
  \href{http://xxx.lanl.gov/abs/1705.04695}{{\normalfont
  [arXiv:astro-ph.HE/1705.04695]}}.
\newblock
  doi:{\changeurlcolor{black}\href{https://doi.org/10.1103/PhysRevD.95.101303}{\detokenize{10.1103/PhysRevD.95.101303}}}.

\bibitem[{Shibata} \em{et~al.}(2003){Shibata}, {Taniguchi}, and
  {Ury{\=u}}]{Shibata:2003ga}
{Shibata}, M.; {Taniguchi}, K.; {Ury{\=u}}, K.
\newblock {Merger of binary neutron stars of unequal mass in full general
  relativity}.
\newblock {\em Phys. Rev. D} {\bf 2003}, {\em 68},~084020,
  \href{http://xxx.lanl.gov/abs/gr-qc/0310030}{{\normalfont [gr-qc/0310030]}}.
\newblock
  doi:{\changeurlcolor{black}\href{https://doi.org/10.1103/PhysRevD.68.084020}{\detokenize{10.1103/PhysRevD.68.084020}}}.

\bibitem[{Shibata} and {Taniguchi}(2006)]{Shibata06a}
{Shibata}, M.; {Taniguchi}, K.
\newblock {Merger of binary neutron stars to a black hole: Disk mass, short
  gamma-ray bursts, and quasinormal mode ringing}.
\newblock {\em Phys. Rev. D} {\bf 2006}, {\em 73},~064027,
  \href{http://xxx.lanl.gov/abs/astro-ph/0603145}{{\normalfont
  [astro-ph/0603145]}}.
\newblock
  doi:{\changeurlcolor{black}\href{https://doi.org/10.1103/PhysRevD.73.064027}{\detokenize{10.1103/PhysRevD.73.064027}}}.

\bibitem[{Baiotti} \em{et~al.}(2008){Baiotti}, {Giacomazzo}, and
  {Rezzolla}]{Baiotti08}
{Baiotti}, L.; {Giacomazzo}, B.; {Rezzolla}, L.
\newblock {Accurate evolutions of inspiralling neutron-star binaries: Prompt
  and delayed collapse to a black hole}.
\newblock {\em Phys. Rev. D} {\bf 2008}, {\em 78},~084033,
  \href{http://xxx.lanl.gov/abs/0804.0594}{{\normalfont
  [arXiv:gr-qc/0804.0594]}}.
\newblock
  doi:{\changeurlcolor{black}\href{https://doi.org/10.1103/PhysRevD.78.084033}{\detokenize{10.1103/PhysRevD.78.084033}}}.

\bibitem[Rezzolla \em{et~al.}(2010)Rezzolla, Macedo, and
  Jaramillo]{Rezzolla:2010df}
Rezzolla, L.; Macedo, R.P.; Jaramillo, J.L.
\newblock {Understanding the 'anti-kick' in the merger of binary black holes}.
\newblock {\em Phys. Rev. Lett.} {\bf 2010}, {\em 104},~221101,
  \href{http://xxx.lanl.gov/abs/1003.0873}{{\normalfont
  [arXiv:gr-qc/1003.0873]}}.
\newblock
  doi:{\changeurlcolor{black}\href{https://doi.org/10.1103/PhysRevLett.104.221101}{\detokenize{10.1103/PhysRevLett.104.221101}}}.

\bibitem[{Margalit} \em{et~al.}(2015){Margalit}, {Metzger}, and
  {Beloborodov}]{Margalit2015}
{Margalit}, B.; {Metzger}, B.D.; {Beloborodov}, A.M.
\newblock {Does the Collapse of a Supramassive Neutron Star Leave a Debris
  Disk?}
\newblock {\em Phys. Rev. Lett.} {\bf 2015}, {\em 115},~171101,
  \href{http://xxx.lanl.gov/abs/1505.01842}{{\normalfont
  [arXiv:astro-ph.HE/1505.01842]}}.
\newblock
  doi:{\changeurlcolor{black}\href{https://doi.org/10.1103/PhysRevLett.115.171101}{\detokenize{10.1103/PhysRevLett.115.171101}}}.

\bibitem[{Camelio} \em{et~al.}(2018){Camelio}, {Dietrich}, and
  {Rosswog}]{Camelio2018}
{Camelio}, G.; {Dietrich}, T.; {Rosswog}, S.
\newblock {Disk formation in the collapse of supramassive neutron stars}.
\newblock {\em ArXiv e-prints} {\bf 2018},
  \href{http://xxx.lanl.gov/abs/1806.07775}{{\normalfont
  [arXiv:astro-ph.HE/1806.07775]}}.

\bibitem[{Shakura} and {Sunyaev}(1973)]{Shakura1973}
{Shakura}, N.I.; {Sunyaev}, R.A.
\newblock {Black holes in binary systems. Observational appearance.}
\newblock {\em Astron. Astrophys.} {\bf 1973}, {\em 24},~337--355.

\bibitem[{Falcke} and {Rezzolla}(2014)]{Falcke2013}
{Falcke}, H.; {Rezzolla}, L.
\newblock {Fast radio bursts: the last sign of supramassive neutron stars}.
\newblock {\em Astron. Astrophys.} {\bf 2014}, {\em 562},~A137,
  \href{http://xxx.lanl.gov/abs/1307.1409}{{\normalfont
  [arXiv:astro-ph.HE/1307.1409]}}.
\newblock
  doi:{\changeurlcolor{black}\href{https://doi.org/10.1051/0004-6361/201321996}{\detokenize{10.1051/0004-6361/201321996}}}.

\bibitem[{Shibata} and {Ury{\=u}}(2002)]{Shibata02a}
{Shibata}, M.; {Ury{\=u}}, K.
\newblock {Gravitational Waves from Merger of Binary Neutron Stars in Fully
  General Relativistic Simulation}.
\newblock {\em Progress of Theoretical Physics} {\bf 2002}, {\em
  107},~265--303,  \href{http://xxx.lanl.gov/abs/gr-qc/0203037}{{\normalfont
  [gr-qc/0203037]}}.
\newblock
  doi:{\changeurlcolor{black}\href{https://doi.org/10.1143/PTP.107.265}{\detokenize{10.1143/PTP.107.265}}}.

\bibitem[{Hotokezaka} \em{et~al.}(2011){Hotokezaka}, {Kyutoku}, {Okawa},
  {Shibata}, and {Kiuchi}]{Hotokezaka2011}
{Hotokezaka}, K.; {Kyutoku}, K.; {Okawa}, H.; {Shibata}, M.; {Kiuchi}, K.
\newblock {Binary neutron star mergers: Dependence on the nuclear equation of
  state}.
\newblock {\em Phys. Rev. D} {\bf 2011}, {\em 83},~124008,
  \href{http://xxx.lanl.gov/abs/1105.4370}{{\normalfont
  [arXiv:astro-ph.HE/1105.4370]}}.
\newblock
  doi:{\changeurlcolor{black}\href{https://doi.org/10.1103/PhysRevD.83.124008}{\detokenize{10.1103/PhysRevD.83.124008}}}.

\bibitem[{Bauswein} \em{et~al.}(2013){Bauswein}, {Baumgarte}, and
  {Janka}]{Bauswein2013}
{Bauswein}, A.; {Baumgarte}, T.W.; {Janka}, H.T.
\newblock {Prompt Merger Collapse and the Maximum Mass of Neutron Stars}.
\newblock {\em Phys. Rev. Lett.} {\bf 2013}, {\em 111},~131101,
  \href{http://xxx.lanl.gov/abs/1307.5191}{{\normalfont
  [arXiv:astro-ph.SR/1307.5191]}}.
\newblock
  doi:{\changeurlcolor{black}\href{https://doi.org/10.1103/PhysRevLett.111.131101}{\detokenize{10.1103/PhysRevLett.111.131101}}}.

\bibitem[{Shibata} \em{et~al.}(2005){Shibata}, {Taniguchi}, and
  {Ury{\=u}}]{Shibata05c}
{Shibata}, M.; {Taniguchi}, K.; {Ury{\=u}}, K.
\newblock {Merger of binary neutron stars with realistic equations of state in
  full general relativity}.
\newblock {\em Phys. Rev. D} {\bf 2005}, {\em 71},~084021,
  \href{http://xxx.lanl.gov/abs/gr-qc/0503119}{{\normalfont [gr-qc/0503119]}}.
\newblock
  doi:{\changeurlcolor{black}\href{https://doi.org/10.1103/PhysRevD.71.084021}{\detokenize{10.1103/PhysRevD.71.084021}}}.

\bibitem[{Oechslin} \em{et~al.}(2007){Oechslin}, {Janka}, and
  {Marek}]{Oechslin07a}
{Oechslin}, R.; {Janka}, H.T.; {Marek}, A.
\newblock {Relativistic neutron star merger simulations with non-zero
  temperature equations of state. I. Variation of binary parameters and
  equation of state}.
\newblock {\em Astron. Astrophys.} {\bf 2007}, {\em 467},~395--409,
  \href{http://xxx.lanl.gov/abs/astro-ph/0611047}{{\normalfont
  [astro-ph/0611047]}}.
\newblock
  doi:{\changeurlcolor{black}\href{https://doi.org/10.1051/0004-6361:20066682}{\detokenize{10.1051/0004-6361:20066682}}}.

\bibitem[{Studzi{\'n}ska} \em{et~al.}(2016){Studzi{\'n}ska}, {Kucaba},
  {Gondek-Rosi{\'n}ska}, {Villain}, and {Ansorg}]{Studzinska2016}
{Studzi{\'n}ska}, A.M.; {Kucaba}, M.; {Gondek-Rosi{\'n}ska}, D.; {Villain}, L.;
  {Ansorg}, M.
\newblock {Effect of the equation of state on the maximum mass of
  differentially rotating neutron stars}.
\newblock {\em Mon. Not. R. Astron. Soc.} {\bf 2016}, {\em 463},~2667--2679.
\newblock
  doi:{\changeurlcolor{black}\href{https://doi.org/10.1093/mnras/stw2152}{\detokenize{10.1093/mnras/stw2152}}}.

\bibitem[{Zhang, C. M.} \em{et~al.}(2011){Zhang, C. M.}, {Wang, J.}, {Zhao, Y.
  H.}, {Yin, H. X.}, {Song, L. M.}, {Menezes, D. P.}, {Wickramasinghe, D. T.},
  {Ferrario, L.}, and {Chardonnet, P.}]{Zhang2011b}
{Zhang, C. M.}.; {Wang, J.}.; {Zhao, Y. H.}.; {Yin, H. X.}.; {Song, L. M.}.;
  {Menezes, D. P.}.; {Wickramasinghe, D. T.}.; {Ferrario, L.}.; {Chardonnet,
  P.}.
\newblock Study of measured pulsar masses and their possible conclusions.
\newblock {\em A\&A} {\bf 2011}, {\em 527},~A83.
\newblock
  doi:{\changeurlcolor{black}\href{https://doi.org/10.1051/0004-6361/201015532}{\detokenize{10.1051/0004-6361/201015532}}}.

\bibitem[{Ruiz} and {Shapiro}(2017)]{Ruiz2017a}
{Ruiz}, M.; {Shapiro}, S.L.
\newblock {General relativistic magnetohydrodynamics simulations of
  prompt-collapse neutron star mergers: The absence of jets}.
\newblock {\em Phys. Rev. D} {\bf 2017}, {\em 96},~084063.
\newblock
  doi:{\changeurlcolor{black}\href{https://doi.org/10.1103/PhysRevD.96.084063}{\detokenize{10.1103/PhysRevD.96.084063}}}.

\bibitem[{Lorimer} \em{et~al.}(2007){Lorimer}, {Bailes}, {McLaughlin},
  {Narkevic}, and {Crawford}]{Lorimer2007}
{Lorimer}, D.R.; {Bailes}, M.; {McLaughlin}, M.A.; {Narkevic}, D.J.;
  {Crawford}, F.
\newblock {A Bright Millisecond Radio Burst of Extragalactic Origin}.
\newblock {\em Science} {\bf 2007}, {\em 318},~777,
  \href{http://xxx.lanl.gov/abs/0709.4301}{{\normalfont [0709.4301]}}.
\newblock
  doi:{\changeurlcolor{black}\href{https://doi.org/10.1126/science.1147532}{\detokenize{10.1126/science.1147532}}}.

\bibitem[{Rane} and {Lorimer}(2017)]{RaneLorimer2017}
{Rane}, A.; {Lorimer}, D.
\newblock {Fast Radio Bursts}.
\newblock {\em Journal of Astrophysics and Astronomy} {\bf 2017}, {\em 38},~55.
\newblock
  doi:{\changeurlcolor{black}\href{https://doi.org/10.1007/s12036-017-9478-1}{\detokenize{10.1007/s12036-017-9478-1}}}.

\bibitem[{Most} \em{et~al.}(2018){Most}, {Nathanail}, and {Rezzolla}]{Most2017}
{Most}, E.R.; {Nathanail}, A.; {Rezzolla}, L.
\newblock {Electromagnetic emission from blitzars and its impact on
  non-repeating fast radio bursts}.
\newblock {\em arXiv:1801.05705, ApJ, \textit{in press}} {\bf 2018},
  \href{http://xxx.lanl.gov/abs/1801.05705}{{\normalfont
  [arXiv:astro-ph.HE/1801.05705]}}.

\bibitem[{Rowlinson} \em{et~al.}(2013){Rowlinson}, {O'Brien}, {Metzger},
  {Tanvir}, and {Levan}]{Rowlinson2013}
{Rowlinson}, A.; {O'Brien}, P.T.; {Metzger}, B.D.; {Tanvir}, N.R.; {Levan},
  A.J.
\newblock {Signatures of magnetar central engines in short GRB light curves}.
\newblock {\em Mon. Not. R. Astron. Soc.} {\bf 2013}, {\em 430},~1061--1087,
  \href{http://xxx.lanl.gov/abs/1301.0629}{{\normalfont
  [arXiv:astro-ph.HE/1301.0629]}}.
\newblock
  doi:{\changeurlcolor{black}\href{https://doi.org/10.1093/mnras/sts683}{\detokenize{10.1093/mnras/sts683}}}.

\bibitem[{Zhang} and {M{\'e}sz{\'a}ros}(2001)]{Zhang2001}
{Zhang}, B.; {M{\'e}sz{\'a}ros}, P.
\newblock {Gamma-Ray Burst Afterglow with Continuous Energy Injection:
  Signature of a Highly Magnetized Millisecond Pulsar}.
\newblock {\em Astrophys. J.} {\bf 2001}, {\em 552},~L35--L38,
  \href{http://xxx.lanl.gov/abs/astro-ph/0011133}{{\normalfont
  [astro-ph/0011133]}}.
\newblock
  doi:{\changeurlcolor{black}\href{https://doi.org/10.1086/320255}{\detokenize{10.1086/320255}}}.

\bibitem[{Gao} and {Fan}(2006)]{Gao2006}
{Gao}, W.H.; {Fan}, Y.Z.
\newblock {Short-living Supermassive Magnetar Model for the Early X-ray Flares
  Following Short GRBs}.
\newblock {\em Chinese Journal of Astronomy and Astrophysics} {\bf 2006}, {\em
  6},~513--516,  \href{http://xxx.lanl.gov/abs/astro-ph/0512646}{{\normalfont
  [astro-ph/0512646]}}.
\newblock
  doi:{\changeurlcolor{black}\href{https://doi.org/10.1088/1009-9271/6/5/01}{\detokenize{10.1088/1009-9271/6/5/01}}}.

\bibitem[{Fan} and {Xu}(2006)]{Fan2006}
{Fan}, Y.Z.; {Xu}, D.
\newblock {The X-ray afterglow flat segment in short GRB 051221A: Energy
  injection from a millisecond magnetar?}
\newblock {\em Mon. Not. R. Astron. Soc.} {\bf 2006}, {\em 372},~L19--L22,
  \href{http://xxx.lanl.gov/abs/astro-ph/0605445}{{\normalfont
  [astro-ph/0605445]}}.
\newblock
  doi:{\changeurlcolor{black}\href{https://doi.org/10.1111/j.1745-3933.2006.00217.x}{\detokenize{10.1111/j.1745-3933.2006.00217.x}}}.

\bibitem[{Metzger} \em{et~al.}(2008){Metzger}, {Piro}, and
  {Quataert}]{Metzger2008a}
{Metzger}, B.D.; {Piro}, A.L.; {Quataert}, E.
\newblock {Time-dependent models of accretion discs formed from compact object
  mergers}.
\newblock {\em Mon. Not. R. Astron. Soc.} {\bf 2008}, {\em 390},~781--797,
  \href{http://xxx.lanl.gov/abs/0805.4415}{{\normalfont [0805.4415]}}.
\newblock
  doi:{\changeurlcolor{black}\href{https://doi.org/10.1111/j.1365-2966.2008.13789.x}{\detokenize{10.1111/j.1365-2966.2008.13789.x}}}.

\bibitem[{Metzger} \em{et~al.}(2010){Metzger}, {Mart{\'{\i}}nez-Pinedo},
  {Darbha}, {Quataert}, {Arcones}, {Kasen}, {Thomas}, {Nugent}, {Panov}, and
  {Zinner}]{Metzger:2010}
{Metzger}, B.D.; {Mart{\'{\i}}nez-Pinedo}, G.; {Darbha}, S.; {Quataert}, E.;
  {Arcones}, A.; {Kasen}, D.; {Thomas}, R.; {Nugent}, P.; {Panov}, I.V.;
  {Zinner}, N.T.
\newblock {Electromagnetic counterparts of compact object mergers powered by
  the radioactive decay of r-process nuclei}.
\newblock {\em Mon. Not. R. Astron. Soc.} {\bf 2010}, {\em 406},~2650--2662,
  \href{http://xxx.lanl.gov/abs/1001.5029}{{\normalfont
  [arXiv:astro-ph.HE/1001.5029]}}.
\newblock
  doi:{\changeurlcolor{black}\href{https://doi.org/10.1111/j.1365-2966.2010.16864.x}{\detokenize{10.1111/j.1365-2966.2010.16864.x}}}.

\bibitem[{Giacomazzo} and {Perna}(2013)]{Giacomazzo2013}
{Giacomazzo}, B.; {Perna}, R.
\newblock {Formation of Stable Magnetars from Binary Neutron Star Mergers}.
\newblock {\em Astrophys. J.} {\bf 2013}, {\em 771},~L26,
  \href{http://xxx.lanl.gov/abs/1306.1608}{{\normalfont
  [arXiv:astro-ph.HE/1306.1608]}}.
\newblock
  doi:{\changeurlcolor{black}\href{https://doi.org/10.1088/2041-8205/771/2/L26}{\detokenize{10.1088/2041-8205/771/2/L26}}}.

\bibitem[{Dall'Osso} \em{et~al.}(2015){Dall'Osso}, {Giacomazzo}, {Perna}, and
  {Stella}]{DallOsso2014}
{Dall'Osso}, S.; {Giacomazzo}, B.; {Perna}, R.; {Stella}, L.
\newblock {Gravitational Waves from Massive Magnetars Formed in Binary Neutron
  Star Mergers}.
\newblock {\em Astrophys. J.} {\bf 2015}, {\em 798},~25,
  \href{http://xxx.lanl.gov/abs/1408.0013}{{\normalfont
  [arXiv:astro-ph.HE/1408.0013]}}.
\newblock
  doi:{\changeurlcolor{black}\href{https://doi.org/10.1088/0004-637X/798/1/25}{\detokenize{10.1088/0004-637X/798/1/25}}}.

\bibitem[{Rezzolla} and {Kumar}(2015)]{Rezzolla2014b}
{Rezzolla}, L.; {Kumar}, P.
\newblock {A Novel Paradigm for Short Gamma-Ray Bursts With Extended X-Ray
  Emission}.
\newblock {\em Astrophys. J.} {\bf 2015}, {\em 802},~95,
  \href{http://xxx.lanl.gov/abs/1410.8560}{{\normalfont
  [arXiv:astro-ph.HE/1410.8560]}}.
\newblock
  doi:{\changeurlcolor{black}\href{https://doi.org/10.1088/0004-637X/802/2/95}{\detokenize{10.1088/0004-637X/802/2/95}}}.

\bibitem[{Ciolfi} and {Siegel}(2015)]{Ciolfi2014}
{Ciolfi}, R.; {Siegel}, D.M.
\newblock {Short Gamma-Ray Bursts in the ``Time-reversal'' Scenario}.
\newblock {\em Astrophys. J.} {\bf 2015}, {\em 798},~L36,
  \href{http://xxx.lanl.gov/abs/1411.2015}{{\normalfont
  [arXiv:astro-ph.HE/1411.2015]}}.
\newblock
  doi:{\changeurlcolor{black}\href{https://doi.org/10.1088/2041-8205/798/2/L36}{\detokenize{10.1088/2041-8205/798/2/L36}}}.

\bibitem[{Nathanail}(2018)]{Nathanail2018}
{Nathanail}, A.
\newblock {The collapse of the remnant from a binary neutron star merger powers
  a magnetic explosion: implication to short gamma-ray bursts}.
\newblock {\em ArXiv e-prints} {\bf 2018},
  \href{http://xxx.lanl.gov/abs/1801.05680}{{\normalfont
  [arXiv:astro-ph.HE/1801.05680]}}.

\bibitem[{Spitkovsky}(2006)]{Spitkovsky2006}
{Spitkovsky}, A.
\newblock {Time-dependent Force-free Pulsar Magnetospheres: Axisymmetric and
  Oblique Rotators}.
\newblock {\em Astrophys. J. Letters} {\bf 2006}, {\em 648},~L51--L54,
  \href{http://xxx.lanl.gov/abs/astro-ph/0603147}{{\normalfont
  [astro-ph/0603147]}}.
\newblock
  doi:{\changeurlcolor{black}\href{https://doi.org/10.1086/507518}{\detokenize{10.1086/507518}}}.

\bibitem[{Contopoulos} and {Spitkovsky}(2006)]{Contopoulos2006a}
{Contopoulos}, I.; {Spitkovsky}, A.
\newblock {Revised Pulsar Spin-down}.
\newblock {\em Astrophys. J.} {\bf 2006}, {\em 643},~1139--1145,
  \href{http://xxx.lanl.gov/abs/astro-ph/0512002}{{\normalfont
  [astro-ph/0512002]}}.
\newblock
  doi:{\changeurlcolor{black}\href{https://doi.org/10.1086/501161}{\detokenize{10.1086/501161}}}.

\bibitem[{Thompson} \em{et~al.}(2003){Thompson}, {Burrows}, and
  {Pinto}]{Thompson03}
{Thompson}, T.A.; {Burrows}, A.; {Pinto}, P.A.
\newblock {Shock Breakout in Core-Collapse Supernovae and Its Neutrino
  Signature}.
\newblock {\em Astrophys. J.} {\bf 2003}, {\em 592},~434--456,
  \href{http://xxx.lanl.gov/abs/arXiv:astro-ph/0211194}{{\normalfont
  [arXiv:astro-ph/0211194]}}.
\newblock
  doi:{\changeurlcolor{black}\href{https://doi.org/10.1086/375701}{\detokenize{10.1086/375701}}}.

\bibitem[{Komissarov}(2007)]{Komissarov2007}
{Komissarov}, S.S.
\newblock {Multidimensional numerical scheme for resistive relativistic
  magnetohydrodynamics}.
\newblock {\em Mon. Not. R. Astron. Soc.} {\bf 2007}, {\em 382},~995--1004,
  \href{http://xxx.lanl.gov/abs/0708.0323}{{\normalfont [0708.0323]}}.
\newblock
  doi:{\changeurlcolor{black}\href{https://doi.org/10.1111/j.1365-2966.2007.12448.x}{\detokenize{10.1111/j.1365-2966.2007.12448.x}}}.

\bibitem[{Thompson} and {ud-Doula}(2018)]{Thompson2018}
{Thompson}, T.A.; {ud-Doula}, A.
\newblock {High-entropy ejections from magnetized proto-neutron star winds:
  implications for heavy element nucleosynthesis}.
\newblock {\em Mon. Not. R. Astron. Soc.} {\bf 2018}, {\em 476},~5502--5515.
\newblock
  doi:{\changeurlcolor{black}\href{https://doi.org/10.1093/mnras/sty480}{\detokenize{10.1093/mnras/sty480}}}.

\bibitem[{Uzdensky}(2005)]{Uzdensky2005}
{Uzdensky}, D.A.
\newblock {Force-Free Magnetosphere of an Accretion Disk-Black Hole System. II.
  Kerr Geometry}.
\newblock {\em Astrophys. J.} {\bf 2005}, {\em 620},~889--904,
  \href{http://xxx.lanl.gov/abs/arXiv:astro-ph/0410715}{{\normalfont
  [arXiv:astro-ph/0410715]}}.
\newblock
  doi:{\changeurlcolor{black}\href{https://doi.org/10.1086/427180}{\detokenize{10.1086/427180}}}.

\bibitem[{Parfrey} \em{et~al.}(2015){Parfrey}, {Giannios}, and
  {Beloborodov}]{Parfrey2015}
{Parfrey}, K.; {Giannios}, D.; {Beloborodov}, A.M.
\newblock {Black hole jets without large-scale net magnetic flux}.
\newblock {\em Mon. Not. R. Astron. Soc.} {\bf 2015}, {\em 446},~L61--L65,
  \href{http://xxx.lanl.gov/abs/1410.0374}{{\normalfont
  [arXiv:astro-ph.HE/1410.0374]}}.
\newblock
  doi:{\changeurlcolor{black}\href{https://doi.org/10.1093/mnrasl/slu162}{\detokenize{10.1093/mnrasl/slu162}}}.

\bibitem[{Contopoulos} \em{et~al.}(2015){Contopoulos}, {Nathanail}, and
  {Katsanikas}]{Contopoulos2015}
{Contopoulos}, I.; {Nathanail}, A.; {Katsanikas}, M.
\newblock {The Cosmic Battery in Astrophysical Accretion Disks}.
\newblock {\em Astrophys. J.} {\bf 2015}, {\em 805},~105,
  \href{http://xxx.lanl.gov/abs/1501.05784}{{\normalfont
  [arXiv:astro-ph.HE/1501.05784]}}.
\newblock
  doi:{\changeurlcolor{black}\href{https://doi.org/10.1088/0004-637X/805/2/105}{\detokenize{10.1088/0004-637X/805/2/105}}}.

\bibitem[{Nathanail} and {Contopoulos}(2015)]{Nathanail2015}
{Nathanail}, A.; {Contopoulos}, I.
\newblock {Are ultralong gamma-ray bursts powered by black holes spinning
  down?}
\newblock {\em Mon. Not. R. Astron. Soc.} {\bf 2015}, {\em 453},~L1--L5,
  \href{http://xxx.lanl.gov/abs/1504.03906}{{\normalfont
  [arXiv:astro-ph.HE/1504.03906]}}.
\newblock
  doi:{\changeurlcolor{black}\href{https://doi.org/10.1093/mnrasl/slv081}{\detokenize{10.1093/mnrasl/slv081}}}.

\bibitem[{Nathanail} \em{et~al.}(2016){Nathanail}, {Strantzalis}, and
  {Contopoulos}]{Nathanail2016}
{Nathanail}, A.; {Strantzalis}, A.; {Contopoulos}, I.
\newblock {The rapid decay phase of the afterglow as the signature of the
  Blandford-Znajek mechanism}.
\newblock {\em Mon. Not. R. Astron. Soc.} {\bf 2016}, {\em 455},~4479--4486,
  \href{http://xxx.lanl.gov/abs/1507.02143}{{\normalfont
  [arXiv:astro-ph.HE/1507.02143]}}.
\newblock
  doi:{\changeurlcolor{black}\href{https://doi.org/10.1093/mnras/stv2558}{\detokenize{10.1093/mnras/stv2558}}}.

\bibitem[{Nakar} and {Piran}(2018)]{Nakar2018}
{Nakar}, E.; {Piran}, T.
\newblock {Implications of the radio and X-ray emission that followed
  GW170817}.
\newblock {\em Mon. Not. R. Astron. Soc.} {\bf 2018}, {\em 478},~407--415,
  \href{http://xxx.lanl.gov/abs/1801.09712}{{\normalfont
  [arXiv:astro-ph.HE/1801.09712]}}.
\newblock
  doi:{\changeurlcolor{black}\href{https://doi.org/10.1093/mnras/sty952}{\detokenize{10.1093/mnras/sty952}}}.

\bibitem[{Nagakura} \em{et~al.}(2014){Nagakura}, {Hotokezaka}, {Sekiguchi},
  {Shibata}, and {Ioka}]{Nagakura2014}
{Nagakura}, H.; {Hotokezaka}, K.; {Sekiguchi}, Y.; {Shibata}, M.; {Ioka}, K.
\newblock {Jet Collimation in the Ejecta of Double Neutron Star Mergers: A New
  Canonical Picture of Short Gamma-Ray Bursts}.
\newblock {\em Astrophys. J.} {\bf 2014}, {\em 784},~L28,
  \href{http://xxx.lanl.gov/abs/1403.0956}{{\normalfont
  [arXiv:astro-ph.HE/1403.0956]}}.
\newblock
  doi:{\changeurlcolor{black}\href{https://doi.org/10.1088/2041-8205/784/2/L28}{\detokenize{10.1088/2041-8205/784/2/L28}}}.

\bibitem[{Blandford} and {McKee}(1976)]{Blandford-McKee1976}
{Blandford}, R.D.; {McKee}, C.F.
\newblock {Fluid dynamics of relativistic blast waves}.
\newblock {\em Physics of Fluids} {\bf 1976}, {\em 19},~1130--1138.
\newblock
  doi:{\changeurlcolor{black}\href{https://doi.org/10.1063/1.861619}{\detokenize{10.1063/1.861619}}}.

\bibitem[{Ramirez-Ruiz} \em{et~al.}(2002){Ramirez-Ruiz}, {Celotti}, and
  {Rees}]{Ramirez-Ruiz2002}
{Ramirez-Ruiz}, E.; {Celotti}, A.; {Rees}, M.J.
\newblock {Events in the life of a cocoon surrounding a light, collapsar jet}.
\newblock {\em Mon. Not. R. Astron. Soc.} {\bf 2002}, {\em 337},~1349--1356,
  \href{http://xxx.lanl.gov/abs/arXiv:astro-ph/0205108}{{\normalfont
  [arXiv:astro-ph/0205108]}}.
\newblock
  doi:{\changeurlcolor{black}\href{https://doi.org/10.1046/j.1365-8711.2002.05995.x}{\detokenize{10.1046/j.1365-8711.2002.05995.x}}}.

\bibitem[{Morsony} \em{et~al.}(2007){Morsony}, {Lazzati}, and
  {Begelman}]{Morsony2007}
{Morsony}, B.J.; {Lazzati}, D.; {Begelman}, M.C.
\newblock {Temporal and Angular Properties of Gamma-Ray Burst Jets Emerging
  from Massive Stars}.
\newblock {\em Astrophys. J.} {\bf 2007}, {\em 665},~569--598,
  \href{http://xxx.lanl.gov/abs/astro-ph/0609254}{{\normalfont
  [astro-ph/0609254]}}.
\newblock
  doi:{\changeurlcolor{black}\href{https://doi.org/10.1086/519483}{\detokenize{10.1086/519483}}}.

\bibitem[{Lazzati} \em{et~al.}(2010){Lazzati}, {Morsony}, and
  {Begelman}]{Lazzati2010}
{Lazzati}, D.; {Morsony}, B.J.; {Begelman}, M.C.
\newblock {Short-duration Gamma-ray Bursts From Off-axis Collapsars}.
\newblock {\em Astrophys. J.} {\bf 2010}, {\em 717},~239--244,
  \href{http://xxx.lanl.gov/abs/0911.3313}{{\normalfont
  [arXiv:astro-ph.HE/0911.3313]}}.
\newblock
  doi:{\changeurlcolor{black}\href{https://doi.org/10.1088/0004-637X/717/1/239}{\detokenize{10.1088/0004-637X/717/1/239}}}.

\bibitem[{Mizuta} and {Aloy}(2008)]{Mizuta:2008a}
{Mizuta}, A.; {Aloy}, M.A.
\newblock {Angular Energy Distribution of Collapsar-Jets}.
\newblock {\em arXiv:0812.4813} {\bf 2008},
  \href{http://xxx.lanl.gov/abs/0812.4813}{{\normalfont [0812.4813]}}.

\bibitem[{L{\'o}pez-C{\'a}mara} \em{et~al.}(2013){L{\'o}pez-C{\'a}mara},
  {Morsony}, {Begelman}, and {Lazzati}]{LopezCamara2013}
{L{\'o}pez-C{\'a}mara}, D.; {Morsony}, B.J.; {Begelman}, M.C.; {Lazzati}, D.
\newblock {Three-dimensional Adaptive Mesh Refinement Simulations of
  Long-duration Gamma-Ray Burst Jets inside Massive Progenitor Stars}.
\newblock {\em Astrophys. J.} {\bf 2013}, {\em 767},~19,
  \href{http://xxx.lanl.gov/abs/1212.0539}{{\normalfont
  [arXiv:astro-ph.HE/1212.0539]}}.
\newblock
  doi:{\changeurlcolor{black}\href{https://doi.org/10.1088/0004-637X/767/1/19}{\detokenize{10.1088/0004-637X/767/1/19}}}.

\bibitem[{Mizuta} and {Ioka}(2013)]{Mizuta2013}
{Mizuta}, A.; {Ioka}, K.
\newblock {Opening Angles of Collapsar Jets}.
\newblock {\em Astrophys. J.} {\bf 2013}, {\em 777},~162,
  \href{http://xxx.lanl.gov/abs/1304.0163}{{\normalfont
  [arXiv:astro-ph.HE/1304.0163]}}.
\newblock
  doi:{\changeurlcolor{black}\href{https://doi.org/10.1088/0004-637X/777/2/162}{\detokenize{10.1088/0004-637X/777/2/162}}}.

\bibitem[{Murguia-Berthier} \em{et~al.}(2014){Murguia-Berthier}, {Montes},
  {Ramirez-Ruiz}, {De Colle}, and {Lee}]{Murguia-Berthier2014}
{Murguia-Berthier}, A.; {Montes}, G.; {Ramirez-Ruiz}, E.; {De Colle}, F.;
  {Lee}, W.H.
\newblock {Necessary Conditions for Short Gamma-Ray Burst Production in Binary
  Neutron Star Mergers}.
\newblock {\em Astrophys. J.} {\bf 2014}, {\em 788},~L8,
  \href{http://xxx.lanl.gov/abs/1404.0383}{{\normalfont
  [arXiv:astro-ph.HE/1404.0383]}}.
\newblock
  doi:{\changeurlcolor{black}\href{https://doi.org/10.1088/2041-8205/788/1/L8}{\detokenize{10.1088/2041-8205/788/1/L8}}}.

\bibitem[{Murguia-Berthier} \em{et~al.}(2017){Murguia-Berthier},
  {Ramirez-Ruiz}, {Montes}, {De Colle}, {Rezzolla}, {Rosswog}, {Takami},
  {Perego}, and {Lee}]{Murguia-Berthier2016}
{Murguia-Berthier}, A.; {Ramirez-Ruiz}, E.; {Montes}, G.; {De Colle}, F.;
  {Rezzolla}, L.; {Rosswog}, S.; {Takami}, K.; {Perego}, A.; {Lee}, W.H.
\newblock {The Properties of Short Gamma-Ray Burst Jets Triggered by Neutron
  Star Mergers}.
\newblock {\em Astrophys. J. Lett.} {\bf 2017}, {\em 835},~L34,
  \href{http://xxx.lanl.gov/abs/1609.04828}{{\normalfont
  [arXiv:astro-ph.HE/1609.04828]}}.
\newblock
  doi:{\changeurlcolor{black}\href{https://doi.org/10.3847/2041-8213/aa5b9e}{\detokenize{10.3847/2041-8213/aa5b9e}}}.

\bibitem[{Nagakura} \em{et~al.}(2011){Nagakura}, {Ito}, {Kiuchi}, and
  {Yamada}]{Nagakura2011}
{Nagakura}, H.; {Ito}, H.; {Kiuchi}, K.; {Yamada}, S.
\newblock {Jet Propagations, Breakouts, and Photospheric Emissions in
  Collapsing Massive Progenitors of Long-duration Gamma-ray Bursts}.
\newblock {\em Astrophys. J.} {\bf 2011}, {\em 731},~80,
  \href{http://xxx.lanl.gov/abs/1009.2326}{{\normalfont
  [arXiv:astro-ph.HE/1009.2326]}}.
\newblock
  doi:{\changeurlcolor{black}\href{https://doi.org/10.1088/0004-637X/731/2/80}{\detokenize{10.1088/0004-637X/731/2/80}}}.

\bibitem[{Qian} and {Woosley}(1996)]{Qian1996}
{Qian}, Y.Z.; {Woosley}, S.E.
\newblock {Nucleosynthesis in Neutrino-driven Winds. I. The Physical
  Conditions}.
\newblock {\em Astrophys. J.} {\bf 1996}, {\em 471},~331,
  \href{http://xxx.lanl.gov/abs/astro-ph/9611094}{{\normalfont
  [astro-ph/9611094]}}.
\newblock
  doi:{\changeurlcolor{black}\href{https://doi.org/10.1086/177973}{\detokenize{10.1086/177973}}}.

\bibitem[{Rosswog} and {Ramirez-Ruiz}(2002)]{Rosswog2002b}
{Rosswog}, S.; {Ramirez-Ruiz}, E.
\newblock {Jets, winds and bursts from coalescing neutron stars}.
\newblock {\em Mon. Not. R. Astron. Soc.} {\bf 2002}, {\em 336},~L7--L11,
  \href{http://xxx.lanl.gov/abs/astro-ph/0207576}{{\normalfont
  [astro-ph/0207576]}}.
\newblock
  doi:{\changeurlcolor{black}\href{https://doi.org/10.1046/j.1365-8711.2002.05898.x}{\detokenize{10.1046/j.1365-8711.2002.05898.x}}}.

\bibitem[{Duffell} \em{et~al.}(2015){Duffell}, {Quataert}, and
  {MacFadyen}]{Duffell2015}
{Duffell}, P.C.; {Quataert}, E.; {MacFadyen}, A.I.
\newblock {A Narrow Short-duration GRB Jet from a Wide Central Engine}.
\newblock {\em Astrophys. J.} {\bf 2015}, {\em 813},~64,
  \href{http://xxx.lanl.gov/abs/1505.05538}{{\normalfont
  [arXiv:astro-ph.HE/1505.05538]}}.
\newblock
  doi:{\changeurlcolor{black}\href{https://doi.org/10.1088/0004-637X/813/1/64}{\detokenize{10.1088/0004-637X/813/1/64}}}.

\bibitem[{Gehrels} \em{et~al.}(2009){Gehrels}, {Ramirez-Ruiz}, and
  {Fox}]{Gehrels2009}
{Gehrels}, N.; {Ramirez-Ruiz}, E.; {Fox}, D.B.
\newblock {Gamma-Ray Bursts in the Swift Era}.
\newblock {\em Annual Review of Astronomy \& Astrophysics} {\bf 2009}, {\em
  47},~567--617,  \href{http://xxx.lanl.gov/abs/0909.1531}{{\normalfont
  [arXiv:astro-ph.HE/0909.1531]}}.
\newblock
  doi:{\changeurlcolor{black}\href{https://doi.org/10.1146/annurev.astro.46.060407.145147}{\detokenize{10.1146/annurev.astro.46.060407.145147}}}.

\bibitem[{Hotokezaka} \em{et~al.}(2013){Hotokezaka}, {Kyutoku}, {Tanaka},
  {Kiuchi}, {Sekiguchi}, {Shibata}, and {Wanajo}]{Hotokezaka2013d}
{Hotokezaka}, K.; {Kyutoku}, K.; {Tanaka}, M.; {Kiuchi}, K.; {Sekiguchi}, Y.;
  {Shibata}, M.; {Wanajo}, S.
\newblock {Progenitor Models of the Electromagnetic Transient Associated with
  the Short Gamma Ray Burst 130603B}.
\newblock {\em Astrophys. J.} {\bf 2013}, {\em 778},~L16,
  \href{http://xxx.lanl.gov/abs/1310.1623}{{\normalfont
  [arXiv:astro-ph.HE/1310.1623]}}.
\newblock
  doi:{\changeurlcolor{black}\href{https://doi.org/10.1088/2041-8205/778/1/L16}{\detokenize{10.1088/2041-8205/778/1/L16}}}.

\bibitem[{Rosswog} and {Ramirez-Ruiz}(2003)]{Rosswog2003c}
{Rosswog}, S.; {Ramirez-Ruiz}, E.
\newblock {On the diversity of short gamma-ray bursts}.
\newblock {\em Mon. Not. R. Astron. Soc.} {\bf 2003}, {\em 343},~L36--L40,
  \href{http://xxx.lanl.gov/abs/astro-ph/0306172}{{\normalfont
  [astro-ph/0306172]}}.
\newblock
  doi:{\changeurlcolor{black}\href{https://doi.org/10.1046/j.1365-8711.2003.06889.x}{\detokenize{10.1046/j.1365-8711.2003.06889.x}}}.

\bibitem[{Lazzati} \em{et~al.}(2017){Lazzati}, {L{\'o}pez-C{\'a}mara},
  {Cantiello}, {Morsony}, {Perna}, and {Workman}]{Lazzati2017b}
{Lazzati}, D.; {L{\'o}pez-C{\'a}mara}, D.; {Cantiello}, M.; {Morsony}, B.J.;
  {Perna}, R.; {Workman}, J.C.
\newblock {Off-axis Prompt X-Ray Transients from the Cocoon of Short Gamma-Ray
  Bursts}.
\newblock {\em Astrophys. J. Letters} {\bf 2017}, {\em 848},~L6,
  \href{http://xxx.lanl.gov/abs/1709.01468}{{\normalfont
  [arXiv:astro-ph.HE/1709.01468]}}.
\newblock
  doi:{\changeurlcolor{black}\href{https://doi.org/10.3847/2041-8213/aa8f3d}{\detokenize{10.3847/2041-8213/aa8f3d}}}.

\bibitem[{Bucciantini} \em{et~al.}(2012){Bucciantini}, {Metzger}, {Thompson},
  and {Quataert}]{Bucciantini2012}
{Bucciantini}, N.; {Metzger}, B.D.; {Thompson}, T.A.; {Quataert}, E.
\newblock {Short gamma-ray bursts with extended emission from magnetar birth:
  jet formation and collimation}.
\newblock {\em Mon. Not. R. Astron. Soc.} {\bf 2012}, {\em 419},~1537--1545,
  \href{http://xxx.lanl.gov/abs/1106.4668}{{\normalfont
  [arXiv:astro-ph.HE/1106.4668]}}.
\newblock
  doi:{\changeurlcolor{black}\href{https://doi.org/10.1111/j.1365-2966.2011.19810.x}{\detokenize{10.1111/j.1365-2966.2011.19810.x}}}.

\bibitem[{Bromberg} \em{et~al.}(2018){Bromberg}, {Tchekhovskoy}, {Gottlieb},
  {Nakar}, and {Piran}]{Bromberg2018}
{Bromberg}, O.; {Tchekhovskoy}, A.; {Gottlieb}, O.; {Nakar}, E.; {Piran}, T.
\newblock {The {$\gamma$}-rays that accompanied GW170817 and the observational
  signature of a magnetic jet breaking out of NS merger ejecta}.
\newblock {\em Mon. Not. R. Astron. Soc.} {\bf 2018}, {\em 475},~2971--2977,
  \href{http://xxx.lanl.gov/abs/1710.05897}{{\normalfont
  [arXiv:astro-ph.HE/1710.05897]}}.
\newblock
  doi:{\changeurlcolor{black}\href{https://doi.org/10.1093/mnras/stx3316}{\detokenize{10.1093/mnras/stx3316}}}.

\bibitem[{Perego} \em{et~al.}(2014){Perego}, {Rosswog}, {Cabez{\'o}n},
  {Korobkin}, {K{\"a}ppeli}, {Arcones}, and {Liebend{\"o}rfer}]{Perego2014}
{Perego}, A.; {Rosswog}, S.; {Cabez{\'o}n}, R.M.; {Korobkin}, O.;
  {K{\"a}ppeli}, R.; {Arcones}, A.; {Liebend{\"o}rfer}, M.
\newblock {Neutrino-driven winds from neutron star merger remnants}.
\newblock {\em Mon. Not. R. Astron. Soc.} {\bf 2014}, {\em 443},~3134--3156,
  \href{http://xxx.lanl.gov/abs/1405.6730}{{\normalfont
  [arXiv:astro-ph.HE/1405.6730]}}.
\newblock
  doi:{\changeurlcolor{black}\href{https://doi.org/10.1093/mnras/stu1352}{\detokenize{10.1093/mnras/stu1352}}}.

\bibitem[{Nakar} and {Piran}(2011)]{Nakar2011}
{Nakar}, E.; {Piran}, T.
\newblock {Detectable radio flares following gravitational waves from mergers
  of binary neutron stars}.
\newblock {\em Nature} {\bf 2011}, {\em 478},~82--84,
  \href{http://xxx.lanl.gov/abs/1102.1020}{{\normalfont
  [arXiv:astro-ph.HE/1102.1020]}}.
\newblock
  doi:{\changeurlcolor{black}\href{https://doi.org/10.1038/nature10365}{\detokenize{10.1038/nature10365}}}.

\bibitem[{Lazzati} \em{et~al.}(2017){Lazzati}, {Deich}, {Morsony}, and
  {Workman}]{Lazzati2017a}
{Lazzati}, D.; {Deich}, A.; {Morsony}, B.J.; {Workman}, J.C.
\newblock {Off-axis emission of short {$\gamma$}-ray bursts and the
  detectability of electromagnetic counterparts of gravitational-wave-detected
  binary mergers}.
\newblock {\em Mon. Not. R. Astron. Soc.} {\bf 2017}, {\em 471},~1652--1661,
  \href{http://xxx.lanl.gov/abs/1610.01157}{{\normalfont
  [arXiv:astro-ph.HE/1610.01157]}}.
\newblock
  doi:{\changeurlcolor{black}\href{https://doi.org/10.1093/mnras/stx1683}{\detokenize{10.1093/mnras/stx1683}}}.

\bibitem[{Nakar} and {Piran}(2017)]{Nakar2017}
{Nakar}, E.; {Piran}, T.
\newblock {The Observable Signatures of GRB Cocoons}.
\newblock {\em Astrophys. J.} {\bf 2017}, {\em 834},~28,
  \href{http://xxx.lanl.gov/abs/1610.05362}{{\normalfont
  [arXiv:astro-ph.HE/1610.05362]}}.
\newblock
  doi:{\changeurlcolor{black}\href{https://doi.org/10.3847/1538-4357/834/1/28}{\detokenize{10.3847/1538-4357/834/1/28}}}.

\bibitem[{Bromberg} \em{et~al.}(2011){Bromberg}, {Nakar}, {Piran}, and
  {Sari}]{Bromberg2011}
{Bromberg}, O.; {Nakar}, E.; {Piran}, T.; {Sari}, R.
\newblock {The Propagation of Relativistic Jets in External Media}.
\newblock {\em Astrophys. J.} {\bf 2011}, {\em 740},~100,
  \href{http://xxx.lanl.gov/abs/1107.1326}{{\normalfont
  [arXiv:astro-ph.HE/1107.1326]}}.
\newblock
  doi:{\changeurlcolor{black}\href{https://doi.org/10.1088/0004-637X/740/2/100}{\detokenize{10.1088/0004-637X/740/2/100}}}.

\bibitem[{Kathirgamaraju} \em{et~al.}(2018){Kathirgamaraju}, {Barniol Duran},
  and {Giannios}]{Kathirgamaraju2018}
{Kathirgamaraju}, A.; {Barniol Duran}, R.; {Giannios}, D.
\newblock {Off-axis short GRBs from structured jets as counterparts to GW
  events}.
\newblock {\em Mon. Not. R. Astron. Soc.} {\bf 2018}, {\em 473},~L121--L125,
  \href{http://xxx.lanl.gov/abs/1708.07488}{{\normalfont
  [arXiv:astro-ph.HE/1708.07488]}}.
\newblock
  doi:{\changeurlcolor{black}\href{https://doi.org/10.1093/mnrasl/slx175}{\detokenize{10.1093/mnrasl/slx175}}}.

\bibitem[{Gottlieb} \em{et~al.}(2018{\natexlab{a}}){Gottlieb}, {Nakar}, and
  {Piran}]{Gottlieb2018}
{Gottlieb}, O.; {Nakar}, E.; {Piran}, T.
\newblock {The cocoon emission - an electromagnetic counterpart to
  gravitational waves from neutron star mergers}.
\newblock {\em Mon. Not. R. Astron. Soc.} {\bf 2018}, {\em 473},~576--584,
  \href{http://xxx.lanl.gov/abs/1705.10797}{{\normalfont
  [arXiv:astro-ph.HE/1705.10797]}}.
\newblock
  doi:{\changeurlcolor{black}\href{https://doi.org/10.1093/mnras/stx2357}{\detokenize{10.1093/mnras/stx2357}}}.

\bibitem[{Gottlieb} \em{et~al.}(2018{\natexlab{b}}){Gottlieb}, {Nakar},
  {Piran}, and {Hotokezaka}]{Gottlieb2018b}
{Gottlieb}, O.; {Nakar}, E.; {Piran}, T.; {Hotokezaka}, K.
\newblock {A cocoon shock breakout as the origin of the {$\gamma$}-ray emission
  in GW170817}.
\newblock {\em Mon. Not. R. Astron. Soc.} {\bf 2018}, {\em 479},~588--600,
  \href{http://xxx.lanl.gov/abs/1710.05896}{{\normalfont
  [arXiv:astro-ph.HE/1710.05896]}}.
\newblock
  doi:{\changeurlcolor{black}\href{https://doi.org/10.1093/mnras/sty1462}{\detokenize{10.1093/mnras/sty1462}}}.

\bibitem[{Piran} \em{et~al.}(2013){Piran}, {Nakar}, and {Rosswog}]{Piran2013}
{Piran}, T.; {Nakar}, E.; {Rosswog}, S.
\newblock {The electromagnetic signals of compact binary mergers}.
\newblock {\em Mon. Not. R. Astron. Soc.} {\bf 2013}, {\em 430},~2121--2136,
  \href{http://xxx.lanl.gov/abs/1204.6242}{{\normalfont
  [arXiv:astro-ph.HE/1204.6242]}}.
\newblock
  doi:{\changeurlcolor{black}\href{https://doi.org/10.1093/mnras/stt037}{\detokenize{10.1093/mnras/stt037}}}.

\bibitem[{Hotokezaka} and {Piran}(2015)]{Hotokezaka2015MNRAS}
{Hotokezaka}, K.; {Piran}, T.
\newblock {Mass ejection from neutron star mergers: different components and
  expected radio signals}.
\newblock {\em Mon. Not. R. Astron. Soc.} {\bf 2015}, {\em 450},~1430--1440,
  \href{http://xxx.lanl.gov/abs/1501.01986}{{\normalfont
  [arXiv:astro-ph.HE/1501.01986]}}.
\newblock
  doi:{\changeurlcolor{black}\href{https://doi.org/10.1093/mnras/stv620}{\detokenize{10.1093/mnras/stv620}}}.

\bibitem[{van Eerten} and {MacFadyen}(2011)]{vanEerten2011}
{van Eerten}, H.J.; {MacFadyen}, A.I.
\newblock {Synthetic Off-axis Light Curves for Low-energy Gamma-Ray Bursts}.
\newblock {\em Astrophys. J. Letters} {\bf 2011}, {\em 733},~L37,
  \href{http://xxx.lanl.gov/abs/1102.4571}{{\normalfont
  [arXiv:astro-ph.HE/1102.4571]}}.
\newblock
  doi:{\changeurlcolor{black}\href{https://doi.org/10.1088/2041-8205/733/2/L37}{\detokenize{10.1088/2041-8205/733/2/L37}}}.

\bibitem[{van Eerten} \em{et~al.}(2012){van Eerten}, {van der Horst}, and
  {MacFadyen}]{vanEerten2012}
{van Eerten}, H.; {van der Horst}, A.; {MacFadyen}, A.
\newblock {Gamma-Ray Burst Afterglow Broadband Fitting Based Directly on
  Hydrodynamics Simulations}.
\newblock {\em Astrophys. J.} {\bf 2012}, {\em 749},~44,
  \href{http://xxx.lanl.gov/abs/1110.5089}{{\normalfont
  [arXiv:astro-ph.HE/1110.5089]}}.
\newblock
  doi:{\changeurlcolor{black}\href{https://doi.org/10.1088/0004-637X/749/1/44}{\detokenize{10.1088/0004-637X/749/1/44}}}.

\bibitem[{Lazzati} \em{et~al.}(2017){Lazzati}, {Perna}, {Morsony},
  {L{\'o}pez-C{\'a}mara}, {Cantiello}, {Ciolfi}, {giacomazzo}, and
  {Workman}]{Lazzati2017c}
{Lazzati}, D.; {Perna}, R.; {Morsony}, B.J.; {L{\'o}pez-C{\'a}mara}, D.;
  {Cantiello}, M.; {Ciolfi}, R.; {giacomazzo}, B.; {Workman}, J.C.
\newblock {Late time afterglow observations reveal a collimated relativistic
  jet in the ejecta of the binary neutron star merger GW170817}.
\newblock {\em ArXiv e-prints} {\bf 2017},
  \href{http://xxx.lanl.gov/abs/1712.03237}{{\normalfont
  [arXiv:astro-ph.HE/1712.03237]}}.

\bibitem[{Xie} \em{et~al.}(2018){Xie}, {Zrake}, and {MacFadyen}]{Xie2018}
{Xie}, X.; {Zrake}, J.; {MacFadyen}, A.
\newblock {Numerical simulations of the jet dynamics and synchrotron radiation
  of binary neutron star merger event GW170817/GRB170817A}.
\newblock {\em ArXiv e-prints} {\bf 2018},
  \href{http://xxx.lanl.gov/abs/1804.09345}{{\normalfont
  [arXiv:astro-ph.HE/1804.09345]}}.

\bibitem[{Duffell} \em{et~al.}(2018){Duffell}, {Quataert}, {Kasen}, and
  {Klion}]{Duffell2018}
{Duffell}, P.C.; {Quataert}, E.; {Kasen}, D.; {Klion}, H.
\newblock {Jet Dynamics in Compact Object Mergers: GW170817 Likely had a
  Successful Jet}.
\newblock {\em ArXiv e-prints} {\bf 2018},
  \href{http://xxx.lanl.gov/abs/1806.10616}{{\normalfont
  [arXiv:astro-ph.HE/1806.10616]}}.

\bibitem[{Gill} and {Granot}(2018)]{Gill2018}
{Gill}, R.; {Granot}, J.
\newblock {Afterglow Imaging and Polarization of Misaligned Structured GRB Jets
  and Cocoons: Breaking the Degeneracy in GRB 170817A}.
\newblock {\em Mon. Not. R. Astron. Soc.} {\bf 2018},
  \href{http://xxx.lanl.gov/abs/1803.05892}{{\normalfont
  [arXiv:astro-ph.HE/1803.05892]}}.
\newblock
  doi:{\changeurlcolor{black}\href{https://doi.org/10.1093/mnras/sty1214}{\detokenize{10.1093/mnras/sty1214}}}.

\bibitem[{Nakar} \em{et~al.}(2018){Nakar}, {Gottlieb}, {Piran}, {Kasliwal}, and
  {Hallinan}]{Nakar2018b}
{Nakar}, E.; {Gottlieb}, O.; {Piran}, T.; {Kasliwal}, M.M.; {Hallinan}, G.
\newblock {From $\gamma$ to Radio - The Electromagnetic Counterpart of GW
  170817}.
\newblock {\em ArXiv e-prints} {\bf 2018},
  \href{http://xxx.lanl.gov/abs/1803.07595}{{\normalfont
  [arXiv:astro-ph.HE/1803.07595]}}.

\bibitem[{Zrake} \em{et~al.}(2018){Zrake}, {Xie}, and {MacFadyen}]{Zrake2018}
{Zrake}, J.; {Xie}, X.; {MacFadyen}, A.
\newblock {Radio sky maps of the GRB 170817A afterglow from simulations}.
\newblock {\em ArXiv e-prints} {\bf 2018},
  \href{http://xxx.lanl.gov/abs/1806.06848}{{\normalfont
  [arXiv:astro-ph.HE/1806.06848]}}.

\bibitem[{Granot} \em{et~al.}(2018){Granot}, {De Colle}, and
  {Ramirez-Ruiz}]{Granot2018}
{Granot}, J.; {De Colle}, F.; {Ramirez-Ruiz}, E.
\newblock {Off-axis afterglow light curves and images from 2D hydrodynamic
  simulations of double-sided GRB jets in a stratified external medium}.
\newblock {\em ArXiv e-prints} {\bf 2018},
  \href{http://xxx.lanl.gov/abs/1803.05856}{{\normalfont
  [arXiv:astro-ph.HE/1803.05856]}}.

\bibitem[{Salafia} \em{et~al.}(2017){Salafia}, {Ghisellini}, {Ghirlanda}, and
  {Colpi}]{Salafia2017}
{Salafia}, O.S.; {Ghisellini}, G.; {Ghirlanda}, G.; {Colpi}, M.
\newblock {GRB170817A: a giant flare from a jet-less double neutron-star
  merger?}
\newblock {\em ArXiv e-prints} {\bf 2017},
  \href{http://xxx.lanl.gov/abs/1711.03112}{{\normalfont
  [arXiv:astro-ph.HE/1711.03112]}}.

\bibitem[{Salafia} \em{et~al.}(2018){Salafia}, {Ghisellini}, and
  {Ghirlanda}]{Salafia2018}
{Salafia}, O.S.; {Ghisellini}, G.; {Ghirlanda}, G.
\newblock {Jet-driven and jet-less fireballs from compact binary mergers}.
\newblock {\em Mon. Not. R. Astron. Soc.} {\bf 2018}, {\em 474},~L7--L11,
  \href{http://xxx.lanl.gov/abs/1710.05859}{{\normalfont
  [arXiv:astro-ph.HE/1710.05859]}}.
\newblock
  doi:{\changeurlcolor{black}\href{https://doi.org/10.1093/mnrasl/slx189}{\detokenize{10.1093/mnrasl/slx189}}}.

\bibitem[{Tong} \em{et~al.}(2018){Tong}, {Yu}, and {Huang}]{Tong2018}
{Tong}, H.; {Yu}, C.; {Huang}, L.
\newblock {A magnetically driven origin for the low luminosity GRB 170817A
  associated with GW170817}.
\newblock {\em Research in Astronomy and Astrophysics} {\bf 2018}, {\em
  18},~067,  \href{http://xxx.lanl.gov/abs/1711.06593}{{\normalfont
  [arXiv:astro-ph.HE/1711.06593]}}.
\newblock
  doi:{\changeurlcolor{black}\href{https://doi.org/10.1088/1674-4527/18/6/67}{\detokenize{10.1088/1674-4527/18/6/67}}}.

\bibitem[{Lamb} and {Kobayashi}(2017)]{Lamb2017}
{Lamb}, G.P.; {Kobayashi}, S.
\newblock {GRB 170817A as a jet counterpart to gravitational wave trigger GW
  170817}.
\newblock {\em ArXiv e-prints} {\bf 2017},
  \href{http://xxx.lanl.gov/abs/1710.05857}{{\normalfont
  [arXiv:astro-ph.HE/1710.05857]}}.

\bibitem[{Ziaeepour}(2018)]{Ziaeepour2018}
{Ziaeepour}, H.
\newblock {Prompt gamma-ray emission of GRB 170817A associated with GW 170817:
  a consistent picture}.
\newblock {\em Mon. Not. R. Astron. Soc.} {\bf 2018}, {\em 478},~3233--3252,
  \href{http://xxx.lanl.gov/abs/1801.06124}{{\normalfont
  [arXiv:astro-ph.HE/1801.06124]}}.
\newblock
  doi:{\changeurlcolor{black}\href{https://doi.org/10.1093/mnras/sty1246}{\detokenize{10.1093/mnras/sty1246}}}.

\bibitem[{Veres} \em{et~al.}(2018){Veres}, {M{\'e}sz{\'a}ros}, {Goldstein},
  {Fraija}, {Connaughton}, {Burns}, {Preece}, {Hamburg}, {Wilson-Hodge},
  {Briggs}, and {Kocevski}]{Veres2018}
{Veres}, P.; {M{\'e}sz{\'a}ros}, P.; {Goldstein}, A.; {Fraija}, N.;
  {Connaughton}, V.; {Burns}, E.; {Preece}, R.D.; {Hamburg}, R.;
  {Wilson-Hodge}, C.A.; {Briggs}, M.S.; {Kocevski}, D.
\newblock {Gamma-ray burst models in light of the GRB 170817A - GW170817
  connection}.
\newblock {\em ArXiv e-prints} {\bf 2018},
  \href{http://xxx.lanl.gov/abs/1802.07328}{{\normalfont
  [arXiv:astro-ph.HE/1802.07328]}}.

\bibitem[{Hotokezaka} \em{et~al.}(2018){Hotokezaka}, {Kiuchi}, {Shibata},
  {Nakar}, and {Piran}]{Hotokezaka2018}
{Hotokezaka}, K.; {Kiuchi}, K.; {Shibata}, M.; {Nakar}, E.; {Piran}, T.
\newblock {Synchrotron radiation from the fast tail of dynamical ejecta of
  neutron star mergers}.
\newblock {\em ArXiv e-prints} {\bf 2018},
  \href{http://xxx.lanl.gov/abs/1803.00599}{{\normalfont
  [arXiv:astro-ph.HE/1803.00599]}}.

\bibitem[{Yamazaki} \em{et~al.}(2018){Yamazaki}, {Ioka}, and
  {Nakamura}]{Yamazaki2018}
{Yamazaki}, R.; {Ioka}, K.; {Nakamura}, T.
\newblock {Prompt emission from the counter jet of a short gamma-ray burst}.
\newblock {\em Progress of Theoretical and Experimental Physics} {\bf 2018},
  {\em 2018},~033E01,  \href{http://xxx.lanl.gov/abs/1711.06856}{{\normalfont
  [arXiv:astro-ph.HE/1711.06856]}}.
\newblock
  doi:{\changeurlcolor{black}\href{https://doi.org/10.1093/ptep/pty012}{\detokenize{10.1093/ptep/pty012}}}.

\end{thebibliography}



\end{document}